\documentclass[reprint,superscriptaddress,amsmath,amssymb,aps,floatfix]{revtex4-1}

\usepackage{graphicx}
\usepackage{bm}
\usepackage{hyperref}
\usepackage{graphicx}
\usepackage[usenames, dvipsnames]{color}
\usepackage{multirow}
\usepackage{amsmath,amstext,array}
\usepackage[normalem]{ulem}

\newcolumntype{C}{>{$}c<{$}}

\begin{document}
\title{The Constructive Standard Model: Part I}
\author{Neil Christensen}
\email{nchris3@ilstu.edu}
\affiliation{Department of Physics, Illinois State University, Normal, IL 61790}
\author{Bryan Field}%
\email{bryan.field@farmingdale.edu}
\affiliation{Department of Physics, Farmingdale State College, Farmingdale, NY, 11735}

\date{\today}

\begin{abstract}
In this paper, we construct the complete set of minimal $3$-point vertices for the massive Standard Model (SM) based purely on symmetry principles, mass dimension and high-energy behavior and without any recourse to field theory, gauge symmetries or Feynman rules.  Because the gravitational vertices are no more challenging than any other vertices in this constructive method, we include them as well.  We also calculate the high-energy behavior of these vertices and compare with the well-known massless vertices, both as a check and as a way to pin down the normalization constants.  We include all these vertices in tables as a reference for future investigations.  
\end{abstract}

\maketitle

%\section{Introduction}

Perturbative calculations of the S-Matrix have been experiencing an awesome transformation during the last few decades.  Many of us were trained as high-energy physicists to view the field as the fundamental object in nature with the ``local" or ``gauge" symmetry playing the central role of uniting and constraining the field interactions.  From this perspective, the particle was thought to be no more than a quantum fluctuation of the field.  However, this view of fundamental physics is in the process of being overturned.  The field, rather than being viewed as a fundamental object, is now being considered more of a convenient packaging of the particle operators allowing for a manifestly local, Poincare invariant theory~\cite{Weinberg:1995mt}.  Moreover, for some particles, this field packaging requires the addition of extra, unphysical degrees of freedom.  As a simple example, the photon has only two physical degrees of freedom (which correspond to the positive and negative helicity states), yet the photon must be packaged in a four-component Lorentz vector field. Unfortunately, this packaging does not transform as a Lorentz four-vector.  Instead, under a Lorentz transformation, it transforms into a new Lorentz four vector (as expected) \textit{plus} a derivative term.  In order to prevent this non-vector piece from spoiling the Lorentz invariance of our theory, we are forced to ``gauge" our symmetry such that it is insensitive to this change. Over the past couple of decades, it has become increasingly clear that the field formulation may be the source of much, if not all, of the inefficiency in calculations involving Feynman diagrams~\cite{ArkaniHamed:2008gz}.  Nevertheless, simply being aware of this fact is not useful in deceasing the complexity of intermediate expressions unless the field construction can be replaced with a better system, both in terms of its elegance, its simplicity, and its computational power.  Such a new understanding has been slowly emerging over the last few decades and is presently reaching a major tipping point.   

In order to give context to the present situation, we very briefly consider some of the relevant history. Before the introduction of fields and gauge theories, there were several competing approaches that attempted to understand the properties of, and make predictions about, particle interactions solely in terms of groups, symmetries, and most importantly at the time, analyticity~\cite{Mandelstam:1958xc, Mandelstam:1959bc, Mandelstam:1959bd, Eden:1966dnq}. These methods focused on the S-Matrix and its analytic properties as a function of complex variables. Unfortunately, the original analytic S-Matrix approach lacked real predictive power in the modern sense as these methods mostly relied on Regge theory and dispersion relations. Although some of these techniques, when applied to fundamental particles, are still widely taught in field theory courses, for instance, when dealing with partial-wave decompositions, these methods mostly evolved to become the early versions of string theory~\cite{Rickles:2014fha}, and particle theory would go on to adopt quantum field theory as its main focus, built on its incredible predictive power. This would remain the situation until a new understanding of the interactions of particles emerged.

This charge towards a new understanding of particle physics has been led by the development of ``on-shell" methods and ``twistor" techniques.  A few milestones include the early, systematic use of helicity methods developed to calculate massless QED and QCD scattering amplitudes~\cite{Gastmans:1990xh}. These early forays into these techniques were the first serious, systematic attempts to reduce what is often referred to as the ``analytical bottleneck'' encountered in the then-standardized reduction techniques which showed concrete examples of expressions where the size and complexity of the intermediate expression were under better control. The concept of introducing ``color-ordered'' helicity amplitudes, written in the language of spinors alone, also led to great simplifications in the massless QCD sector of the Standard Model (SM)~\cite{Dixon:1996wi}.  The maximally-helicity-violating tree-level amplitudes had been found to simplify to only one term no matter how many thousands of Feynman diagrams were involved~\cite{Parke:1986gb}.  However, it would be fair to say that although these methods were exciting, they were not deeply employed in large scale calculations, but were rather used to study the structure of specific pieces of non-abelian theories. The BCFW recursion relations were developed~\cite{Britto:2005fq} allowing any tree-level helicity amplitude of gluons to be calculated using a simple on-shell recursion relation.  With all these advancements in scattering amplitudes, bypassing fields and Feynman diagrams entirely ever more frequently, a nascent hope has been reemerging that we might come full circle and ``construct" realistic particle theories completely in terms of the properties of the particles and the unitary S-Matrix~\cite{Elvang:2013cua, Benincasa:2007xk, Conde:2016vxs, ArkaniHamed:2008yf}.  Central to this construction have been unitarity, locality and the transformation properties under Wigner's ``little" group, a subgroup of the Lorentz group~\cite{Wigner:1939cj}.  However, the introduction of massive particles had, until recently, caused major problems in realizing many of the advantages given by the massless methods.

This situation advanced in a fundamental way when~\cite{Arkani-Hamed:2017jhn} showed how to extend the helicity-spinor formalism to particle theories of any mass and any spin, making it possible, in principle, to apply these methods to any fundamental theory of particles. This new formalism extends and simplifies previous attempts at creating a helicity-spinor for massive particles, some of which can be found in Ref.~\cite{Kleiss:1985yh, Ballestrero:1994jn, Dittmaier:1998nn, Schwinn:2005pi, Badger:2005jv, Badger:2005zh, Catani:2002hc, Schwinn:2007ee}. It is ``constructive" in the sense that fields and gauge symmetries are never introduced. Rather, the fundamental building blocks, the minimal $3$-point vertices, are determined purely from their transformation properties under the little group.  These $3$-point vertices are constructed not only from the massless helicity-spinors [objects transforming under both the helicity little group and the SL($2,\mathbb{C}$) Lorentz group], but also from spin-spinors [objects transforming under both the SU($2$) ``spin" little group and the SL($2,\mathbb{C}$) Lorentz group].  The $3$-point vertices are connected to form $4$-point amplitudes by use of unitarity with the assumption that the propagator is always on shell, albeit in complex momentum space.  Furthermore, the allowed vertices are restricted by properties such as locality and the requirement that the high-energy limit of these vertices and amplitudes agrees with massless calculations. Notwithstanding this progress, this constructive approach for all masses and spins is not a complete theory yet.  It has only been clearly defined up to $4$-point amplitudes and up to one loop, whereas Feynman diagrams, although often unwieldy, are capable of calculating any multiplicity scattering amplitude at any loop order, at least in principle.  Furthermore, very few actual calculations have been carried out in this new formalism in the full SM or beyond.  In fact, an explicit set of massive $3$-point vertices for the SM within this formalism is not yet present in the literature, nor have all the $4$-point ``contact" terms for the SM been determined.

In this paper, it is our goal to begin to fill this gap.  We construct the complete set of minimal $3$-point couplings for the SM particles.  Our structure is as follows.  In Sec.~\ref{app:masses={m,m,0}}, we construct the SM vertices with one massless particle and two massive particles of the same mass.  This set already covers an impressive fraction of the SM as it includes the interactions of photons with charged particles, gluons with quarks and gravitons with massive particles.  In Sec.~\ref{sec: 2 massless 1 massive}, we describe the $3$-point vertices with two massless particles and one massive particle which is used to describe the interaction of the $Z$-boson with the neutrinos (which we take to be massless for simplicity, but this can easily be extended for massive neutrinos).  In Sec.~\ref{sec: 1 massless and 2 massive with different masses}, we consider vertices with one massless particle and two massive particles of differing masses, applicable to the interactions of the $W$-boson and the leptons.  In Sec.~\ref{sec: 3 massive}, we find the vertex with all three particles massive.  This structure describes the interaction of the $Z$-boson with massive fermions, the $W$-boson with quarks, the interaction between the $Z$-boson and $W$-bosons and the Higgs 3$-$point vertices.  We leave all the $4$-point ``contact" vertices to a follow-up paper, as they are significantly more involved.  This includes the Higgs $4$-point vertices as well as $4$-point vertices involving Goldstone bosons. In Sec.~\ref{sec:conclusions}, we conclude and describe open questions.  In order to make our results more useful, we organize all our vertices into tables and include the high-energy limit for each vertex.  Our tables are organized as follows.  In Table~\ref{tab:qed}, we present the quantum electrodynamics (QED) vertices, in Table~\ref{tab:qcd}, we display the quantum chromodynamics (QCD) vertices, in Table~\ref{tab:gravitational}, we give the gravitational vertices, in Table~\ref{tab:ew gauge}, we give the electroweak vertices, and in Table~\ref{tab:Higgs}, we give the Higgs vertices.  Finally, we include two appendices for the convenience of the reader.  In App.~\ref{app:conventions}, we give the full details of our conventions, including the full structure of our helicity-spinors and spin-spinors as well as their high-energy limit.  In App.~\ref{app:HE 3-point}, we review the massless $3$-point vertices that must match with the high-energy limit of our massive vertices.  

\begin{widetext}

%%%%%%%%%%%%%%%%%%%%%%%%%%%%%%%%%%%%%%%%%%%%%%%%%%%%%%%%%%
%       Massive - Massive - Massless
%%%%%%%%%%%%%%%%%%%%%%%%%%%%%%%%%%%%%%%%%%%%%%%%%%%%%%%%%%

\section{One Massless and Two Massive Particles of the Same Mass}
\label{app:masses={m,m,0}}
In this section, we construct all the vertices of the SM that contain one massless particle (such as a photon, gluon, or graviton) and two massive particles of the same mass. There are many QED, QCD and gravitational vertices that fall into this category.  Here, we will not concern ourselves with the internal symmetry structures (such as the color coefficients $T^a_{i j}$ and $f^{abc}$ within QCD) as they are already well known.  Instead, we will focus on describing the SL$(2,\mathbb{C})$ spinor part of these vertices and their high-energy limit.  For the reader's convenience, we have included our complete conventions, including our massless helicity-spinors and massive spin-spinors in Appendix~\ref{app:conventions}.

Before we analyze the vertices individually, we briefly comment on the so-called ``$x$"-factor introduced by \cite{Arkani-Hamed:2017jhn}.  In general, in order to build each vertex, we need two linearly independent helicity-spinors. However, in the current case we are considering (two equal mass particles and one massless particle), there is only one linearly independent helicity-spinor among the considered particles -- that of the massless particle.   To overcome this challenge, \cite{Arkani-Hamed:2017jhn} constructs the minimal vertex as,
\begin{equation}
    x^h\frac{\langle\mathbf{12}\rangle^{2S}}{m^{2S-1}},
    \label{eq:app:{m,m,0} vertex def}
\end{equation}
where both of the massive particles (here labelled $1$ and $2$) have the same mass $m$ and spin $S$ (where the spin indices for each are completely symmetrized). They note that $x$ is not uniquely defined, but can be written conveniently as,
\begin{equation}
    x=\frac{\langle \xi | p_2 | 3 \rbrack}
           {m \langle \xi 3 \rangle},
    \label{eq:app:x definition}
\end{equation}
where particle~$3$ will be our massless particle, although either massive particle momenta, $p_1$ or $p_2$, could be used, and $\langle \xi |$ is a helicity-spinor that must be linearly independent from $|3 \rangle$.  Since there are not any linearly independent helicity-spinors among the properties of the particles currently under consideration, $|\xi \rangle$ must be chosen independently of the current particle configuration.  For this reason, it is a spurious degree of freedom and the final scattering amplitude cannot depend on it, but should be chosen judiciously for ease of computation. Ref.~\cite{Arkani-Hamed:2017jhn} points out that it is often convenient to choose it as a helicity-spinor from an external leg that is not directly connected to this vertex, but is part of a larger scattering calculation, such as on the other side of a 4-point amplitude, because then it leads to convenient factorization properties.  We note that ``$x$'' transforms under the little group the same as if particle~$3$ is helcity $+1$.  Therefore, we find that the vertex given in Eq.~(\ref{eq:app:{m,m,0} vertex def}) was constructed to have the right transformation properties since $x$ is raised to the $h$ power and $\langle \mathbf{12} \rangle$ is raised to the $2S$ power. 

Although we see that Eq.~(\ref{eq:app:{m,m,0} vertex def}) has the right transformation properties whether the helicity of the massless particle is positive or negative, we find it convenient to only use it for positive helicity particles.  For negative helicity particles, we find it more convenient to introduce,
\begin{equation}
    \tilde{x} = \frac{\lbrack\xi|p_2|3\rangle}{m\lbrack\xi3\rbrack},
    \label{eq:app:xtilde definition}
\end{equation}
and define the minimal vertex as,
\begin{equation}
    \tilde{x}^{-h}\frac{\lbrack\mathbf{12}\rbrack^{2S}}{m^{2S-1}}.
\end{equation}
As we see, $\tilde{x}$ transforms like particle~$3$ with a helicity of $-1$, therefore this vertex transforms properly as well.  When these vertices are used to construct larger scattering amplitudes, it is important to remember that $\xi$ can be chosen independently for each vertex.  It does not need to be the same for any two vertices in the same amplitude.  But, again, their dependence must cancel at the end of the calculation.

In the vertices below, in order to find the high-energy limit, we will need to expand $x$ and $\tilde{x}$ to linear order in the mass, this will also allow us to identify useful $\xi$ spinors to simplify our results. In other situations, it will become necessary to expand beyond linear order in the mass.  We will do our linear expansion by first inserting $p_2 = |\mathbf{2}\rangle^J\lbrack\mathbf{2}|_J$ [where $J$ is the SU($2$) spin index] to obtain,
\begin{displaymath}
x = \frac{\langle\xi\mathbf{2}\rangle^J\lbrack\mathbf{2}3\rbrack_J}
{m\langle\xi3\rangle}.
\end{displaymath}
Using Eqs.~(\ref{eq:|i>^I expanded in m}) and (\ref{eq:[i|^I expanded in m}) gives,
\begin{eqnarray}
\langle\xi\mathbf{2}\rangle^J\lbrack\mathbf{2}3\rbrack_J &=&
\left[\left(1-\frac{m^2}{4E_2^2}\right)\langle\xi2\rangle\zeta^{-J}+\frac{m}{\sqrt{2E_2}}\langle\xi\zeta_2^-\rangle\zeta^{+J}\right]\nonumber
\times\left[\left(1-\frac{m^2}{4E_2^2}\right)\lbrack23\rbrack\zeta^+_J+\frac{m}{\sqrt{2E_2}}\lbrack\tilde{\zeta}_2^+3\rbrack\zeta^-_J\right] + \mathcal{O}\left(m^3\right)\nonumber\\
&=& \left(1-\frac{m^2}{2E_2^2}\right)\langle\xi2\rangle\lbrack23\rbrack - 
\frac{m^2}{2E_2}\langle\xi\zeta_2^-\rangle\lbrack\tilde{\zeta}_2^+3\rbrack + \mathcal{O}\left(m^3\right), 
\label{eq:<xi2>[23]}
\end{eqnarray}
where $\zeta^{\pm J}=\zeta^{\pm J}(k)$ (see App.~\ref{app:conventions} for the appropriate definitions).
Therefore,
\begin{equation}
x =  \left(1-\frac{m^2}{2E_2^2}\right)\frac{\langle\xi2\rangle\lbrack23\rbrack}
{m\langle\xi3\rangle}
-\frac{m}{2E_2}\frac{\langle\xi\zeta_2^-\rangle\lbrack\tilde{\zeta}_2^+3\rbrack}{\langle\xi3\rangle}
+\mathcal{O}\left(m^2\right),
\label{eq:x expansion}
\end{equation} 
and a similar calculation gives,
\begin{equation}
\tilde{x} =  \left(1-\frac{m^2}{2E_2^2}\right)\frac{\lbrack\xi2\rbrack\langle23\rangle}
{m\lbrack\xi3\rbrack}
-\frac{m}{2E_2}\frac{\lbrack\xi\zeta_2^-\rbrack\langle\tilde{\zeta}_2^+3\rangle}{\lbrack\xi3\rbrack}
+\mathcal{O}\left(m^2\right).
\label{eq:xtilde expansion}
\end{equation}

If it is not already obvious, we can start to see a typical identity where replacing $x \leftrightarrow \tilde{x}$ means replacing $\langle \, \rangle \leftrightarrow \lbrack \, \rbrack$ within each expression leads to a valid expression. We will also need to apply momentum conservation in the following calculations.  It is straight forward in terms of the momenta, which we take to be all incoming.  For example, since all our momenta sum to zero, when we remember that $p_3$ is our massless momentum we find,
\begin{equation}
    |\mathbf{2}\rangle^J\lbrack\mathbf{2}|_J = 
    - |\mathbf{1}\rangle^J\lbrack\mathbf{1}|_J 
    - |3\rangle\lbrack3|.
\end{equation}
However, what we will actually need is the high-energy expansion of this expression to quadratic order in $m$.  Rather than doing this in complete generality, we will expand $\langle 12 \rangle \lbrack 23 \rbrack$, which will appear in the calculations below.  We note that if we only kept terms up to zeroth order in $m$, the result would be identically zero (since $\langle12\rangle\lbrack23\rbrack=-\langle11\rangle\lbrack13\rbrack-\langle13\rangle\lbrack33\rbrack$ but $\langle11\rangle=\lbrack33\rbrack=0$).  However, the quadratic term gives us the desired results.  Thus, the momentum conservation expression can be simplified,
\begin{eqnarray}
    \langle1\mathbf{2}\rangle^J\lbrack\mathbf{2}3\rbrack_J &=& - \langle1\mathbf{1}\rangle^J\lbrack\mathbf{1}3\rbrack_J - \langle13\rangle\lbrack33\rbrack\nonumber\\
    &=& - \langle1\mathbf{1}\rangle^J\lbrack\mathbf{1}3\rbrack_J,
\end{eqnarray}
where we have used $\lbrack33\rbrack=0$.  Next, we expand both sides using Eq. (\ref{eq:<xi2>[23]}) where we identify $\xi$ with $1$.  We obtain,
\begin{equation}
    \langle12\rangle\lbrack23\rbrack = \frac{m_1^2}{2E_1}\langle1\zeta_1^-\rangle\lbrack\tilde{\zeta}_1^+3\rbrack + \frac{m_2^2}{2E_2}\langle1\zeta_2^-\rangle\lbrack\tilde{\zeta}_2^+3\rbrack + \mathcal{O}\left(m^3\right),
    \label{eq:<12>[23] expanded}
\end{equation}
which clearly shows that the expansion starts with terms quadratic in mass as expected. 

\subsection{Massless Spin One Boson With Massive Spin-One-Half Fermions}

\bgroup
\renewcommand{\arraystretch}{2.5}
\begin{table*}
    \centering
    \begin{tabular}{|C|C|c|c|}
    \hline
        \text{Particles} 
      & \text{Coupling} 
      & Vertex 
      & High-Energy Limit (Helicity Signature) \\ \hline
        f \bar{f} \gamma^+
      &  i \sqrt{2}\ e Q_f
      & $\displaystyle
         x \langle \mathbf{12} \rangle$ 
      & $\displaystyle
         \frac{\lbrack 23 \rbrack^2}
              {\lbrack 12 \rbrack} \, 
              \scalebox{0.75}{(\textendash\ +)}, \qquad
        -\frac{\lbrack 31 \rbrack^2}
              {\lbrack 12 \rbrack} \, 
              \scalebox{0.75}{(+ \textendash)}$ \\[5pt] \hline
        f \bar{f} \gamma^-
      &  i \sqrt{2}\ e Q_f
      & $\displaystyle
         \tilde{x} \lbrack \mathbf{12} \rbrack$ 
      & $\displaystyle
         \frac{\langle 31 \rangle^2}
              {\langle 12 \rangle} \, 
              \scalebox{0.75}{(\textendash\ +)}, \qquad
        -\frac{\langle 23 \rangle^2}
              {\langle 12 \rangle} \, 
              \scalebox{0.75}{(+ \textendash)}$ \\[5pt] \hline
        W \bar{W} \gamma^+
      & i \sqrt{2}\ e 
      & $\displaystyle
         \frac{x}
              {M_W} \langle \mathbf{12} \rangle^2$ 
      & $\displaystyle
         \frac{\lbrack 23 \rbrack^3}
              {\lbrack 12 \rbrack \lbrack 31 \rbrack} 
              \scalebox{0.75}{(\textendash\ +)}, \quad
        2\frac{\langle 12 \rangle^3}
              {\langle 23 \rangle \langle 31 \rangle} 
              \scalebox{0.75}{(\textendash\ \textendash)}, \quad
         \frac{\lbrack 31 \rbrack^3}
              {\lbrack 12 \rbrack \lbrack 23 \rbrack} 
              \scalebox{0.75}{(+ \textendash)}, \quad
        -\!\frac{1}{2}\!
         \frac{\lbrack 31 \rbrack \lbrack 23 \rbrack}
              {\lbrack 12 \rbrack} 
              \scalebox{0.75}{(0 0)}$ \\[5pt] \hline
        W \bar{W} \gamma^-
      & i \sqrt{2}\ e 
      & $\displaystyle
         \frac{\tilde{x}}
              {M_W} \lbrack \mathbf{12} \rbrack^2$ 
      & $\displaystyle
         \frac{\langle 31 \rangle^3}
              {\langle 12 \rangle \langle 23 \rangle} 
              \scalebox{0.75}{(\textendash\ +)}, \quad
        2\frac{\lbrack 12 \rbrack^3}
              {\lbrack 23 \rbrack \lbrack 31 \rbrack} 
              \scalebox{0.75}{(+ +)}, \quad
        \frac{\langle 23 \rangle^3}
              {\langle 12 \rangle \langle 31 \rangle} 
              \scalebox{0.75}{(+ \textendash)}, \quad
        -\!\frac{1}{2}\!
         \frac{\langle 31 \rangle \langle 23 \rangle}
              {\langle 12 \rangle} 
              \scalebox{0.75}{(0 0)}$ \\[5pt] \hline
    \end{tabular}
    \caption{QED vertices within the Standard Model and their high-energy limit.  Here, $f$ stands for a fermion while $\bar{f}$ stands for an anti-fermion. The superscript in the first column gives the helicity of the massless particles. The position of the particle determines the number in the last two columns.  Definitions for $x$ and $\tilde{x}$ are in Eqs.~(\ref{eq:app:x definition}) and (\ref{eq:app:xtilde definition}).  See Sec.~\ref{app:masses={m,m,0}} for further details. In the high-energy limit, we show all the terms (including Goldstone-boson terms) that do not vanish at order $(m/E)^0$. The helicity signature is for the massive particles in the high-energy limit.}
    \label{tab:qed}
\end{table*}
\egroup

\bgroup
\renewcommand{\arraystretch}{2.5}
\begin{table}
    \centering
    \begin{tabular}{|C|C|c|c|}
    \hline
        \text{Particles} 
      & \text{Coupling} 
      & Vertex 
      & High-Energy Limit (Helicity Signature) \\[5pt] \hline
        q \bar{q} g^+
      & i \sqrt{2}\ g_s (T^{a_3})^{i_2}_{i_1}
      & $\displaystyle
         x \langle \mathbf{12} \rangle$ 
      & $\displaystyle
         \frac{\lbrack 23 \rbrack^2}
              {\lbrack 12 \rbrack} \, 
              \scalebox{0.75}{(\textendash\ +)}, \qquad
        -\frac{\lbrack 31 \rbrack^2}
              {\lbrack 12 \rbrack} \, 
              \scalebox{0.75}{(+ \textendash)}$ \\[5pt] \hline
        q \bar{q} g^-
      & i \sqrt{2}\ g_s (T^{a_3})^{i_2}_{i_1}
      & $\displaystyle
         \tilde{x} \lbrack \mathbf{12} \rbrack$
      & $\displaystyle
         \frac{\langle 31 \rangle^2}
              {\langle 12 \rangle} \, 
              \scalebox{0.75}{(\textendash\ +)}, \qquad
        -\frac{\langle 23 \rangle^2}
              {\langle 12 \rangle} \, 
              \scalebox{0.75}{(+ \textendash)}$ \\[5pt] \hline
        g^- g^- g^+ 
      & i \sqrt{2}g_s f^{a_1 a_2 a_3}
      & $\displaystyle
         \frac{\langle 12 \rangle^3}
              {\langle 23 \rangle \langle 31 \rangle}$ 
      & Already massless \\[5pt] \hline
        g^+ g^+ g^-
      & i \sqrt{2}g_s f^{a_1 a_2 a_3}
      & $\displaystyle
         \frac{\lbrack 12 \rbrack^3}
              {\lbrack 23 \rbrack \lbrack 31 \rbrack}$ 
      & Already massless \\[5pt] \hline
    \end{tabular}
    \caption{QCD vertices within the Standard Model along with their high-energy limit.  Here, $q$ stands for a quark while $\bar{q}$ stands for an anti-quark. The superscript in the first column gives the helicity of the massless particles.  The position of the particle determines the number in the last two columns. Definitions for $x$ and $\tilde{x}$ are in Eqs.~(\ref{eq:app:x definition}) and (\ref{eq:app:xtilde definition}).  See Sec.~\ref{app:masses={m,m,0}} for further details. In the high-energy limit, we show all the terms that do not vanish at order $(m/E)^0$. The helicity signature is for the massive particles in the high-energy limit.}
    \label{tab:qcd}
\end{table}
\egroup

We now move on to constructing a specific case.  We begin with a massless $\pm 1$-helicity boson interacting with two $1/2$-spin fermions which will give us the vertex structure for a photon interacting with two charged fermions and also for a gluon interacting with two quarks.  We first consider the case of particle~$3$ being $+1$-helicity, therefore the vertex is simply given by,
\begin{equation}
    x\langle\mathbf{12}\rangle.
    \label{eq:QQg:x<12>}
\end{equation}
We will check this result by comparing with the vertex when all three particles are massless by taking the high-energy limit and then keeping terms up to zeroth order in the mass (the massless limit).  For convenience, we have included a review of the massless vertices in App.~\ref{app:HE 3-point}.  The expression $\langle \mathbf{12} \rangle$ has two SU($2$) indices and can be written as a matrix.  We have already described and expanded this matrix in the high-energy limit in Appendix~\ref{app:conventions}.  Therefore using Eqs. (\ref{eq:<ij> expanded in m}), (\ref{eq:x expansion}) and (\ref{eq:<12>[23] expanded}), we obtain,
\begin{equation} \displaystyle
x\langle\mathbf{12}\rangle =
  \left(
  \begin{array}{cc} \displaystyle
    0 & \displaystyle
        \frac{1}{\sqrt{2E_2}}
        \frac{\langle 1\zeta^-_2 \rangle \langle \xi 2 \rangle \lbrack 23 \rbrack}
             {\langle \xi 3 \rangle} \\ \displaystyle
    \frac{1}{\sqrt{2E_1}}
    \frac{\langle \zeta^-_1 2 \rangle \langle \xi 2 \rangle \lbrack 23 \rbrack}
         {\langle\xi3\rangle} &  0
  \end{array}
  \right)  + \mathcal{O}(m_f).
\end{equation}
We can further simplify these terms by applying momentum conservation.  In the top right entry, we can set $\langle\xi2\rangle\lbrack23\rbrack=-\langle\xi1\rangle\lbrack13\rbrack +\mathcal{O}(m_f^2)$.  We then follow this with $\langle\zeta_2^-1\rangle\lbrack13\rbrack=-\langle\zeta_2^-2\rangle\lbrack23\rbrack+\mathcal{O}(m_f^2)$, again on the top-right term, along with $\langle\zeta_1^-2\rangle\lbrack23\rbrack=-\langle\zeta_1^-1\rangle\lbrack13\rbrack+\mathcal{O}(m_f^2)$ on the bottom-left term to obtain,
\begin{equation}
x\langle\mathbf{12}\rangle =
   \left(
   \begin{array}{cc}
    0 & \displaystyle 
        \frac{1}{\sqrt{2E_2}}
        \frac{\langle 2\zeta^-_2\rangle\langle\xi1\rangle\lbrack23\rbrack}
             {\langle\xi3\rangle} \\ \displaystyle
    \frac{1}{\sqrt{2E_1}}
    \frac{\langle \zeta^-_1 1\rangle\langle\xi2\rangle\lbrack31\rbrack}
         {\langle\xi3\rangle} & 0
   \end{array}
   \right)
   +\mathcal{O}(m_f).
\end{equation}
Now, using the fact that $\langle2\zeta_2^-\rangle=\sqrt{2E_2}$ and similarly for particle~$1$, we have,
\begin{equation}
x\langle\mathbf{12}\rangle =
\left(\begin{array}{cc}
    0 & \displaystyle 
    \frac{\langle\xi1\rangle\lbrack23\rbrack}{\langle\xi3\rangle} \\ \displaystyle 
    \frac{\langle\xi2\rangle\lbrack31\rbrack}{\langle\xi3\rangle} &
    0
    \end{array}\right)
    +\mathcal{O}(m_f).
\end{equation}
Next, multiplying by $\lbrack32\rbrack/\lbrack32\rbrack$ on the top right and $\lbrack31\rbrack/\lbrack31\rbrack$ on the bottom left, and using momentum conservation in the denominator, $\langle\xi3\rangle\lbrack32\rbrack=-\langle\xi1\rangle\lbrack12\rbrack+\mathcal{O}(m_f^2)$ and $\langle\xi3\rangle\lbrack31\rbrack=-\langle\xi2\rangle\lbrack21\rbrack+\mathcal{O}(m_f^2)$, we finally cancel the dependence on $\xi$ to obtain,
\begin{equation}
x\langle\mathbf{12}\rangle =
\left(\begin{array}{cc}
    0 & \displaystyle 
    \frac{\lbrack23\rbrack^2}{\lbrack12\rbrack} \\ \displaystyle 
    -\frac{\lbrack31\rbrack^2}{\lbrack12\rbrack} &
    0
    \end{array}\right)
    +\mathcal{O}(m_f).
    \label{eq:x<12> HE limit}
\end{equation}
We remind the reader that $|i \rbrack$ transforms as a $+1/2$-helicity particle while $1/|i\rbrack$ transforms as a $-1/2$-helicity particle (since the transformation is a simple phase).  The angle brackets have the opposite transformation properties.  From this, we can see that the upper right of this matrix corresponds with the quark having $-1/2$-helicity  and the anti-quark having $+1/2$-helicity.  The bottom left of this matrix correspond with the opposite helicities for the fermions (but, of course, the same $+1$-helicity for the massless photon or gluon).  Both these high-energy-limit results agree with the massless vertices given on the left of Eq.~(\ref{eq:massless:1/2,-1/2,1}).  We note that although the signs of the massless vertices are not fixed by the transformation properties alone, here the relative signs of these two vertices \textit{is} fixed by the transformation properties of the massive vertex.  Furthermore, we note that the top-left entry corresponds with both fermions having $-1/2$-helicity while the photon or gluon has $+1$-helicity.  We find zero in the high-energy limit which is exactly what we expect from the massless vertices as reviewed in Appendix~\ref{app:HE 3-point}.  Altogether, we obtain,
\begin{equation}
x\langle\mathbf{12}\rangle =
\left[\begin{array}{cc}\displaystyle
    \mathcal{A}\left(-\frac{1}{2},-\frac{1}{2},+1\right) & \displaystyle 
    \mathcal{A}\left(-\frac{1}{2},+\frac{1}{2},+1\right) \\ \displaystyle 
    -\mathcal{A}\left(+\frac{1}{2},-\frac{1}{2},+1\right) & \displaystyle
    \mathcal{A}\left(+\frac{1}{2},+\frac{1}{2},+1\right)
    \end{array}\right]
    +\mathcal{O}(m_f),
\end{equation}
where the massless vertices $\mathcal{A}(h_1,h_2,h_3)$ are given explicitly in Appendix~\ref{app:HE 3-point}.

On the other hand, when the massless particle has $-1$-helicity, we have the vertex,
\begin{equation}
    \tilde{x}\lbrack\mathbf{12}\rbrack
    \label{eq:QQg:xtilde[12]}
\end{equation}
and the high-energy limit is obtained by an analogous set of steps, exchanging angle and square brackets and raising the spin indices leading to the final result,
\begin{equation}
\tilde{x}\lbrack\mathbf{12}\rbrack =
\left(\begin{array}{cc}
    0 & \displaystyle 
    \frac{\langle31\rangle^2}{\langle12\rangle} \\ \displaystyle 
    -\frac{\langle23\rangle^2}{\langle12\rangle} &
    0
    \end{array}\right)
    +\mathcal{O}(m_f).
    \label{eq:xtilde[12] HE limit}
\end{equation}
The upper-right term corresponds with a $-1/2$-helicity fermion and a $+1/2$-helicity anti-fermion, the bottom-left term is the opposite case, and both have a $-1$-helicity photon or gluon.  These high-energy-limit expressions agree with the massless vertices given on the right of Eq.~(\ref{eq:massless:1/2,-1/2,1}).  Again, the relative sign is fixed by the massive vertex structure whereas it is not fixed by the massless vertices themselves.  Making the helicity structure explicit as we did for the $+1$-helicity case, we have here,
\begin{equation}
\tilde{x}\lbrack\mathbf{12}\rbrack =
\left[\begin{array}{cc}\displaystyle
    \mathcal{A}\left(-\frac{1}{2},-\frac{1}{2},-1\right) & \displaystyle 
    \mathcal{A}\left(-\frac{1}{2},+\frac{1}{2},-1\right) \\ \displaystyle 
    -\mathcal{A}\left(+\frac{1}{2},-\frac{1}{2},-1\right) & \displaystyle
    \mathcal{A}\left(+\frac{1}{2},+\frac{1}{2},-1\right)
    \end{array}\right]
    +\mathcal{O}(m_f),
\end{equation}
where the massless vertices are zero when the sum of the helicities is not $\pm1$ as expected. We see that the helicity combinations of the fermions are in the same spin locations as the $-1$-helicity photon as they should.

We have included these vertices, along with their high-energy limit, in Table~\ref{tab:qed} for QED and Table~\ref{tab:qcd} for QCD.

\subsection{Massless Spin One Boson With Massive Spin One Bosons}
We next work out the $3$-point vertex for one massless $\pm 1$-helicity particle with two massive $1$-spin particles.  This vertex is appropriate for the $W\bar{W}$-photon vertex.  We begin with the $W\bar{W}\gamma^+$ vertex connecting a $+$-helicity photon and two $1$-spin particles of the same mass.  We have,
\begin{equation}
    x\frac{\langle\mathbf{12}\rangle^2}{M_W}.
    \label{eq:x<12>^2/MW}
\end{equation}
There are two independent indices on this vertex.  They are the spin index of particle~$1$ and particle~$2$.  Each is a symmetric combination of two $1/2$-spin indices on each spin-spinor.  There are three symmetric combinations.  For example, for the first $W$-boson, the indices could take the values $\langle\mathbf{1}|^1\langle\mathbf{1}|^1$, $\left(\langle\mathbf{1}|^1\langle\mathbf{1}|^2+\langle\mathbf{1}|^2\langle\mathbf{1}|^1\right)/2$, or $\langle\mathbf{1}|^2\langle\mathbf{1}|^2$.  As a result, we can write this vertex as a 3x3 matrix.  We have already worked out this matrix in Appendix~\ref{app:conventions} as well as its high-energy limit.
Using Eqs. (\ref{eq:<12>^2 expanded in m}), (\ref{eq:x expansion}) and (\ref{eq:<12>[23] expanded}), we have to leading order in $M_W$,
\begin{equation} \nonumber
x\frac{\langle\mathbf{12}\rangle^2}{M_W} =
  \left(
  \begin{array}{ccc} \displaystyle
    \frac{ \langle \xi 2 \rangle
           \langle 12 \rangle
           \langle 1\zeta_1^- \rangle
           \lbrack \tilde{\zeta}_1^+ 3 \rbrack}
         {2E_1\langle\xi3\rangle} 
   +\frac{\langle\xi2\rangle
          \langle12\rangle\langle1\zeta_2^-\rangle
          \lbrack\tilde{\zeta}_2^+3\rbrack}
          {2E_2\langle\xi3\rangle} &
    0 & \displaystyle
    \frac{\langle1\zeta_2^-\rangle^2\langle\xi2\rangle\lbrack23\rbrack}
                          {E_2\langle\xi3\rangle} \\
    0 & \displaystyle
    \frac{\langle1\zeta_2^-\rangle\langle\zeta_1^-2\rangle\langle\xi2\rangle\lbrack23\rbrack}
         {2\sqrt{4E_1E_2}\langle\xi3\rangle} & 0 \\
    \displaystyle
    \frac{\langle\zeta_1^-2\rangle^2\langle\xi2\rangle\lbrack23\rbrack}
         {E_1\langle\xi3\rangle} & 0 & 0
  \end{array}
  \right)
  +\mathcal{O}(M_W).
\end{equation}
Now, focusing on the top-right term, we use that $\langle\xi2\rangle\lbrack23\rbrack=-\langle\xi1\rangle\lbrack13\rbrack+\mathcal{O}(M_W^2)$ followed by $-\langle1\zeta_2^-\rangle\lbrack13\rbrack=\langle\zeta_2^-1\rangle\lbrack13\rbrack=-\langle\zeta_2^-2\rangle\lbrack23\rbrack+\mathcal{O}(M_W^2)=\sqrt{2E_2}\lbrack23\rbrack+\mathcal{O}(M_W^2)$.  Next, we multiply by $\lbrack13\rbrack/\lbrack13\rbrack$ and use $\langle1\zeta_2^-\rangle\lbrack13\rbrack=-\sqrt{2E_2}\lbrack23\rbrack+\mathcal{O}(M_W^2)$.  Finally, we multiply by $\lbrack23\rbrack/\lbrack23\rbrack$ and use $\langle\xi3\rangle\lbrack23\rbrack=-\langle\xi1\rangle\lbrack21\rbrack+\mathcal{O}(M_W^2)$ in the denominator so that the $\langle\xi1\rangle$ cancels between the numerator and denominator.  With this, we have, $\lbrack23\rbrack^3/(\lbrack12\rbrack\lbrack31\rbrack)$ for the top-right entry.  The other entries are simplified by a similar set of steps.  We will go through one more in detail for the convenience of the reader.  Let's consider the center term.  We use $\langle\zeta_1^-2\rangle\lbrack23\rbrack=-\langle\zeta_1^-1\rangle\lbrack13\rbrack+\mathcal{O}(M_W^2)=\sqrt{2E_1}\lbrack13\rbrack+\mathcal{O}(M_W^2)$, followed by $\langle1\zeta_2^-\rangle\lbrack13\rbrack=-\langle2\zeta_2^-\rangle\lbrack23\rbrack+\mathcal{O}(M_W^2)=-\sqrt{2E_2}\lbrack23\rbrack+\mathcal{O}(M_W^2)$.  We are now left with $-\langle\xi2\rangle\lbrack23\rbrack/(2\langle\xi3\rangle)+\mathcal{O}(M_W^2)$.  We multiply this by $\lbrack31\rbrack/\lbrack31\rbrack$ and use $\langle\xi3\rangle\lbrack31\rbrack=-\langle\xi2\rangle\lbrack21\rbrack+\mathcal{O}(M_W^2)$ at which point the $\langle\xi2\rangle$ cancels between the numerator and denominator and we are left with $-\lbrack23\rbrack\lbrack31\rbrack/(2\lbrack12\rbrack)+\mathcal{O}(M_W^2)$.  The top-left and bottom-left are obtained through a similar set of manipulations that include conservation of momentum and multiplication by appropriate forms of $1$.  Finally, we obtain,
\begin{equation}
    x\frac{\langle\mathbf{12}\rangle^2}{M_W} =
    \left(\begin{array}{ccc} \displaystyle
    2\frac{\langle12\rangle^3}{\langle23\rangle\langle31\rangle} & 0 &
    \displaystyle
    \frac{\lbrack23\rbrack^3}{\lbrack12\rbrack\lbrack31\rbrack} \\
    0 & \displaystyle
        -\frac{1}{2}\frac{\lbrack23\rbrack\lbrack31\rbrack}{\lbrack12\rbrack} & 0 \\
    \displaystyle
    \frac{\lbrack31\rbrack^3}{\lbrack12\rbrack\lbrack23\rbrack} & 0 & 0
    \end{array}\right)
    +\mathcal{O}(M_W).
    \label{eq:x<12>^2/MW HE}
\end{equation}
We see that the top-right term contributes when the helicity of $W_1$ is $-1$ and that of $W_2$ is $+1$ (where the sub-script obviously refers to the multiplicity of the $W$-bosons), while the bottom-left term is for exactly the opposite helicity combination for the $W$-bosons.  The top-left term is for both $W$'s having helicity $-1$.  Interestingly, the middle term corresponds with both $W$'s having helicity-$0$, namely, scattering of the Goldstone bosons, as it must. Explicitly seeing the contributions from the Goldstone bosons speaks to the power and simplicity of this method.  

All of these vertices agree perfectly with the massless vertices of Appendix~\ref{app:HE 3-point} up to an overall factor.  As for the fermion interactions of the previous subsection, we see that although the massless vertices do not fix the relative signs and factors of 2 purely based on transformation properties, they are fixed by their inclusion in the massive vertices.  The top-right and bottom-left entries are given by the left side of Eq.~(\ref{eq:3-point gluon amplitude}), the top-left entry is given by the right side of Eq.~(\ref{eq:3-point gluon amplitude}) and the center entry is given by the left side of Eq.~(\ref{eq:massless:0,0,1}). Writing this in terms of the massless vertices of App.~\ref{app:HE 3-point} gives,
\begin{equation}
    x\frac{\langle\mathbf{12}\rangle^2}{M_W} =
    \left[\begin{array}{ccc} \displaystyle
      2\mathcal{A}(-1,-1,+1) &  \displaystyle
       \mathcal{A}(-1,0,+1)  &  \displaystyle
       \mathcal{A}(-1,+1,+1) \\ \displaystyle
       \mathcal{A}(0,-1,+1)  & \displaystyle
      -\frac{1}{2}
       \mathcal{A}(0,0,+1)   & \displaystyle
       \mathcal{A}(0,+1,+1)   \\ \displaystyle
       \mathcal{A}(+1,-1,+1)  &  \displaystyle
       \mathcal{A}(+1,0,+1)   &  \displaystyle
      2\mathcal{A}(+1,+1,+1)
     \end{array}
    \right] + \mathcal{O}(M_W).
\end{equation}
As expected, all the helicity combinations that do not sum to $\pm1$ are zero in this limit.  

On the other hand, the $WW\gamma^-$ vertex is,
\begin{equation}
    \tilde{x}\frac{\lbrack\mathbf{12}\rbrack^2}{M_W}.
    \label{eq:xtilde[12]^2/MW}
\end{equation}
Following a similar set of steps and after raising the spin indices, we find that at leading order in the high-energy limit, this reduces to
\begin{eqnarray}
    \tilde{x}\frac{\lbrack\mathbf{12}\rbrack^2}{M_W} &=&
    \left(\begin{array}{ccc} \displaystyle
    0 & 0 & 
    \displaystyle
    \frac{\langle31\rangle^3}{\langle12\rangle\langle23\rangle} \\
    0 & \displaystyle
        -\frac{1}{2}
         \frac{\langle23\rangle\langle31\rangle}{\langle12\rangle} & 0 \\
    \displaystyle
    \frac{\langle23\rangle^3}{\langle12\rangle\langle31\rangle} & 0 & \displaystyle
    2\frac{\lbrack12\rbrack^3}{\lbrack23\rbrack\lbrack31\rbrack}
    \end{array}\right)
    +\mathcal{O}(M_W)
    \label{eq:xtilde[12]^2/MW HE}\\
    &=& 
    \left[
    \begin{array}{ccc} \displaystyle
     2\mathcal{A}(-1,-1,-1) &  \displaystyle
      \mathcal{A}(-1,0,-1)  &  \displaystyle
      \mathcal{A}(-1,+1,-1) \\ \displaystyle
      \mathcal{A}(0,-1,-1)  &  \displaystyle
     -\frac{1}{2}
      \mathcal{A}(0,0,-1)   &  \displaystyle
      \mathcal{A}(0,+1,-1)  \\ \displaystyle
      \mathcal{A}(+1,-1,-1) &  \displaystyle
      \mathcal{A}(+1,0,-1)  &  \displaystyle
      2\mathcal{A}(+1,+1,-1)
    \end{array}
    \right] + \mathcal{O}(M_W).
\end{eqnarray}
Once again, all the vertices agree with the massless vertices in the massless limit.  Once again, the relative factors are fixed by the massive vertex that they come from. Once again, the helicity structure of the high-energy limit is the same as for the $+1$-helicity-photon case.  In fact, we see a general trend that after the spin indices have been raised, the helicities begin at their lowest value at the top left and increase to their highest values at the bottom right. In fact, we see that in the convention we are following, the spin component is the component along the direction of motion.  It is the helicity of the particle.

We have included these vertices and their high-energy limit in Table~\ref{tab:qed}.

\subsection{Gravitational Vertices}

\bgroup
\renewcommand{\arraystretch}{2.5}
\begin{table}
    \centering
    \begin{tabular}{|C|C|c|c|}
    \hline
        \text{Particles} 
      & \text{Coupling} 
      & Vertex 
      & High-Energy Limit (Helicity Signature) \\[5pt] \hline
        h h G^+
      & \displaystyle \frac{i}{M_P}
      & $\displaystyle
         x^2 m_h^2$
      & $\displaystyle
         \left(\frac{\lbrack 23 \rbrack \lbrack 31 \rbrack}
                    {\lbrack 12 \rbrack}
         \right)^2$ \\[5pt] \hline
        h h G^-
      & \displaystyle \frac{i}{M_P}
      & $\displaystyle
         \tilde{x}^2 m_h^2$
      & $\displaystyle
         \left(\frac{\langle 23 \rangle \langle 31 \rangle}
                    {\langle 12 \rangle}
         \right)^2$ \\[5pt] \hline
        f \bar{f} G^+
      & \displaystyle \frac{i}{M_P}
      & $\displaystyle
         x^2 m_f \langle \mathbf{12} \rangle$
      & $\displaystyle
         \frac{\lbrack 23 \rbrack^3  \lbrack 31 \rbrack}
              {\lbrack 12 \rbrack^2} 
              \scalebox{0.75}{(\textendash\ +)}, \quad 
        -\frac{\lbrack 31 \rbrack^3  \lbrack 23 \rbrack}
              {\lbrack 12 \rbrack^2} 
              \scalebox{0.75}{(+ \textendash)}$ \\[5pt] \hline
        f \bar{f} G^-
      & \displaystyle \frac{i}{M_P}
      & $\displaystyle
         \tilde{x}^2 m_f \lbrack \mathbf{12} \rbrack$
      & $\displaystyle
         \frac{\langle 31 \rangle^3  \langle 23 \rangle}
              {\langle 12 \rangle^2} 
              \scalebox{0.75}{(\textendash\ +)}, \quad 
        -\frac{\langle 23 \rangle^3  \langle 31 \rangle}
              {\langle 12 \rangle^2} 
              \scalebox{0.75}{(+ \textendash)}$ \\[5pt] \hline
        V \bar{V} G^+
      & \displaystyle \frac{i}{M_P}
      & $\displaystyle
         x^2 \langle \mathbf{12} \rangle^2$
      & $\displaystyle
         \frac{\lbrack 23 \rbrack^4}
              {\lbrack 12 \rbrack^2} 
              \scalebox{0.75}{(\textendash\ +)}, \quad 
        -\!\frac{1}{2}\!
         \frac{\lbrack 31 \rbrack^2  \lbrack 23 \rbrack^2}
              {\lbrack 12 \rbrack^2} 
              \scalebox{0.75}{(0 0)}, \quad 
         \frac{\lbrack 31 \rbrack^4}
              {\lbrack 12 \rbrack^2} 
              \scalebox{0.75}{(+ \textendash)}$ \\[5pt] \hline
        V \bar{V} G^-
      & \displaystyle \frac{i}{M_P}
      & $\displaystyle
         \tilde{x}^2 \lbrack \mathbf{12} \rbrack^2$
      & $\displaystyle
         \frac{\langle 31 \rangle^4}
              {\langle 12 \rangle^2} 
              \scalebox{0.75}{(\textendash\ +)}, \quad 
        -\frac{1}{2}
         \frac{\langle 31 \rangle^2  \langle 23 \rangle^2}
              {\langle 12 \rangle^2} 
              \scalebox{0.75}{(0 0)}, \quad 
        \frac{\langle 23 \rangle^4}
              {\langle 12 \rangle^2} 
              \scalebox{0.75}{(+ \textendash)}$ \\[5pt] \hline
        \gamma^+ \gamma^- G^+
      & \multirow{2}{*}{$\displaystyle \frac{i}{M_P}$}
      & \multirow{2}{*}{$\displaystyle
         \frac{\lbrack31\rbrack^4}{\lbrack12\rbrack^2}$}
      & \multirow{2}{*}{Already massless} \\
      g^+ g^- G^+ &&&\\[5pt]\hline
        \gamma^+ \gamma^- G^-
      & \multirow{2}{*}{$\displaystyle \frac{i}{M_P}$}
      & \multirow{2}{*}{$\displaystyle
         \frac{\langle31\rangle^4}{\langle12\rangle^2}$}
      & \multirow{2}{*}{Already massless} \\
      g^+ g^- G^- &&&\\[5pt] \hline
        G^- G^- G^+
      & \displaystyle \frac{i}{M_P}
      & $\displaystyle
         \frac{\langle 12 \rangle^6}
              {\langle 23 \rangle^2 \langle 31 \rangle^2}$
      & Already massless \\[5pt] \hline
        G^+ G^+ G^-
      & \displaystyle \frac{i}{M_P}
      & $\displaystyle
         \frac{\lbrack 12 \rbrack^6}
              {\lbrack 23 \rbrack^2 \lbrack 31 \rbrack^2}$ 
      & Already massless \\[5pt] \hline
    \end{tabular}
    \caption{Gravitational Vertices along with their high-energy limit. Here, $f$ stands for a fermion while $\bar{f}$ stands for an anti-fermion, and $V$ and $\bar{V}$ stand for a $1$-spin boson and its anti-particle, respectively.  Also, $h$ stands for the Higgs boson. The superscript in the first column gives the helicity of the massless particles.  The position of the particle determines the number in the last two columns. Definitions for $x$ and $\tilde{x}$ are in Eqs.~(\ref{eq:app:x definition}) and (\ref{eq:app:xtilde definition}).  See Sec.~\ref{app:masses={m,m,0}} for further details. The helicity signature is for the massive particles in the high-energy limit.}
    \label{tab:gravitational}
\end{table}
\egroup

One of the truly great things about the constructive formalism is that gravitational interactions are no more complicated than any other interactions (at least before renormalization is considered).  In this subsection, we work the gravitational vertices of the SM out.  There is now an extra factor of $m/M_P$ in each vertex, where $M_P$ is the Planck mass.  Therefore, the coupling of two fermions to a graviton is given by,
\begin{equation}
    \frac{1}{M_P}x^2m_f\langle\mathbf{12}\rangle 
    \quad\mbox{and}\quad 
    \frac{1}{M_P}\tilde{x}^2m_f\lbrack\mathbf{12}\rbrack
    \label{eq:x^2mf<12>}
\end{equation}
for helicity $+2$ and $-2$, respectively.  We must now determine the high-energy behavior of this vertex.  We already know the behavior of $x\langle\mathbf{12}\rangle$ [see Eq.~(\ref{eq:x<12> HE limit})].  We must now calculate the high-energy behavior of $m_f x$,
\begin{equation}
    m_fx = \frac{\langle\xi2\rangle\lbrack23\rbrack}
{\langle\xi3\rangle}.
\end{equation}
Multiplying by $\lbrack21\rbrack/\lbrack21\rbrack$ and using momentum conservation, $\langle\xi2\rangle\lbrack21\rbrack=-\langle\xi3\rangle\lbrack31\rbrack+\mathcal{O}(m_f^2)$, we have,
\begin{equation}
    m_fx = \frac{\lbrack23\rbrack\lbrack31\rbrack}{\lbrack12\rbrack}
    +\mathcal{O}(m_f^2),
\end{equation}
resulting in,
\begin{equation}
    x^2m_f\langle\mathbf{12}\rangle =
    \left(\begin{array}{cc}
         0 & \displaystyle
        \frac{\lbrack23\rbrack^3\lbrack31\rbrack}{\lbrack12\rbrack^2} \\
        \displaystyle
        -\frac{\lbrack31\rbrack^3\lbrack23\rbrack}{\lbrack12\rbrack^2} & 0 
    \end{array}\right)
    +\mathcal{O}(m_f),
    \label{eq:x^2mf<12> HE}
\end{equation}
which has the right little-group transformation properties.  Both of these transform like a $+2$-helicity particle~$3$ (the graviton) while the top right has a $-1/2$-helicity particle~$1$ and $+1/2$-helicity particle~$2$.  The bottom left has the opposite helicities for particles $1$ and $2$.  They both agree with the massless vertices found on the left side of Eq.~(\ref{eq:A(1/2,-1/2,2)}).  As expected, when the helicities do not add up to $\pm2$ (for a graviton), the massless vertex is zero.
\begin{equation}
    \frac{x^2m_f}{M_P}\langle\mathbf{12}\rangle = 
    \left[\begin{array}{cc} \displaystyle
         \mathcal{A}\left(-\frac{1}{2},-\frac{1}{2},+2\right) & \displaystyle
        \mathcal{A}\left(-\frac{1}{2},+\frac{1}{2},+2\right) \\
        \displaystyle
        -\mathcal{A}\left(+\frac{1}{2},-\frac{1}{2},+2\right) & \displaystyle
        \mathcal{A}\left(+\frac{1}{2},+\frac{1}{2},+2\right)
    \end{array}\right]
    +\mathcal{O}(m_f).
\end{equation}

Similarly, for a $-2$-helicity graviton, after raising the spin indices we find,
\begin{eqnarray}
    \frac{\tilde{x}^2m_f}{M_P}\lbrack\mathbf{12}\rbrack &=&
    \frac{1}{M_P}\left(\begin{array}{cc}
         0 & \displaystyle
        \frac{\langle31\rangle^3\langle23\rangle}{\langle12\rangle^2} \\
        \displaystyle
        -\frac{\langle23\rangle^3\langle31\rangle}{\langle12\rangle^2} & 0 
    \end{array}\right)
    +\mathcal{O}(m_f)
    \label{eq:xtilde^2mf[12] HE}\\
    &=& \left[\begin{array}{cc}\displaystyle
         \mathcal{A}\left(-\frac{1}{2},-\frac{1}{2},-2\right) & \displaystyle
        \mathcal{A}\left(-\frac{1}{2},+\frac{1}{2},-2\right) \\
        \displaystyle
        -\mathcal{A}\left(+\frac{1}{2},-\frac{1}{2},-2\right) & \displaystyle
        \mathcal{A}\left(+\frac{1}{2},+\frac{1}{2},-2\right)
    \end{array}\right]
    +\mathcal{O}(m_f),
\end{eqnarray}
in agreement with the massless vertex on the right side of Eq.~(\ref{eq:A(1/2,-1/2,2)}).

For $1$-spin bosons (with mass $M_V$), the vertices are given by,
\begin{equation}
    \frac{1}{M_P}x^2\langle\mathbf{12}\rangle^2
    \quad\mbox{and}\quad
    \frac{1}{M_P}\tilde{x}^2\lbrack\mathbf{12}\rbrack^2,
    \label{eq:x^2<12>^2}
\end{equation}
for a graviton with helicity $+2$ and $-2$, respectively.  We have already determined the high-energy behavior of $x\langle\mathbf{12}\rangle$.  All we need to do is square it and symmetrize the indices on particles 1 and 2 to obtain,
\begin{eqnarray}
    \frac{x^2}{M_P}\langle\mathbf{12}\rangle^2 &=&
    \frac{1}{M_P}\left(\begin{array}{ccc} \displaystyle
         0 & 0 & \displaystyle
                 \frac{\lbrack23\rbrack^4}{\lbrack12\rbrack^2}  \\
         0 & \displaystyle
             -\frac{1}{2}
              \frac{\lbrack31\rbrack^2\lbrack23\rbrack^2}{\lbrack12\rbrack^2} & 0 \\
         \displaystyle
         \frac{\lbrack31\rbrack^4}{\lbrack12\rbrack^2} & 0 & 0
    \end{array}\right) 
    +\mathcal{O}(M_V)
    \label{eq:x^2<12>^2 HE}\\
    &=& \left[\begin{array}{ccc} \displaystyle
         \mathcal{A}\left(-1,-1,+2\right) & \displaystyle
         \mathcal{A}\left(-1,0,+2\right) & \displaystyle
         \mathcal{A}\left(-1,+1,+2\right)  \\ \displaystyle
         \mathcal{A}\left(0,-1,+2\right) & \displaystyle
         -\frac{1}{2}\mathcal{A}\left(0,0,+2\right) & \displaystyle
         \mathcal{A}\left(0,+1,+2\right) \\
         \displaystyle
         \mathcal{A}\left(+1,-1,+2\right) & \displaystyle
         \mathcal{A}\left(+1,0,+2\right)& \displaystyle
         \mathcal{A}\left(+1,+1,+2\right)
    \end{array}\right]
    +\mathcal{O}(M_V),
\end{eqnarray}
where we have included the reference to the massless vertices of Appendix~\ref{app:HE 3-point}.  We see that the top-right term corresponds with a $-1$-helicity particle~$1$ and a $+1$-helicity particle~$2$ while the bottom-left term has the opposite helicities for particles 1 and 2.  Both agree with the massless vertices on the left side of Eq.~(\ref{eq:A(1,-1,2)}).   The center term corresponds with the helicity-$0$ components of the $1$-spin particles, the Goldstone bosons.  It also agrees with the massless vertices as seen on the left side of  Eq.~(\ref{eq:A(0,0,2)}).  If the sum of the helicities is not $\pm 2$, we find zero as we must.  Once again, we see that the relative factor, including both a sign and a factor of 2, are fixed by the inclusion of the massless vertices in the complete massive vertex.

Similarly, for a $-2$-helicity graviton, after raising the spin indices we have,
\begin{eqnarray}
    \frac{\tilde{x}^2}{M_P}\lbrack\mathbf{12}\rbrack^2 &=&
    \frac{1}{M_P}\left(\begin{array}{ccc} \displaystyle
         0 & 0 & \displaystyle
                 \frac{\langle31\rangle^4}{\langle12\rangle^2} \\
         0 & \displaystyle
             -\frac{\langle31\rangle^2\langle23\rangle^2}{2\langle12\rangle^2} & 0 \\
         \displaystyle
         \frac{\langle23\rangle^4}{\langle12\rangle^2} & 0 & 0
    \end{array}\right) 
    +\mathcal{O}(M_V)
    \label{eq:xtilde^2[12]^2 HE}\\
    &=&
    \left[
      \begin{array}{ccc} \displaystyle
         \mathcal{A}(-1,-1,-2)  & \displaystyle
         \mathcal{A}(-1,0,-2)   & \displaystyle
         \mathcal{A}(-1,+1,-2)  \\ \displaystyle
         \mathcal{A}(0,-1,-2)   & \displaystyle
        -\frac{1}{2}
         \mathcal{A}(0,0,-2)    & \displaystyle
         \mathcal{A}(0,+1,-2)   \\ \displaystyle
         \mathcal{A}(+1,-1,-2)  & \displaystyle
         \mathcal{A}(+1,0,-2)   & \displaystyle
         \mathcal{A}(+1,+1,-2)
      \end{array}
    \right] + \mathcal{O}(M_V).
\end{eqnarray}
As before, the high-energy limit of this vertex agrees with the massless vertices expected on symmetry grounds.  In particular, the upper-right and lower-left terms correspond with the right side of Eq.~(\ref{eq:A(1,-1,2)}) while the center term (for the Goldstone bosons) agrees with the right side of Eq.~(\ref{eq:A(0,0,2)}).

We end this section with the gravitational coupling to the Higgs boson which, for helicity $+2$, has the form,
\begin{equation}
    \frac{1}{M_P}x^2M_h^2.
    \label{eq:x^2Mh^2}
\end{equation}
Inserting the definition of $x$, we obtain,
\begin{equation}
    \frac{1}{M_P}(x M_h)^2 =\frac{1}{M_P}\left(\frac{\langle\xi2\rangle\lbrack23\rbrack}{\langle\xi3\rangle}\right)^2.
\end{equation}
We next multiply $xM_h$ by $\lbrack31\rbrack/\lbrack31\rbrack$ and use conservation of momentum, $\langle\xi3\rangle\lbrack31\rbrack=-\langle\xi2\rangle\lbrack21\rbrack+\mathcal{O}(M_h^2)$, to obtain,
\begin{equation}
    \frac{1}{M_P}(x M_h)^2 =
    \frac{1}{M_P}\left(\frac{\lbrack23\rbrack\lbrack31\rbrack}{\lbrack12\rbrack}\right)^2 + \mathcal{O}(M_h^2).
    \label{eq:x^2Mh^2 HE}
\end{equation}
The leading term is exactly $\mathcal{A}(0,0,+2)$ from the left side of Eq.~(\ref{eq:A(0,0,2)}) as seen above for the Goldstone bosons.  

The form for a $-2$-helicity graviton has $x$ replaced with $\tilde{x}$ and the square and angle brackets interchanged.
\begin{equation}
    \frac{1}{M_P}(\tilde{x} M_h)^2 =
    \frac{1}{M_P}\left(\frac{\langle23\rangle\langle31\rangle}{\langle12\rangle}\right)^2 + \mathcal{O}(M_h^2),
    \label{eq:xtilde^2Mh^2}
\end{equation}
and agrees with $\mathcal{A}(0,0,-2)$ from the right side of Eq.~(\ref{eq:A(0,0,2)}).

All of the gravitational vertices, along with their high-energy limit can be seen in Table~\ref{tab:gravitational}.

%%%%%%%%%%%%%%%%%%%%%%%%%%%%%%%%%%%%%%%%%%%%%%%%%%%%%%%%%%
%       Massless - Massless - Massive
%%%%%%%%%%%%%%%%%%%%%%%%%%%%%%%%%%%%%%%%%%%%%%%%%%%%%%%%%%
\section{\label{sec: 2 massless 1 massive}Two Massless and One Massive}

\bgroup
\renewcommand{\arraystretch}{2.5}
\begin{table*}
    \centering
    \begin{tabular}{|C|C|c|c|}
    \hline
        \text{Particles} 
      & \text{Coupling} 
      & Vertex 
      & High-Energy Limit \\[5pt] \hline
      \nu^-\bar{\nu}^+ Z
      & \displaystyle - \frac{i\sqrt{2}\ e}{\sin 2\theta_w} 
      & $\displaystyle \frac{\langle\mathbf{3}1\rangle\lbrack2\mathbf{3}\rbrack}{M_Z}$
      & $\displaystyle
       -\frac{\langle31\rangle^2}{\langle12\rangle} 
       \scalebox{0.75}{(\textendash)}\ ,\quad
       -\frac{\lbrack23\rbrack^2}{\lbrack12\rbrack} 
       \scalebox{0.75}{(+)}$ \\[5pt] \hline
      f\bar{f} Z
      & \displaystyle - \frac{i\sqrt{2}\ e}{\sin 2\theta_w}
      & \begin{tabular}{c}
      $\displaystyle 
         \frac{g_L\langle\mathbf{31}\rangle\lbrack\mathbf{23}\rbrack +
               g_R\lbrack\mathbf{31}\rbrack\langle\mathbf{23}\rangle}
              {M_Z}$ \\
        $+ \, \mathcal{N}_{Zff}
         \left(\tilde{g}_L\langle\mathbf{31}\rangle\langle\mathbf{23}\rangle +
               \tilde{g}_R\lbrack\mathbf{31}\rbrack\lbrack\mathbf{23}\rbrack
         \right)$
         \end{tabular}
      & \begin{tabular}{c}
        $\displaystyle
          g_R\frac{\langle23\rangle^2}{\langle12\rangle} 
          \scalebox{0.75}{(+ \textendash\ \textendash)}\ , \quad
         -g_L\frac{\langle31\rangle^2}{\langle12\rangle} 
          \scalebox{0.75}{(\textendash\ + \textendash)}$\ , \\
        $\displaystyle
          g_R\frac{\lbrack31\rbrack^2}{\lbrack12\rbrack} 
          \scalebox{0.75}{(+ \textendash\ +)}\ , \quad
         -g_L\frac{\lbrack23\rbrack^2}{\lbrack12\rbrack} 
          \scalebox{0.75}{(\textendash\ + +)}$ , \\
        $\displaystyle
         \frac{m_1 g_L - m_2 g_R}{2M_Z} \lbrack12\rbrack 
         \scalebox{0.75}{(+ + 0)}\ , \quad
         \frac{m_2 g_L - m_1 g_R}{2M_Z} \langle12\rangle 
         \scalebox{0.75}{(\textendash\ \textendash\ 0)}$
        \end{tabular} \\ \hline
      l \bar{\nu}_l^+ W
      & \displaystyle \frac{ie}{\sin\theta_w}
      & $\displaystyle
         \frac{\langle\mathbf{31}\rangle\lbrack2\mathbf{3}\rbrack}
              {M_W}
        +\mathcal{N}_{Wl\nu}
         \lbrack\mathbf{31}\rbrack\lbrack2\mathbf{3}\rbrack$
      & \begin{tabular}{c}
        $\displaystyle
        \frac{\langle31\rangle^2}{\langle12\rangle} 
         \scalebox{0.75}{(\textendash\ \textendash)}$ , \quad
        $\displaystyle
        -\frac{\lbrack23\rbrack^2}{\lbrack12\rbrack} 
         \scalebox{0.75}{(\textendash\ +)}$ , \quad
        $\displaystyle
         \frac{m_l}{2M_W}\lbrack12\rbrack 
         \scalebox{0.75}{(+ 0)}$
        \end{tabular} \\[5pt] \hline
      \bar{l} \nu_l^- \bar{W}
      & \displaystyle  \frac{ie}{\sin\theta_w}
      & $\displaystyle
         \frac{\lbrack\mathbf{31}\rbrack\langle2\mathbf{3}\rangle}
              {M_W}
        +\mathcal{N}^*_{Wl\nu}
         \langle\mathbf{31}\rangle\langle2\mathbf{3}\rangle$
      & \begin{tabular}{c}
        $\displaystyle
         \frac{\langle23\rangle^2}{\langle12\rangle} 
         \scalebox{0.75}{(+ \textendash)}$ ,\quad
        $\displaystyle
         \frac{\lbrack31\rbrack^2}{\lbrack12\rbrack} 
         \scalebox{0.75}{(+ +)}$ , \quad 
        $\displaystyle
         -\frac{m_l}{2M_W}\langle12\rangle 
         \scalebox{0.75}{(\textendash\ 0)} \, $
        \end{tabular}\\[5pt] \hline
      f_i\bar{f}_j W
      & \displaystyle \frac{ie}{\sin\theta_w} V_{ij}
      & \begin{tabular}{c}
        $\displaystyle 
         \frac{\langle\mathbf{31}\rangle\lbrack\mathbf{23}\rbrack}{M_W}$
        $+ \, \mathcal{N}_{Wff}
         \langle\mathbf{31}\rangle\langle\mathbf{23}\rangle$
        \end{tabular}
      & \begin{tabular}{c}
        $\displaystyle
         -\frac{\langle31\rangle^2}{\langle12\rangle} 
          \scalebox{0.75}{(\textendash\ + \textendash)}\ , \quad
         -\frac{\lbrack23\rbrack^2}{\lbrack12\rbrack} 
          \scalebox{0.75}{(\textendash\ + +)}$ ,\\
        $\displaystyle
         \frac{m_i}{2M_W} \lbrack12\rbrack 
         \scalebox{0.75}{(+ + 0)}\ , \quad
         \frac{m_j}{2M_W} \langle12\rangle 
         \scalebox{0.75}{(\textendash\ \textendash\ 0)}$
        \end{tabular} \\[5pt] \hline
      \bar{f}_i f_j \overline{W}
      & \displaystyle  \frac{ie}{\sin\theta_w} V^*_{ij}
      & \begin{tabular}{c}
        $\displaystyle 
         \frac{\lbrack\mathbf{31}\rbrack\langle\mathbf{23}\rangle}{M_W}$
        $+ \, \mathcal{N}^*_{Wff}
         \lbrack\mathbf{31}\rbrack\lbrack\mathbf{23}\rbrack$
        \end{tabular}
      & \begin{tabular}{c}
        $\displaystyle
         \frac{\langle23\rangle^2}{\langle12\rangle} 
         \scalebox{0.75}{(+ \textendash\ \textendash)}\ , \quad
         \frac{\lbrack31\rbrack^2}{\lbrack12\rbrack} 
         \scalebox{0.75}{(+ \textendash\ +)}$ ,\\
        $\displaystyle
         \frac{-m_j}{2M_W} \lbrack 12 \rbrack 
         \scalebox{0.75}{(+ + 0)}\ , \quad
         \frac{-m_i}{2M_W} \langle 12 \rangle 
         \scalebox{0.75}{(\textendash\ \textendash\ 0)}$
        \end{tabular} \\[5pt] \hline
        W \overline{W} Z
      & \displaystyle \frac{i e \cot \theta_w}{\sqrt{2}}
      & \begin{tabular}{c}
        $\displaystyle
         -\frac{\langle\mathbf{12}\rangle\langle\mathbf{23}\rangle
                \lbrack\mathbf{31}\rbrack+\lbrack\mathbf{12}\rbrack
                \lbrack\mathbf{23}\rbrack\langle\mathbf{31}\rangle}
               {M_W M_Z}$ \\
        $\displaystyle
         -\frac{\langle\mathbf{12}\rangle\lbrack\mathbf{23}\rbrack
                \langle\mathbf{31}\rangle+\lbrack\mathbf{12}\rbrack
                \langle\mathbf{23}\rangle\lbrack\mathbf{31}\rbrack}
               {M_W M_Z}$ \\
        $\displaystyle
         -\frac{\lbrack\mathbf{12}\rbrack\langle\mathbf{23}\rangle
                \langle\mathbf{31}\rangle+\langle\mathbf{12}\rangle
                \lbrack\mathbf{23}\rbrack\lbrack\mathbf{31}\rbrack}
               {M_W^2}$ \\
        $\displaystyle
         +\frac{\langle\mathbf{12}\rangle\langle\mathbf{23}\rangle
                \langle\mathbf{31}\rangle+\lbrack\mathbf{12}\rbrack
                \lbrack\mathbf{23}\rbrack\lbrack\mathbf{31}\rbrack}
               {\mathcal{N}_{WWZ}}$
        \end{tabular}
      & \begin{tabular}{l}
        $\displaystyle
         \frac{\langle23\rangle^3}{\langle12\rangle\langle31\rangle} 
         \scalebox{0.75}{(+ \textendash\ \textendash)}$, \quad
         $\displaystyle
         \frac{\langle31\rangle^3}{\langle12\rangle\langle23\rangle}
         \scalebox{0.75}{(\textendash\ + \textendash)}$, \quad
         $\displaystyle
         \frac{\langle12\rangle^3}{\langle23\rangle\langle31\rangle}
         \scalebox{0.75}{(\textendash\ \textendash\ +)}$, \\
        $\displaystyle
         \frac{\lbrack23\rbrack^3}{\lbrack12\rbrack\lbrack31\rbrack} 
         \scalebox{0.75}{(\textendash\ + +)}$, \quad
        $\displaystyle
         \frac{\lbrack31\rbrack^3}{\lbrack12\rbrack\lbrack23\rbrack}
         \scalebox{0.75}{(+ \textendash\ +)}$, \quad
        $\displaystyle
         \frac{\lbrack12\rbrack^3}{\lbrack23\rbrack\lbrack31\rbrack}
         \scalebox{0.75}{(+ \textendash\ \textendash)}$, \\
        $\displaystyle \quad
        -\frac{M_Z}{4M_W}
         \frac{\langle12\rangle\langle31\rangle}{\langle23\rangle}
         \scalebox{0.75}{(\textendash\ 0 0)}$, \quad
        $\displaystyle
        -\frac{M_Z}{4M_W}
         \frac{\langle12\rangle\langle23\rangle}{\langle31\rangle}
         \scalebox{0.75}{(0 \textendash\ 0)}$, \\
        $\displaystyle
         \frac{1}{4}\left(\frac{M_Z^2}{M_W^2}-2\right)
         \frac{\langle23\rangle\langle31\rangle}{\langle12\rangle}
         \scalebox{0.75}{(0 0 \textendash)}$,\\
        $\displaystyle \quad
        -\frac{M_Z}{4M_W}
         \frac{\lbrack12\rbrack\lbrack31\rbrack}{\lbrack23\rbrack}
         \scalebox{0.75}{(+ 0 0)}$, \quad
        $\displaystyle
        -\frac{M_Z}{4M_W}
         \frac{\lbrack12\rbrack\lbrack23\rbrack}{\lbrack31\rbrack}
         \scalebox{0.75}{(0 + 0)}$,\\
        $\displaystyle
         \frac{1}{4}\left(\frac{M_Z^2}{M_W^2}-2\right)
         \frac{\lbrack23\rbrack\lbrack31\rbrack}{\lbrack12\rbrack}
         \scalebox{0.75}{(0 0 +)}$
        \end{tabular} \\[5pt] \hline
    \end{tabular}
    \caption{Standard Model Weak Boson Sector Vertices along with their high-energy limit. The superscript in the first column gives the helicity of the massless particles. The CKM matrix is represented by $V_{ij}$. The $Z$-boson couples differently to left- and right-handed fermions as made explicit in the $g_L$ and $g_R$ couplings, which we write as $g_L = T_3 - Q_f \sin^2 \theta_w$ and $g_R = - Q_f \sin^2 \theta_w$ which leaves the Goldstone boson modes for the $f\bar{f}Z$ interactions non-zero in the limit of identical fermion masses, as expected, where $T_3$ is the isospin and $Q_f$ is the electric charge of the fermion. The terms in the high-energy limit with a ratio of masses are Goldstone boson interactions.  In some cases, the masses in the ratio canceled.  See sections \ref{sec: 2 massless 1 massive}, \ref{sec: 1 massless and 2 massive with different masses} and \ref{sec: 3 massive} for further details. The helicity signature is for the massive particles in the high-energy limit.}
    \label{tab:ew gauge}
\end{table*}
\egroup

In this section, we consider $3$-point amplitudes with one massive particle and two massless particles.  This is appropriate for the $Z$-boson vertex with two neutrinos.  (Since we do not know the full structure of the massive neutrino sector, we are limiting ourselves to the massless neutrinos of the SM.)  In \cite{Arkani-Hamed:2017jhn}, the authors write this vertex purely in terms of products of $\langle\mathbf{3}1\rangle$, $\langle2\mathbf{3}\rangle$ and $\lbrack12\rbrack$, which is convenient for their objectives.  However, we find it more convenient to write these vertices in a more minimal way, with only a single power of $m_3$ in the denominator for every term.  In particular, we find that for a $-1/2$-helicity neutrino (particle~$1$) and $+1/2$-helicity anti-neutrino (particle~$2$), the vertex is given by,
\begin{equation}
    \frac{1}{M_Z}\langle\mathbf{3}1\rangle\lbrack2\mathbf{3}\rbrack.
    \label{eq:<3b1>[23b]}
\end{equation}
This amplitude has one index, symmetrized over the two $\mathbf{3}$'s.  We can write this in vector notation and expand in the high-energy limit as,
\begin{equation}
    \frac{1}{M_Z}\langle\mathbf{3}1\rangle\lbrack2\mathbf{3}\rbrack
   =\frac{1}{M_Z}
    \left(\begin{array}{c} 
      \displaystyle
     -\frac{M_Z}{\sqrt{2E_3}}
      \langle31\rangle\lbrack2\tilde{\zeta}_3^+\rbrack \\
      \displaystyle
      \frac{1}{2}\langle31\rangle\lbrack23\rbrack \\ 
      \displaystyle
      \frac{M_Z}{\sqrt{2E_3}}\langle\zeta_3^-1\rangle\lbrack23\rbrack
    \end{array} \right)
    + \mathcal{O}(M_Z).
\end{equation}
The middle term is $0$ (at this order) by momentum conservation, $\langle31\rangle\lbrack23\rbrack=-\langle11\rangle\lbrack21\rbrack-\langle21\rangle\lbrack22\rbrack+\mathcal{O}(M_Z^2)=\mathcal{O}(M_Z^2)$ since $\langle11\rangle=0$ and $\lbrack22\rbrack=0$.  For the first term, we can multiply by $\langle12\rangle/\langle12\rangle$ and use momentum conservation in the numerator, $\langle12\rangle\lbrack2\tilde{\zeta}_3^+\rbrack=-\langle13\rangle\lbrack3\tilde{\zeta}_3^+\rbrack+\mathcal{O}(M_Z^2)=\sqrt{2E_3}\langle31\rangle+\mathcal{O}(M_Z^2)$.  For the third term, we can multiply by $\lbrack12\rbrack/\lbrack12\rbrack$ and use momentum conservation in the numerator, $\langle\zeta_3^-1\rangle\lbrack12\rbrack=-\langle\zeta_3^-3\rangle\lbrack32\rbrack+\mathcal{O}(M_Z^2) = -\sqrt{2E_3}\lbrack23\rbrack+\mathcal{O}(M_Z^2)$ and , to obtain,
\begin{equation}
    \frac{1}{M_Z}\langle\mathbf{3}1\rangle\lbrack2\mathbf{3}\rbrack =
    \left(\begin{array}{c} \displaystyle
        -\frac{\langle31\rangle^2}{\langle12\rangle} \\ \displaystyle
        0 \\ \displaystyle
        -\frac{\lbrack23\rbrack^2}{\lbrack12\rbrack}
    \end{array}\right)
    +\mathcal{O}(M_Z).
    \label{eq:<3b1>[23b] HE}
\end{equation}
The first term has a $-1$-helicity $Z$-boson, while the third term has a $+1$-helicity $Z$-boson.  The middle term, on the other hand, is for the helicity-$0$ component of the $Z$-boson, the Goldstone boson.  In fact, we find,
\begin{equation}
    \frac{1}{M_Z}\langle\mathbf{3}1\rangle\lbrack2\mathbf{3}\rbrack =
    \left[\begin{array}{c}
        \displaystyle
        -\mathcal{A}\left(-\frac{1}{2},+\frac{1}{2},-1\right) \\
        \displaystyle
        \mathcal{A}\left(-\frac{1}{2},+\frac{1}{2},0\right)  \\
        \displaystyle
        -\mathcal{A}\left(-\frac{1}{2},+\frac{1}{2},+1\right)
    \end{array}\right]
    +\mathcal{O}(M_Z),
\end{equation}
where the helicity of the $Z$~boson increases down the column vector as expected.
The middle term agrees with the massless vertex being zero since the helicities do not add to $\pm 1$.  The top and bottom expressions agree with the massless vertices found in Eq.~(\ref{eq:massless:1/2,-1/2,1}), giving the relative sign of these contributions.  
This vertex can be seen in Table \ref{tab:ew gauge}.

Although this is all we need for the SM, it will be useful to fill in the rest of the vertex structures for a $1$-spin particle interacting with two massless $\pm 1/2$-helicity fermions so that when we consider vertices with three massive particles, we can check this as one of the special limits.  We first give the vertex with all angle and square brackets interchanged,
\begin{eqnarray}
    \frac{1}{m_3}\lbrack\mathbf{3}1\rbrack\langle2\mathbf{3}\rangle &=& 
    \left(\begin{array}{c} \displaystyle
        \frac{\langle23\rangle^2}{\langle12\rangle} \\ \displaystyle
        0\\ \displaystyle
        \frac{\lbrack31\rbrack^2}{\lbrack12\rbrack}
    \end{array}\right)
    +\mathcal{O}(m_3)
    =
    \left[\begin{array}{c} \displaystyle
        \mathcal{A}\left(+\frac{1}{2},-\frac{1}{2},-1\right) \\ \displaystyle
        \mathcal{A}\left(+\frac{1}{2},-\frac{1}{2},0\right)
         \\ \displaystyle
        \mathcal{A}\left(+\frac{1}{2},-\frac{1}{2},+1\right)
    \end{array}\right]
    +\mathcal{O}(M_Z),
    \label{eq:1/m3[31]<23>}
\end{eqnarray}
for a $+1/2$-helicity fermion and $-1/2$-helicity anti-fermion.

If both the fermion and anti-fermion have $-1/2$-helicity, we obtain,
\begin{equation}
    \frac{1}{m_3}\langle\mathbf{3}1\rangle\langle2\mathbf{3}\rangle = 
    \frac{1}{m_3}\left(\begin{array}{c}
        \langle31\rangle\langle23\rangle\\ \displaystyle
        \frac{m_3}{2\sqrt{2E_3}}
        \left(
          \langle\zeta_3^-1\rangle\langle23\rangle
          +\langle31\rangle\langle2\zeta_3^-\rangle
        \right) \\
        0
    \end{array}\right)
    +\mathcal{O}(m_3).
    %\label{eq:1/m3<31>[23]}
\end{equation}
Multiplying this expression by $\lbrack12\rbrack/\lbrack12\rbrack$ sets the top term to zero at this order, by conservation of momentum in two ways, $\langle31\rangle\lbrack12\rbrack=\mathcal{O}(m_3)$ or $\lbrack12\rbrack\langle23\rangle=\mathcal{O}(m_3)$.  The middle term simplifies by use of $\langle\zeta_3^-1\rangle\lbrack12\rbrack=-\langle\zeta_3^-3\rangle\lbrack32\rbrack+\mathcal{O}(m_3^2)=\sqrt{2E_3}\lbrack32\rbrack+\mathcal{O}(m_3^2)$ and $\lbrack12\rbrack\langle2\zeta_3^-\rangle=-\lbrack13\rbrack\langle3\zeta_3^-\rangle+\mathcal{O}(m_3^2)=-\sqrt{2E_3}\lbrack13\rbrack+\mathcal{O}(m_3^2)$, giving us,
\begin{eqnarray}
    \frac{1}{m_3}\langle\mathbf{3}1\rangle\langle2\mathbf{3}\rangle &=& 
    \left(\begin{array}{c}
        0\\ \displaystyle
        \frac{1}{2}\frac{\lbrack32\rbrack\langle23\rangle}{\lbrack12\rbrack} -
        \frac{1}{2}\frac{\langle31\rangle\lbrack13\rbrack}{\lbrack12\rbrack}
        \\
        0
    \end{array}\right)
    +\mathcal{O}(m_3).
\end{eqnarray}
The numerator of the nonzero term is actually the difference between momenta, $\lbrack32\rbrack\langle23\rangle=2p_2\cdot p_3 +\mathcal{O}(m_3^2)$ and $\langle31\rangle\lbrack13\rbrack=2p_1\cdot p_3 +\mathcal{O}(m_3^2)$  so that the numerator is $2(p_2-p_1)\cdot p_3$.  However, remembering that conservation of momentum tells us that $p_3=-(p_1+p_2)$, we obtain,
\begin{equation}
    \frac{\lbrack32\rbrack\langle23\rangle}{\lbrack12\rbrack} -
    \frac{\langle31\rangle\lbrack13\rbrack}{\lbrack12\rbrack} =
    2\frac{p_1^2-p_2^2}{\lbrack12\rbrack} + \mathcal{O}(m_3^2) = 
    0 + \mathcal{O}(m_3^2),
\end{equation}
since particles 1 and 2 are assumed massless in this section.  Therefore, we have,
\begin{equation}
    \frac{1}{m_3}\langle\mathbf{3}1\rangle\langle2\mathbf{3}\rangle = 
    \mathcal{O}(m_3)
\end{equation}
and similarly,
\begin{equation}
    \frac{1}{m_3}\lbrack\mathbf{3}1\rbrack\lbrack2\mathbf{3}\rbrack = 
    \mathcal{O}(m_3).
\end{equation}

%%%%%%%%%%%%%%%%%%%%%%%%%%%%%%%%%%%%%%%%%%%%%%%%%%%%%%%%%%
%       Massless - Massive - Massive  (Different Masses)
%%%%%%%%%%%%%%%%%%%%%%%%%%%%%%%%%%%%%%%%%%%%%%%%%%%%%%%%%%
\section{\label{sec: 1 massless and 2 massive with different masses}One Massless Particle, Two Massive Particles with Different Masses}
In this section, we consider vertices with one massless particle and two massive particles of differing mass.  This vertex is appropriate to the interaction of the $W$-boson and leptons in the SM.  We begin with a lepton, a $+1/2$-helicity anti-neutrino and a $W$-boson.  The simplest vertex for this contains one spin-spinor for particle~$1$ of either type, one square helicity-spinor for particle~$2$ and two spin-spinors for particle~$3$ of either type.  The vertex is a linear combination of all possibilities satisfying these requirements, namely,
\begin{equation}
    \mathcal{N}_1\langle\mathbf{31}\rangle\lbrack2\mathbf{3}\rbrack
    +\mathcal{N}_2\lbrack\mathbf{31}\rbrack\lbrack2\mathbf{3}\rbrack.
    \label{eq:Wlnubar}
\end{equation}
for some $\mathcal{N}_1$ and $\mathcal{N}_2$, where we have not included $\langle\mathbf{33}\rangle$ or $\lbrack\mathbf{33}\rbrack$ since they would be antisymmetric in the SU($2$) indices that we are symmetrizing.  The form given in \cite{Arkani-Hamed:2017jhn} is equivalent to this choice, though we write a more minimal form for this SM vertex.  On the other hand, the vertex with an anti-lepton, neutrino (of $-1/2$-helicity), and anti-$W$-boson is given by,
\begin{equation}
    \mathcal{N}^*_1\lbrack\mathbf{31}\rbrack\langle2\mathbf{3}\rangle
    +\mathcal{N}^*_2\langle\mathbf{31}\rangle\langle2\mathbf{3}\rangle.
    \label{eq:Wlnu}
\end{equation}
$\mathcal{N}_1$ and $\mathcal{N}_2$ have inverse mass dimension of 1, therefore, we need to expand the products of spinors to linear order in the masses.  

These vertices have two independent indices, so we can write them as a rectangular matrix, with the index on particle~$1$ giving the row.  We begin our high-energy expansion with,
\begin{equation}
    \lbrack\mathbf{31}\rbrack\lbrack2\mathbf{3}\rbrack =
    \left(\begin{array}{ccc}
        0 & 0 &
        \displaystyle
        -\frac{m_1}{\sqrt{2E_1}}\lbrack3\tilde{\zeta}_1^+\rbrack\lbrack23\rbrack \\ 0 &
        \displaystyle
        -\frac{m_3}{2\sqrt{2E_3}}
        \left(
          \lbrack31\rbrack\lbrack2\tilde{\zeta}_3^+\rbrack
          +\lbrack\tilde{\zeta}_3^+1\rbrack\lbrack23\rbrack
        \right) &
        \lbrack31\rbrack\lbrack23\rbrack
    \end{array}\right)
    +\mathcal{O}(m^2).
\end{equation}
We multiply the bottom row by $\langle12\rangle/\langle12\rangle$.  The bottom-right term vanishes at this order by momentum conservation.  On the middle term, we use, $\langle12\rangle\lbrack2\tilde{\zeta}_3^+\rbrack=-\langle13\rangle\lbrack3\tilde{\zeta}_3^+\rbrack+\mathcal{O}(m^2) = \sqrt{2E_3}\langle31\rangle+\mathcal{O}(m^2)$ and $\lbrack\tilde{\zeta}_3^+1\rbrack\langle12\rangle = -\lbrack\tilde{\zeta}_3^+3\rbrack\langle32\rangle+\mathcal{O}(m^2) = -\sqrt{2E_3}\langle23\rangle+\mathcal{O}(m^2)$.  On the first row, we multiply by $\langle23\rangle/\langle23\rangle$ and use $\langle23\rangle\lbrack3\tilde{\zeta}_1^+\rbrack = -\langle21\rangle\lbrack1\tilde{\zeta}_1^+\rbrack + \mathcal{O}(m^2) = \sqrt{2E_1}\langle12\rangle+\mathcal{O}(m^2)$.  However, we see that with this, we obtain $\langle12\rangle\lbrack23\rbrack$ for the numerator of the top-right term, which is zero at this order by conservation of momentum.  Therefore, we have,
\begin{equation}
    \lbrack\mathbf{31}\rbrack\lbrack2\mathbf{3}\rbrack =
    \left(\begin{array}{ccc}
        0 & \displaystyle
        0 &
        0 \\
        0 & \displaystyle
        -\frac{m_3}{2}
        \left(
          \frac{\lbrack31\rbrack\langle31\rangle}{\langle12\rangle}
          -\frac{\lbrack23\rbrack\langle23\rangle}{\langle12\rangle}
        \right) &
        0
    \end{array}\right)
    +\mathcal{O}(m^2).
\end{equation}
However, we now see that the numerator contains $\lbrack31\rbrack\langle31\rangle=-2p_1\cdot p_3$ and $\lbrack23\rbrack\langle23\rangle=-2p_2\cdot p_3$  and, therefore, is proportional to $2(p_1-p_2)\cdot p_3+\mathcal{O}(m^2)$, which by conservation of momentum is $2(p_2^2-p_1^2)+\mathcal{O}(m^2)=\mathcal{O}(m^2)$.  Therefore, our final result is,
\begin{equation}
    \lbrack\mathbf{31}\rbrack\lbrack2\mathbf{3}\rbrack = \mathcal{O}(m^2)
    \ \mbox{and}\ 
    \langle\mathbf{31}\rangle\langle2\mathbf{3}\rangle = \mathcal{O}(m^2),
    \label{eq:Wlnu HE}
\end{equation}
and $\mathcal{N}_2$ does not appear to be restricted by this high-energy limit.  

We now move on to the other term.  We find,
\begin{equation}
    \langle\mathbf{31}\rangle\lbrack2\mathbf{3}\rbrack =
    \left(\begin{array}{ccc} \displaystyle
         -\frac{m_3}{\sqrt{2E_3}}\langle31\rangle\lbrack2\tilde{\zeta}_3^+\rbrack & \displaystyle
         \frac{1}{2}\langle31\rangle\lbrack23\rbrack & \displaystyle
        \frac{m_3}{\sqrt{2E_3}}\langle\zeta_3^-1\rangle\lbrack23\rbrack \\ \displaystyle
        0 & \displaystyle
        \frac{m_1}{2\sqrt{2E_1}}\langle3\zeta_1^-\rangle\lbrack23\rbrack &
        0
    \end{array}\right)
    + \mathcal{O}(m^2)
\end{equation}
The top-middle term is zero at this order due to momentum conservation.  For the top-left term, we multiply by $\langle12\rangle/\langle12\rangle$ and use $\langle12\rangle\lbrack2\tilde{\zeta}_3^+\rbrack=-\langle13\rangle\lbrack3\tilde{\zeta}_3^+\rbrack+\mathcal{O}(m^2)=\sqrt{2E_3}\langle31\rangle+\mathcal{O}(m^2)$.  For the top-right term, we multiply by $\lbrack12\rbrack/\lbrack12\rbrack$.   We then use $\langle\zeta_3^-1\rangle\lbrack12\rbrack=-\langle\zeta_3^-3\rangle\lbrack32\rbrack+\mathcal{O}(m^2)=-\sqrt{2E_3}\lbrack23\rbrack+\mathcal{O}(m^2)$.  For the bottom-middle term, we use that $\lbrack23\rbrack\langle3\zeta_1^-\rangle=-\lbrack21\rbrack\langle1\zeta_1^-\rangle+\mathcal{O}(m^2)=\sqrt{2E_1}\lbrack12\rbrack+\mathcal{O}(m^2)$.  Finally, we obtain,
\begin{equation}
    \langle\mathbf{31}\rangle\lbrack2\mathbf{3}\rbrack =
    \left(\begin{array}{ccc} \displaystyle
        -m_3\frac{\langle31\rangle^2}{\langle12\rangle} & 0 & \displaystyle
        -m_3\frac{\lbrack23\rbrack^2}{\lbrack12\rbrack} \\
        0 & \displaystyle
        \frac{m_1}{2}\lbrack12\rbrack &
        0
    \end{array}\right)
    + \mathcal{O}(m^2).
    \label{eq:Wlnubar HE 2}
\end{equation}
The top-left term gives the interaction of the $-1$-helicity $W$-boson while the top-right term gives the interaction of the $+1$-helicity $W$-boson.   Both have a $-1/2$-helicity charged lepton and a $+1/2$-helicity anti-neutrino.  The bottom-middle term contains the interaction of the $0$-helicity Goldstone boson with the $+1/2$-helicity charged lepton and anti-neutrino.  Similarly, we find for the neutrino,
\begin{equation}
    \lbrack\mathbf{31}\rbrack\langle2\mathbf{3}\rangle =
    \left(\begin{array}{ccc}
        0 & \displaystyle
        -\frac{m_1}{2}\langle12\rangle &
        0 \\ \displaystyle
        m_3\frac{\langle23\rangle^2}{\langle12\rangle} & 0 & \displaystyle
        m_3\frac{\lbrack31\rbrack^2}{\lbrack12\rbrack} 
    \end{array}\right)
    + \mathcal{O}(m^2).
    \label{eq:Wlnu HE 2}
\end{equation}
We must now decide what the normalization factor $\mathcal{N}_1$ should be.  We know that it is an inverse mass, but we do not yet know whether it is $1/m_1$, $1/m_3$, $1/(m_1+m_3)$, $1/\sqrt{m_1m_3}$ or something else, entirely.  However, it is not difficult to determine.  The top-middle term corresponds with the $\pm 1$-helicity part of the $1$-spin object, the $W$-boson.  It should be present whether the masses are non-zero or not.  On the other hand, the bottom-left term corresponds with the ``eaten" Goldstone boson and should contain the discontinuity.  Therefore, we find that $\mathcal{N}_1=1/m_3$ giving us the vertices,
\begin{equation}
\frac{1}{m_3}\langle\mathbf{31}\rangle\lbrack2\mathbf{3}\rbrack =
\left[\begin{array}{ccc} \displaystyle
    -\mathcal{A}\left(-\frac{1}{2},+\frac{1}{2},-1\right) & \displaystyle
    \mathcal{A}\left(-\frac{1}{2},+\frac{1}{2},0\right) & \displaystyle
    -\mathcal{A}\left(-\frac{1}{2},+\frac{1}{2},+1\right) \\ \displaystyle
    \mathcal{A}\left(+\frac{1}{2},+\frac{1}{2},-1\right) & \displaystyle
    \frac{m_1}{2m_3}
        \mathcal{A}\left(+\frac{1}{2},+\frac{1}{2},0\right) & \displaystyle
    \mathcal{A}\left(+\frac{1}{2},+\frac{1}{2},+1\right)
\end{array}\right]+\mathcal{O}(m),
\label{eq:1/m3<31>[23]}
\end{equation}
and,
\begin{equation}
\frac{1}{m_3}\lbrack\mathbf{31}\rbrack\langle2\mathbf{3}\rangle =
\left[\begin{array}{ccc} \displaystyle
    \mathcal{A}\left(-\frac{1}{2},-\frac{1}{2},-1\right) & \displaystyle
    -\frac{m_1}{2m_3}\mathcal{A}\left(-\frac{1}{2},-\frac{1}{2},0\right) & \displaystyle
    \mathcal{A}\left(-\frac{1}{2},-\frac{1}{2},+1\right) \\ \displaystyle
    \mathcal{A}\left(+\frac{1}{2},-\frac{1}{2},-1\right) & \displaystyle
    \mathcal{A}\left(+\frac{1}{2},-\frac{1}{2},0\right) & \displaystyle
    \mathcal{A}\left(+\frac{1}{2},-\frac{1}{2},+1\right)
\end{array}\right]+\mathcal{O}(m),
%\label{eq:1/m3[31]<23>}
\end{equation}
which agrees with the massless vertices given in Eq.~(\ref{eq:massless:1/2,-1/2,1}) for the $\pm 1$-helicity $W$-boson and Eq.~(\ref{eq:A(0,1/2,1/2)}) for the Goldstone boson interactions.  It also agrees with a zero massless amplitude when the helicities do not add to $\pm1$.  Once again, the relative signs and sizes of the amplitude are determined by their incorporation in the larger massive vertex.  In particular, the $m_1/m_3$ ratio between the $\pm 1$-helicity vertex and the Goldstone vertex is determined.  Furthermore, in the limit $m_1 \to 0$, the Goldstone interactions vanish and we are left with the 3-point amplitudes,
\begin{equation}
    \frac{1}{m_3}\langle\mathbf{3}1\rangle\lbrack2\mathbf{3}\rbrack
    \quad \mbox{and}\quad 
    \frac{1}{m_3}\lbrack\mathbf{3}1\rbrack\langle2\mathbf{3}\rangle,
\end{equation}
(where we have unbolded the $1$) agreeing with Eqs.~(\ref{eq:1/m3[31]<23>}) and (\ref{eq:1/m3<31>[23]}) of the previous section.  We see that we can simply unbold the $\mathbf{1}$ for the $1/2$-spin fermions.  However, it is not so simple for the $1$-spin boson because of the Goldstone boson living in it.  We must carefully expand the $W$-boson spin-spinors in the high-energy limit.

Coming briefly back to the normalization constant $\mathcal{N}_2$, we see that not only is it not determined by the high-energy limit, its term is not necessary to achieve any of the massless vertices for the $W$-lepton interactions.  At this point, we do not know what it should be based purely on symmetry principles, including both the symmetry principles at low and high energy.  We consider the determination of $\mathcal{N}_2$ in the absence of field theory an open question.  

We include this vertex along with its high-energy limit in Table \ref{tab:ew gauge}.

%%%%%%%%%%%%%%%%%%%%%%%%%%%%%%%%%%%%%%%%%%%%%%%%%%%%%%%%%%
%       Massive - Massive - Massive
%%%%%%%%%%%%%%%%%%%%%%%%%%%%%%%%%%%%%%%%%%%%%%%%%%%%%%%%%%
\section{\label{sec: 3 massive}Three Massive Particles}

\bgroup
\renewcommand{\arraystretch}{2.5}
\begin{table*}
    \centering
    \begin{tabular}{|c|c|c|c|}
    \hline
    Particles & Coupling & Vertex & High-Energy Limit  \\
    \hline
        $f\bar{f} h$
      & $-i m_f/v$
      & $\displaystyle
        \langle\mathbf{12}\rangle+\lbrack\mathbf{12}\rbrack
      $
      & $\displaystyle
        \langle12\rangle \scalebox{0.75}{(\textendash\ \textendash)}$,
        $\displaystyle
        \lbrack12\rbrack \scalebox{0.75}{(+ +)}$ \\[5pt] \hline
        $W \overline{W} h$ 
      & $-2i$ %  (2 i M_W^2 / v)g_(mu)(nu) is the SM answer, which is part of the HE limit
      & $\displaystyle
        \frac{\langle\mathbf{12}\rangle\lbrack\mathbf{12}\rbrack}{v} +
        \mathcal{N}_{WWh}
        \left(
           \langle\mathbf{12}\rangle\langle\mathbf{12}\rangle
          +\lbrack\mathbf{12}\rbrack\lbrack\mathbf{12}\rbrack
        \right)$
      & \begin{tabular}{c} 
        $\displaystyle
         \frac{M_W}{2v} 
         \frac{\langle12\rangle\langle31\rangle}{\langle23\rangle}
         \scalebox{0.75}{(\textendash\ 0)},$
        $\displaystyle
        -\frac{M_W}{2v}
         \frac{\langle12\rangle\langle23\rangle}{\langle31\rangle}
         \scalebox{0.75}{(0 \textendash)},$\\
        $\displaystyle
        -\frac{M_W}{2v} 
         \frac{\lbrack12\rbrack\lbrack31\rbrack}{\lbrack23\rbrack}
         \scalebox{0.75}{(+ 0)},$
        $\displaystyle
         \frac{M_W}{2v}
         \frac{\lbrack12\rbrack\lbrack23\rbrack}{\lbrack31\rbrack}
         \scalebox{0.75}{(0 +)},$
        \end{tabular} \\[5pt] \hline
        $Z Z h$ 
      & $-2i$ %  (2 i M_Z^2 / v)g_(mu)(nu) is the SM answer, which is part of the HE limit
      & $\displaystyle
         \frac{\langle\mathbf{12}\rangle\lbrack\mathbf{12}\rbrack}{v} 
        +\mathcal{N}_{ZZh}
         \left(
            \langle\mathbf{12}\rangle\langle\mathbf{12}\rangle
           +\lbrack\mathbf{12}\rbrack\lbrack\mathbf{12}\rbrack
         \right)
      $
      & \begin{tabular}{c}
        $\displaystyle
         \frac{M_Z}{2v} 
         \frac{\langle12\rangle\langle31\rangle}{\langle23\rangle}
         \scalebox{0.75}{(\textendash\ 0)},$
        $\displaystyle
        -\frac{M_Z}{2v}
         \frac{\langle12\rangle\langle23\rangle}{\langle31\rangle}
         \scalebox{0.75}{(0 \textendash)},$\\
        $\displaystyle
        -\frac{M_Z}{2v} 
         \frac{\lbrack12\rbrack\lbrack31\rbrack}{\lbrack23\rbrack}
         \scalebox{0.75}{(+ 0)},$
        $\displaystyle
         \frac{M_Z}{2v}
         \frac{\lbrack12\rbrack\lbrack23\rbrack}{\lbrack31\rbrack}
         \scalebox{0.75}{(0 +)},$
        \end{tabular} \\ \hline
        $h h h$ 
      & $-3 i M_h^2/v^2$ % this one needs more thought (was -6 i \lambda)
      & $v$
      & 0 \\ \hline
    \end{tabular}
    \caption{Standard Model Higgs Sector Vertices along with their high-energy limit. See Section \ref{sec: 3 massive} for further details. The helicity signatures of the particles is given in parentheses for all but the Higgs boson in the high-energy limit.}
    \label{tab:Higgs}
\end{table*}
\egroup

As pointed out in \cite{Arkani-Hamed:2017jhn}, there are no massless spinors to span the SL($2,\mathbb{C}$) space for these vertices so we have to employ tensors.  We can use the antisymmetric tensor $\epsilon_{\alpha\beta}$ and the symmetric tensor $\mathcal{O}_{\alpha\beta}=\epsilon^{\dot{\alpha}\dot{\beta}}\left(p_{1\alpha\dot{\alpha}}p_{2\beta\dot{\beta}}+p_{1\beta\dot{\alpha}}p_{2\alpha\dot{\beta}}\right)$.  Ref.~\cite{Arkani-Hamed:2017jhn} points out that products of $\epsilon$ tensors can be replaced with momentum tensors $\mathcal{O}$, which suits their purpose.  However, we find that the simplest form of the minimal SM coupling only uses the epsilon tensors.  Therefore, we will construct our vertices out of terms such as $\langle\mathbf{ij}\rangle$ and $\lbrack\mathbf{ij}\rbrack$ for particles $i$ and $j$.  

\subsection{\label{sec:3massive 1/2,1/2,1}Two Spin-1/2 Fermions and One Spin-1 Boson}
We begin with a vertex for two $1/2$-spin fermions and one $1$-spin boson.  This vertex will apply to the massive $f\bar{f}Z$ and $f\bar{f}'W$ vertices. There will be a single spin-spinor for each fermion of either type and two spin-spinors for the $W$ or $Z$ of either type.  There are six possible combinations, however two of them are proportional to either $\langle\mathbf{33}\rangle$ or $\lbrack\mathbf{33}\rbrack$, where the $W$ or $Z$-boson is the third particle.  As we see in Eq.~(\ref{eq:<jj>}), this is antisymmetric in the SU($2$) indices.  However, we remember that all the vertices are totally symmetrized over the SU($2$) indices for each particle.  Therefore, this combination does not contribute.  As a result, there are four terms that potentially contribute to this vertex.  They are,
\begin{equation}
    \langle\mathbf{31}\rangle\langle\mathbf{23}\rangle\ ,\ 
    \langle\mathbf{31}\rangle\lbrack\mathbf{23}\rbrack\ ,\ 
    \lbrack\mathbf{31}\rbrack\langle\mathbf{23}\rangle\ \mbox{and}\ 
    \lbrack\mathbf{31}\rbrack\lbrack\mathbf{23}\rbrack\ .
\end{equation}
Since all three particles are massive, if our theory were parity symmetric, we would expect our vertex to be symmetric between angle and square brackets. However, the interaction of the $Z$-boson and fermions is not parity symmetric, therefore, we expect each term to have its own unique coupling.  With some foresight, we will call these $g_R$ and $g_L$, giving us
\begin{equation}
    \mathcal{N}_1\left(
    g_L\lbrack\mathbf{23}\rbrack\langle\mathbf{31}\rangle +
    g_R\langle\mathbf{23}\rangle\lbrack\mathbf{31}\rbrack
    \right)
    +\mathcal{N}_2\left(
    \tilde{g}_L\langle\mathbf{31}\rangle\langle\mathbf{23}\rangle +
    \tilde{g}_R\lbrack\mathbf{31}\rbrack\lbrack\mathbf{23}\rbrack
    \right)
    \label{eq:1/2,1/2,1 general}
\end{equation}
for some $\mathcal{N}_1$ and $\mathcal{N}_2$, both of which have inverse mass dimension of $1$.  From the previous two sections, we expect $\mathcal{N}_1=1/m_3$ and $\mathcal{N}_2$ to be unconstrained by the high-energy limit.  Furthermore, we expect to be able to simply unbold the $\mathbf{1}$ and $\mathbf{2}$ in the high-energy limit for the fermions as long as we take $m_1\to0$ and $m_2\to0$ before $m_3$.  The $\mathbf{3}$ for the $W$-boson, on the other hand, can not be simply unbolded in the high-energy limit, due to its associated Goldstone boson.  Nonetheless, we will go through the high-energy limit in detail.  

We need to expand these vertex structures to linear order in the masses.  We note that since these structures have three independent SU($2$) indices, we cannot write them as matrices.  We must explicitly label their indices as we enumerate them.  In principle, we need to consider each index value of these spin-spinor products.  However, most of them begin at higher than linear order.  So, there are actually only a few that we need to explicitly calculate.  We begin with $\langle\mathbf{23}\rangle\lbrack\mathbf{31}\rbrack$ since, as we will see, it contains all the leading high-energy terms.  We use Eqs. (\ref{eq:<ij> expanded in m}) and (\ref{eq:[ij] expanded in m}), throughout this section.  We begin with $g_R\langle\mathbf{23}\rangle^{11} \lbrack\mathbf{31}\rbrack^{11} + g_L\lbrack\mathbf{23}\rbrack^{11} \langle\mathbf{31}\rangle^{11} = \mathcal{O}(m^2) = \mathcal{A}(-1/2,-1/2,-1) + \mathcal{O}(m^2)$.  We next consider,
\begin{equation}
    \langle\mathbf{23}\rangle^{11}\lbrack\mathbf{31}\rbrack^{12} =
    -\frac{m_3}{\sqrt{2E_3}}\langle23\rangle\lbrack\tilde{\zeta}^+_31\rbrack 
    + \mathcal{O}(m^3).
\end{equation}
We multiply this by $\langle12\rangle/\langle12\rangle$ and use momentum conservation $\lbrack\tilde{\zeta}_3^+1\rbrack\langle12\rangle=-\lbrack\tilde{\zeta}_3^+3\rbrack\langle32\rangle+\mathcal{O}(m^2)=-\sqrt{2E_3}\langle23\rangle+\mathcal{O}(m^2)$ to obtain,
\begin{equation}
    \langle\mathbf{23}\rangle^{11}\lbrack\mathbf{31}\rbrack^{12} =
    m_3\frac{\langle23\rangle^2}{\langle12\rangle} + \mathcal{O}(m^2).
\end{equation}
On the other hand, $\lbrack\mathbf{23}\rbrack^{11}\langle\mathbf{31}\rangle^{12} = \mathcal{O}(m^2)$, therefore we have,
\begin{equation}
\frac{1}{m_3}\left(g_R\langle\mathbf{23}\rangle^{11} \lbrack\mathbf{31}\rbrack^{12} + g_L\lbrack\mathbf{23}\rbrack^{11}\langle\mathbf{31}\rangle^{12}\right) = 
g_R\frac{\langle23\rangle^2}{\langle12\rangle} + \mathcal{O}(m^2) =
g_R\mathcal{A}\left(+\frac{1}{2},-\frac{1}{2},-1\right) + \mathcal{O}(m^2),
\label{eq:ffZ:A+--}
\end{equation}
where we have inserted the expected $\mathcal{N}_1=1/m_3$. As we see, this gives us the vertex for a $+1/2$-helicity fermion, a $-1/2$-helicity anti-fermion and a $-1$-helicity boson as expected. This agrees with the massless vertices of Eq.~(\ref{eq:massless:1/2,-1/2,1}).
We next increment the index on particle~$2$ to obtain $\langle\mathbf{23}\rangle^{21}\lbrack\mathbf{31}\rbrack^{11} = \mathcal{O}(m^3)$ and,
\begin{equation}
    \lbrack\mathbf{23}\rbrack^{21}\langle\mathbf{31}\rangle^{11} =
    -\frac{m_3}{\sqrt{2E_3}}\langle31\rangle\lbrack2\tilde{\zeta}^+_3\rbrack + \mathcal{O}(m^3).
\end{equation}
We multiply this by $\langle12\rangle/\langle12\rangle$ and use momentum conservation $\langle12\rangle\lbrack2\tilde{\zeta}^+_3\rbrack=-\langle13\rangle\lbrack3\tilde{\zeta}^+_3\rbrack+\mathcal{O}(m^2)=\sqrt{2E_3}\langle31\rangle+\mathcal{O}(m^2)$, we obtain
\begin{equation}
    \lbrack\mathbf{23}\rbrack^{21}\langle\mathbf{31}\rangle^{11} = 
    -m_3\frac{\langle31\rangle^2}{\langle12\rangle}+\mathcal{O}(m^3).
\end{equation}
Therefore,
\begin{equation}
    \frac{1}{m_3}\left(g_R\langle\mathbf{23}\rangle^{21}\lbrack\mathbf{31}\rbrack^{11}+
    g_L\lbrack\mathbf{23}\rbrack^{21}\langle\mathbf{31}\rangle^{11}\right) =
   -g_L\frac{\langle31\rangle^2}{\langle12\rangle}+\mathcal{O}(m^2)
    =-g_L\mathcal{A}\left(-\frac{1}{2},+\frac{1}{2},-1\right) + \mathcal{O}(m^2),
    \label{eq:ffZ:A-+-}
\end{equation}
where we have again inserted the expected $\mathcal{N}_1=1/m_3$ and will throughout the remainder of this section.  Once again, the vertex for the expected helicities appears and agrees with the massless vertices of Eq.~(\ref{eq:massless:1/2,-1/2,1}).
We next increment both particles 1 and 2 to obtain $g_R\langle\mathbf{23}\rangle^{21}\lbrack\mathbf{31}\rbrack^{12} + g_L\lbrack\mathbf{23}\rbrack^{21}\langle\mathbf{31}\rangle^{12} = \mathcal{O}(m^2) = \mathcal{A}(+1/2,+1/2,-1)+\mathcal{O}(m^2)$.

Moving on to incrementing the spin of the $1$-spin boson, we begin with,
\begin{equation}
    \langle\mathbf{23}\rangle^{11}\lbrack\mathbf{31}\rbrack^{21} =
    -\frac{m_1}{\sqrt{2E_1}}\langle23\rangle\lbrack3\tilde{\zeta}^+_1\rbrack + \mathcal{O}(m^3) = -m_1\langle12\rangle+\mathcal{O}(m^3),
\end{equation}
where we used momentum conservation $\langle23\rangle\lbrack3\tilde{\zeta}^+_1\rbrack = -\langle21\rangle\lbrack1\tilde{\zeta}^+_1\rbrack + \mathcal{O}(m^2) = \sqrt{2E_1}\langle12\rangle+\mathcal{O}(m^2)$.  We also need $\langle\mathbf{23}\rangle^{12}\lbrack\mathbf{31}\rbrack^{11} = \mathcal{O}(m^2)$, $\lbrack\mathbf{23}\rbrack^{11}\langle\mathbf{31}\rangle^{21} = \mathcal{O}(m^2)$ and, 
\begin{equation}
    \lbrack\mathbf{23}\rbrack^{12}\langle\mathbf{31}\rangle^{11} = 
    -\frac{m_2}{\sqrt{2E_2}}\lbrack\tilde{\zeta}^+_23\rbrack\langle31\rangle + \mathcal{O}(m^3) =
    m_2\langle12\rangle+\mathcal{O}(m^3),
\end{equation}
where we have used momentum conservation $\lbrack\tilde{\zeta}^+_23\rbrack\langle31\rangle = -\lbrack\tilde{\zeta}^+_22\rbrack\langle21\rangle+\mathcal{O}(m^2) = -\sqrt{2E_2}\langle12\rangle + \mathcal{O}(m^2)$.  Putting this together we have,
\begin{eqnarray}
    \frac{1}{2m_3}
    \Big(
        g_R\langle\mathbf{23}\rangle^{11} \lbrack\mathbf{31}\rbrack^{21}
       +g_R\langle\mathbf{23}\rangle^{12} \lbrack\mathbf{31}\rbrack^{11} \\ \nonumber
       +g_L\lbrack\mathbf{23}\rbrack^{11} \langle\mathbf{31}\rangle^{21} 
       +g_L\lbrack\mathbf{23}\rbrack^{12} \langle\mathbf{31}\rangle^{11}  
    \Big) &=& 
    \Big(
       \frac{m_2 g_L - m_1 g_R}{2m_3}
    \Big) \langle12\rangle + \mathcal{O}(m^2) \nonumber\\
    &=&
    \left(
       \frac{m_2 g_L - m_1 g_R}{2m_3}
    \right) \mathcal{A} \left(-\frac{1}{2},-\frac{1}{2},0\right) + \mathcal{O}(m^2),
    \label{eq:ffZ:A--0}
\end{eqnarray}
where the factor of $1/2$ is due to the symmetrization of the spin indices on particle-$3$. This gives the high-energy vertex for a $-1/2$-helicity fermion and anti-fermion and a helicity-$0$ boson, the Goldstone boson of the $1$-spin particle.  This high-energy result agrees with the massless vertices given in Eq.~(\ref{eq:A(0,1/2,1/2)}).  We now hold the spin index on particle-$3$ fixed and increment the indices on particles $1$ and $2$.  We begin by incrementing them separately and expect to find zero at this order since the helicities will not add to $+1$ until we increment them both.  This is what we find.  First, $\langle\mathbf{23}\rangle^{11} \lbrack\mathbf{31}\rbrack^{22} = \langle23\rangle \lbrack31\rbrack + \mathcal{O}(m^2) = \mathcal{O}(m^2)$ due to momentum conservation, $\langle\mathbf{23}\rangle^{12} \lbrack\mathbf{31}\rbrack^{12} = \mathcal{O}(m^2)$, $\lbrack\mathbf{23}\rbrack^{11} \langle\mathbf{31}\rangle^{22} = \mathcal{O}(m^2)$ and $\lbrack\mathbf{23}\rbrack^{12} \langle\mathbf{31}\rangle^{12} = \mathcal{O}(m^2)$.  Putting these together, we find $\frac{1}{2m_3} \left( g_R\langle\mathbf{23}\rangle^{11} \lbrack\mathbf{31}\rbrack^{22} + g_R\langle\mathbf{23}\rangle^{12} \lbrack\mathbf{31}\rbrack^{12} + g_L\lbrack\mathbf{23}\rbrack^{11} \langle\mathbf{31}\rangle^{22} + g_L\lbrack\mathbf{23}\rbrack^{12} \langle\mathbf{31}\rangle^{12}  \right) = \mathcal{O}(m^2) = \mathcal{A}(+1/2,-1/2,0) + \mathcal{O}(m^2)$, as expected.  Similarly, we find $\frac{1}{2m_3} \left(g_R\langle\mathbf{23}\rangle^{21} \lbrack\mathbf{31}\rbrack^{21} + g_R\langle\mathbf{23}\rangle^{22} \lbrack\mathbf{31}\rbrack^{11} + g_L\lbrack\mathbf{23}\rbrack^{21} \langle\mathbf{31}\rangle^{21} + g_L\lbrack\mathbf{23}\rbrack^{22} \langle\mathbf{31}\rangle^{11}  \right) = \mathcal{O}(m^2) = \mathcal{A}(-1/2,+1/2,0)+\mathcal{O}(m^2)$. 

Our next non-zero result occurs when both fermions have positive helicity,
\begin{equation}
    \langle\mathbf{23}\rangle^{21}\lbrack\mathbf{31}\rbrack^{22} = 
    \frac{m_2}{\sqrt{2E_2}}\langle\zeta^-_23\rangle\lbrack31\rbrack+\mathcal{O}(m^3) = -m_2\lbrack12\rbrack+\mathcal{O}(m^2) ,
\end{equation}
where we have used $\langle\zeta^-_23\rangle \lbrack31\rbrack = -\langle\zeta^-_22\rangle \lbrack21\rbrack + \mathcal{O}(m^2) = -\sqrt{2E_2}\lbrack12\rbrack + \mathcal{O}(m^2), \langle\mathbf{23}\rangle^{22} \lbrack\mathbf{31}\rbrack^{12} = \mathcal{O}(m^3), \lbrack\mathbf{23}\rbrack^{21} \langle\mathbf{31}\rangle^{22} = \mathcal{O}(m^3)$ and 
\begin{equation}
    \lbrack\mathbf{23}\rbrack^{22}\langle\mathbf{31}\rangle^{12}= 
    \frac{m_1}{\sqrt{2E_1}}\lbrack23\rbrack\langle3\zeta^-_1\rangle+\mathcal{O}(m^3) =
    m_1\lbrack12\rbrack+\mathcal{O}(m^3) ,
\end{equation}
where we have used $\lbrack23\rbrack\langle3\zeta^-_1\rangle=-\lbrack21\rbrack\langle1\zeta^-_1\rangle+\mathcal{O}(m^2)=\sqrt{2E_1}\lbrack12\rbrack+\mathcal{O}(m^2)$.  Putting all of this together, we have
\begin{eqnarray}
    \frac{1}{2m_3}
    \Big(
       g_R\langle\mathbf{23}\rangle^{21} \lbrack\mathbf{31}\rbrack^{22}
      +g_R\langle\mathbf{23}\rangle^{22} \lbrack\mathbf{31}\rbrack^{12} \\ \nonumber
      +g_L\lbrack\mathbf{23}\rbrack^{21} \langle\mathbf{31}\rangle^{22}
      +g_L\lbrack\mathbf{23}\rbrack^{22} \langle\mathbf{31}\rangle^{12}
    \Big) &=&
    \left(
      \frac{m_1 g_L - m_2 g_R}{2m_3}\right)\lbrack12\rbrack + \mathcal{O}(m^2) \nonumber\\
    &=& 
    \Big(
       \frac{m_1 g_L - m_2 g_R}{2m_3}
    \Big)\mathcal{A}\left(+\frac{1}{2},+\frac{1}{2},0\right) + \mathcal{O}(m^2) .
    \label{eq:ffZ:A++0}
\end{eqnarray}
Once again, this gives the Goldstone boson interaction, this time with $+1/2$-helicity fermion and anti-fermion agreeing with the massless vertices given in Eq.~(\ref{eq:A(0,1/2,1/2)}).

We now move on to the highest helicity for the $1$-spin boson by incrementing the spin index on particle-$3$ one more time.  We begin with $g_R\langle\mathbf{23}\rangle^{12} \lbrack\mathbf{31}\rbrack^{21} + g_L\lbrack\mathbf{23}\rbrack^{12} \langle\mathbf{31}\rangle^{21} = \mathcal{O}(m^2) = \mathcal{A}(-1/2,-1/2,+1) + \mathcal{O}(m^2)$.  Next we consider,
\begin{equation}
    \langle\mathbf{23}\rangle^{12}\lbrack\mathbf{31}\rbrack^{22} =
    \frac{m_3}{\sqrt{2E_3}}\langle2\zeta^-_3\rangle\lbrack31\rbrack+\mathcal{O}(m^3) .
\end{equation}
We multiply this by $\lbrack12\rbrack/\lbrack12\rbrack$ and use momentum conservation $\lbrack12\rbrack\langle2\zeta^-_3\rangle = -\lbrack13\rbrack\langle3\zeta^-_3\rangle + \mathcal{O}(m^2)=\sqrt{2E_3}\lbrack31\rbrack + \mathcal{O}(m^2)$.  We also note that $\lbrack\mathbf{23}\rbrack^{12} \langle\mathbf{31}\rangle^{22} = \mathcal{O}(m^3)$.  From this we obtain,
\begin{equation}
    \frac{1}{m_3}\left(g_R\langle\mathbf{23}\rangle^{12}\lbrack\mathbf{31}\rbrack^{22} + g_L\lbrack\mathbf{23}\rbrack^{12}\langle\mathbf{31}\rangle^{22}\right) =
    g_R\frac{\lbrack31\rbrack^2}{\lbrack12\rbrack} + \mathcal{O}(m^2) =
    g_R\mathcal{A}\left(+\frac{1}{2},-\frac{1}{2},+1\right) + \mathcal{O}(m^2).
    \label{eq:ffZ:A+-+}
\end{equation}
Similarly, $\langle\mathbf{23}\rangle^{22}\lbrack\mathbf{31}\rbrack^{21}=\mathcal{O}(m^3)$ and
\begin{equation}
    \lbrack\mathbf{23}\rbrack^{22}\langle\mathbf{31}\rangle^{21} =
    \frac{m_3}{\sqrt{2E_3}}\lbrack23\rbrack\langle\zeta^-_31\rangle+\mathcal{O}(m^3).
\end{equation}
We multiply this by $\lbrack12\rbrack/\lbrack12\rbrack$ and use momentum conservation $\langle\zeta^-_31\rangle\lbrack12\rbrack=-\langle\zeta^-_33\rangle\lbrack32\rbrack+\mathcal{O}(m^2)=-\sqrt{2E_3}\lbrack23\rbrack+\mathcal{O}(m^2)$.  Therefore, we have
\begin{equation}
    \frac{1}{m_3}\left(
        g_R\langle\mathbf{23}\rangle^{22}\lbrack\mathbf{31}\rbrack^{21} +
        g_L\lbrack\mathbf{23}\rbrack^{22}\langle\mathbf{31}\rangle^{21} 
    \right) =-g_L\frac{\lbrack23\rbrack^2}{\lbrack12\rbrack} + \mathcal{O}(m^2) =
    -g_L\mathcal{A}\left(-\frac{1}{2},+\frac{1}{2},+1\right)+\mathcal{O}(m^2).
    \label{eq:ffZ:A-++}
\end{equation}
Finally, we have $\left(g_R\langle\mathbf{23}\rangle^{22} \lbrack\mathbf{31}\rbrack^{22} + g_L\lbrack\mathbf{23}\rbrack^{22} \langle\mathbf{31}\rangle^{22}\right) = \mathcal{O}(m^2) = \mathcal{A}(+1/2,+1/2,+1) + \mathcal{O}(m^2)$.  As before, all these high-energy limits agree with the massless vertices given in Eq.~(\ref{eq:massless:1/2,-1/2,1}).  Furthermore, as in the previous sections, we learn that $\mathcal{N}_1=1/m_3$ and that, in the high-energy limit, we must take $m_1 \to 0$ and $m_2 \to 0$ before taking $m_3 \to 0$.  This removes the Goldstone interactions before taking the final $m_3 \to 0$ limit and agrees with the results of the previous sections.

For the $W$-boson vertices, we find the usual chiral results, $g_L=1$ while $g_R=0$ since the $W$-boson only interacts with the left-handed fermions.  For the $Z$-boson vertices, $g_L$ and $g_R$ are different for each flavor of fermion. We use $g_L = T_3 - Q_f \sin^2 \theta_w$ and $g_R= - Q_f \sin^2 \theta_w$ where $Q_f$ is the electric charge of the fermion and $T_3$ is its isospin.

We must now consider the other vertex structure,  $\langle\mathbf{31}\rangle\langle\mathbf{23}\rangle$ and its partner interchanged in angle and square brackets.  We begin with all indices equal to $1$, and keep up to linear order,
\begin{equation}
    \langle\mathbf{31}\rangle^{11}\langle\mathbf{23}\rangle^{11} =
    \langle31\rangle\langle23\rangle + \mathcal{O}(m^2).
\end{equation}
However, by multiplying by $\lbrack12\rbrack/\lbrack12\rbrack$ and using momenum conservation in the numerator, we immediately see that the first term vanishes to quadratic order and we are left with,
\begin{equation}
    \langle\mathbf{31}\rangle^{11}\langle\mathbf{23}\rangle^{11} =
    \mathcal{O}(m^2).
    \label{eq:1/2,1/2,1 general vanish 1}
\end{equation}
So, this term does not contribute to the high-energy limit.  

We next look at,
\begin{equation}
    \langle\mathbf{31}\rangle^{12}\langle\mathbf{23}\rangle^{11} = 
    \frac{m_1}{\sqrt{2E_1}}\langle3\zeta_1^-\rangle\langle23\rangle + \mathcal{O}(m^3).
\end{equation}
In order to simplify this, we need to multiply by $\lbrack23\rbrack/\lbrack23\rbrack$ and use conservation of momentum.  After we do this, we are left with,
\begin{equation}
    \langle\mathbf{31}\rangle^{12}\langle\mathbf{23}\rangle^{11} = m_1
    \frac{\lbrack12\rbrack\langle23\rangle}{\lbrack23\rbrack}
    +\mathcal{O}(m^3).
\end{equation}
However, $\lbrack12\rbrack\langle23\rangle=\mathcal{O}(m^2)$ by conservation of momentum.  Therefore,
\begin{equation}
    \langle\mathbf{31}\rangle^{12}\langle\mathbf{23}\rangle^{11} = 
    \mathcal{O}(m^3).
    \label{eq:1/2,1/2,1 general vanish 2}
\end{equation}
And, similarly, if we take the particle-2 index to be 2, we obtain,
\begin{equation}
    \langle\mathbf{31}\rangle^{11}\langle\mathbf{23}\rangle^{21} = 
    \mathcal{O}(m^3).
    \label{eq:1/2,1/2,1 general vanish 3}
\end{equation}
We have one more to check.  It is,
\begin{equation}
\frac{1}{2}\left(
    \langle\mathbf{31}\rangle^{11}\langle\mathbf{23}\rangle^{12} +
    \langle\mathbf{31}\rangle^{21}\langle\mathbf{23}\rangle^{11} 
\right) =
    \frac{m_3}{2\sqrt{2E_3}}\left(
        \langle31\rangle\langle2\zeta_3^-\rangle +
        \langle\zeta_3^-1\rangle\langle23\rangle
    \right) + \mathcal{O}(m^3).
\end{equation}
Multiplying by $\lbrack12\rbrack/\lbrack12\rbrack$ and using momentum conservation, we obtain,
\begin{equation}
\frac{1}{2}\left(
    \langle\mathbf{31}\rangle^{11}\langle\mathbf{23}\rangle^{12} +
    \langle\mathbf{31}\rangle^{21}\langle\mathbf{23}\rangle^{11} 
\right) =
    \frac{m_3}{2}\left(
        \frac{\langle31\rangle\lbrack31\rbrack}{\lbrack12\rbrack} -
        \frac{\lbrack23\rbrack\langle23\rangle}{\lbrack12\rbrack}
    \right) + \mathcal{O}(m^3).
\end{equation}
As in the previous section, we see that the numerator is $\langle31\rangle\lbrack31\rbrack-\langle23\rangle\lbrack23\rbrack=2(p_2-p_1)\cdot p_3 +\mathcal{O}(m^2) = 2(p_1-p_2)\cdot (p_2+p_1) + \mathcal{O}(m^2)=2p_1^2-2p_2^2 + \mathcal{O}(m^2)=\mathcal{O}(m^2)$, leaving us with,
\begin{equation}
\frac{1}{2}\left(
    \langle\mathbf{31}\rangle^{11}\langle\mathbf{23}\rangle^{12} +
    \langle\mathbf{31}\rangle^{21}\langle\mathbf{23}\rangle^{11} 
\right) =
    \mathcal{O}(m^3),
    \label{eq:1/2,1/2,1 general vanish 4}
\end{equation}
and similarly,
\begin{equation}
\frac{1}{2}\left(
    \lbrack\mathbf{31}\rbrack^{11}\lbrack\mathbf{23}\rbrack^{12} +
    \lbrack\mathbf{31}\rbrack^{21}\lbrack\mathbf{23}\rbrack^{11} 
\right) =
    \mathcal{O}(m^3).
\end{equation}
So, once again, we find that $\mathcal{N}_2$ is unconstrained by the high-energy limit.  We do not yet know what it should be.  From this, we can determine the vertices for the $Z$-boson and the fermions as well as the $W$-boson and the quarks.  We have included these in Table \ref{tab:ew gauge}.

\subsection{\label{sec:3 spin-1}Three Spin-1 Bosons}
We now consider the case where all three particles are $1$-spin bosons.  This is appropriate to the $WWZ$ vertex.  Now, each vertex has two spin-spinors of either type for each particle.  We must combine them in all possible ways, but we remember that we do not have any terms with $\langle\mathbf{ii}\rangle$ or $\lbrack\mathbf{ii}\rbrack$ for particle~$i$ because its SU($2$) index is symmetrized taking this term to zero.  Therefore, we construct our vertices out of $\langle\mathbf{12}\rangle, \langle\mathbf{23}\rangle, \langle\mathbf{31}\rangle$ and their partners with square brackets.   We expect our vertex to be symmetric between angle and square brackets, therefore, we expect our vertex to take the form,
\begin{eqnarray}
    &\mathcal{N}_1\left(\langle\mathbf{12}\rangle\langle\mathbf{23}\rangle\langle\mathbf{31}\rangle+\lbrack\mathbf{12}\rbrack\lbrack\mathbf{23}\rbrack\lbrack\mathbf{31}\rbrack\right) + 
    \mathcal{N}_2\left(\langle\mathbf{12}\rangle\langle\mathbf{23}\rangle\lbrack\mathbf{31}\rbrack+
    \lbrack\mathbf{12}\rbrack\lbrack\mathbf{23}\rbrack\langle\mathbf{31}\rangle\right) +&\nonumber\\
    &\mathcal{N}_3\left(\langle\mathbf{12}\rangle\lbrack\mathbf{23}\rbrack\langle\mathbf{31}\rangle+
    \lbrack\mathbf{12}\rbrack\langle\mathbf{23}\rangle\lbrack\mathbf{31}\rbrack\right) +
    \mathcal{N}_4\left(\lbrack\mathbf{12}\rbrack\langle\mathbf{23}\rangle\langle\mathbf{31}\rangle+
    \langle\mathbf{12}\rangle\lbrack\mathbf{23}\rbrack\lbrack\mathbf{31}\rbrack\right),&
    \label{eq:3 spin-1 general}
\end{eqnarray}
where $\mathcal{N}_1$, $\mathcal{N}_2$, $\mathcal{N}_3$ and $\mathcal{N}_4$ are distinct from the normalization constants in previous sections and have inverse mass dimension of 2.  This means we have to expand these structures to quadratic order in the masses.  All of these have three indices, therefore, we will not be able to write them in matrix form.  The last three are related by interchange of the particles.  Computing one gives the others.  But, the coefficients may be different combinations of masses, so we separate them.  We use Eqs. (\ref{eq:|i>^I expanded in m}) and (\ref{eq:[i|^I expanded in m}) to expand in high energy, keeping up to quadratic order in the masses.

We begin with,
\begin{equation}
    \langle\mathbf{12}\rangle^{11}\langle\mathbf{23}\rangle^{11}\langle\mathbf{31}\rangle^{11} = 
    \left(1-\frac{m_1^2}{4E_1^2}-\frac{m_2^2}{4E_2^2}-\frac{m_3^2}{4E_3^2}\right)
    \langle12\rangle\langle23\rangle\langle31\rangle + \mathcal{O}(m^4).
\end{equation}
However, we can multiply this by $\lbrack12\rbrack^2/\lbrack12\rbrack^2$ and use conservation of momentum twice in the numerator, $\lbrack12\rbrack\langle23\rangle=\mathcal{O}(m^2)$ and $\langle31\rangle\lbrack12\rbrack=\mathcal{O}(m^2)$, to obtain,
\begin{equation}
    \langle\mathbf{12}\rangle^{11}\langle\mathbf{23}\rangle^{11}\langle\mathbf{31}\rangle^{11} = 
    \mathcal{O}(m^4).
    \label{eq:3 spin-1 general HE 1}
\end{equation}
We will see this sort of simplification in many of the terms.

We next consider the index on 1 incremented,
\begin{equation}
    \langle\mathbf{12}\rangle^{11}\langle\mathbf{23}\rangle^{11}\langle\mathbf{31}\rangle^{12} +
    \langle\mathbf{12}\rangle^{21}\langle\mathbf{23}\rangle^{11}\langle\mathbf{31}\rangle^{11} = 
    \frac{m_1}{\sqrt{2E_1}}\left(
        \langle12\rangle\langle23\rangle\langle3\zeta_1^-\rangle +
        \langle\zeta_1^-2\rangle\langle23\rangle\langle31\rangle
    \right) 
    +\mathcal{O}(m^3).
\end{equation}
Once again, we multiply by $\lbrack23\rbrack/\lbrack23\rbrack$ and use conservation of momentum.  We begin by simplifying the $\zeta$ terms.  For example, on the first term, we use $\lbrack23\rbrack\langle3\zeta_1^-\rangle=-\lbrack21\rbrack\langle1\zeta_1^-\rangle+\mathcal{O}(m^2)=\sqrt{2E_1}\lbrack12\rbrack+\mathcal{O}(m^2)$.  We then follow this with $\lbrack12\rbrack\langle23\rangle=\mathcal{O}(m^2)$, showing that this term begins at $\mathcal{O}(m^3)$.  The second term is the same order by a similar set of steps,
\begin{equation}
    \langle\mathbf{12}\rangle^{11}\langle\mathbf{23}\rangle^{11}\langle\mathbf{31}\rangle^{12} +
    \langle\mathbf{12}\rangle^{21}\langle\mathbf{23}\rangle^{11}\langle\mathbf{31}\rangle^{11} = 
    \mathcal{O}(m^3).
\end{equation}
Incrementing, instead, the index on particle~$2$ or $3$ would obtain the same result.

We next try incrementing two of the indices, for example on particles 1 and 2, to obtain,
\begin{align}
    &\langle\mathbf{12}\rangle^{11}\langle\mathbf{23}\rangle^{21}\langle\mathbf{31}\rangle^{12} 
    +\langle\mathbf{12}\rangle^{21}\langle\mathbf{23}\rangle^{21}\langle\mathbf{31}\rangle^{11} 
    +\langle\mathbf{12}\rangle^{12}\langle\mathbf{23}\rangle^{11}\langle\mathbf{31}\rangle^{12} 
    +\langle\mathbf{12}\rangle^{22}\langle\mathbf{23}\rangle^{11}\langle\mathbf{31}\rangle^{11} 
    \nonumber \\ 
    &\displaystyle \qquad
    =\frac{m_1m_2}{\sqrt{4E_1E_2}}
     \left(
        \langle12\rangle\langle\zeta_2^-3\rangle\langle3\zeta_1^-\rangle +
        \langle\zeta_1^-2\rangle\langle\zeta_2^-3\rangle\langle31\rangle +
        \langle1\zeta_2^-\rangle\langle23\rangle\langle3\zeta_1^-\rangle +
        \langle\zeta_1^-\zeta_2^-\rangle\langle23\rangle\langle31\rangle
     \right) 
    +\mathcal{O}(m^4)
\end{align}
To check whether this expression is zero at this order, we must first simplify the $\zeta$ terms.  We multiply the first, second and third by $\lbrack31\rbrack\lbrack23\rbrack/\lbrack31\rbrack\lbrack23\rbrack$.  On the first, we use $\langle\zeta_2^-3\rangle\lbrack31\rbrack=-\sqrt{2E_2}\lbrack12\rbrack+\mathcal{O}(m^2)$ along with $\lbrack23\rbrack\langle3\zeta_1^-\rangle=\sqrt{2E_1}\lbrack12\rbrack+\mathcal{O}(m^2)$.  For the second, we use $\langle\zeta_2^-3\rangle\lbrack31\rbrack=-\sqrt{2E_2}\lbrack12\rbrack+\mathcal{O}(m^2)$ and $\langle\zeta_1^-2\rangle\lbrack23\rbrack=-\sqrt{2E_1}\lbrack31\rbrack+\mathcal{O}(m^2)$.  For the third, we use $\lbrack31\rbrack\langle1\zeta_2^-\rangle=\sqrt{2E_2}\lbrack23\rbrack+\mathcal{O}(m^2)$ and $\lbrack23\rbrack\langle3\zeta_1^-\rangle=\sqrt{2E_1}\lbrack12\rbrack+\mathcal{O}(m^2)$.  We do not know of a way to simplify the $\langle\zeta_1^-\zeta_2^-\rangle$, but, since we still have $\langle23\rangle\langle31\rangle$ in the last term, we can multiply it by $\lbrack23\rbrack/\lbrack23\rbrack$ and use $\lbrack23\rbrack\langle31\rangle=\mathcal{O}(m^2)$ to show that this term is higher order.  Putting all of this together, we have,
\begin{align}
   &\langle\mathbf{12}\rangle^{11}\langle\mathbf{23}\rangle^{21}\langle\mathbf{31}\rangle^{12}
   +\langle\mathbf{12}\rangle^{21}\langle\mathbf{23}\rangle^{21}\langle\mathbf{31}\rangle^{11}
   +\langle\mathbf{12}\rangle^{12}\langle\mathbf{23}\rangle^{11}\langle\mathbf{31}\rangle^{12}
   +\langle\mathbf{12}\rangle^{22}\langle\mathbf{23}\rangle^{11}\langle\mathbf{31}\rangle^{11} \nonumber \\
   &\displaystyle \qquad
   = m_1 m_2
    \left(
        \frac{-\langle12\rangle\lbrack12\rbrack^2+\langle31\rangle\lbrack31\rbrack\lbrack12\rbrack + \langle23\rangle\lbrack23\rbrack\lbrack12\rbrack}{\lbrack31\rbrack\lbrack23\rbrack} 
    \right) 
    +\mathcal{O}(m^4).
\end{align}
However, each of these terms is zero at this order.  There are several ways to see this, but we note that $\langle12\rangle\lbrack12\rbrack=-2p_1\cdot p_2=-(p_1+p_2)^2+\mathcal{O}(m^2)=-p_3^2+\mathcal{O}(m^2)=\mathcal{O}(m^2)$, and similarly for the other terms.  Therefore, this term does not contribute at quadratic order.  

Incrememting all three indices once will begin at cubic order and does not need to be considered.   The only case left to consider is incrementing one of the indices to $22$.  Let's try,
\begin{equation}
    \langle\mathbf{12}\rangle^{21}\langle\mathbf{23}\rangle^{11}\langle\mathbf{31}\rangle^{12} =
    \frac{m_1^2}{2E_1} \langle\zeta_1^-2\rangle\langle23\rangle\langle3\zeta_1^-\rangle +
    \mathcal{O}(m^4).
\end{equation}
We simplify the $\zeta$ terms by multiplying by $\lbrack23\rbrack^2/\lbrack23\rbrack^2$ and using conservation of momentum, $\langle\zeta_1^-2\rangle\lbrack23\rbrack=-\sqrt{2E_1}\lbrack31\rbrack+\mathcal{O}(m^2)$ and $\lbrack23\rbrack\langle3\zeta_1^-\rangle=\sqrt{2E_1}\lbrack12\rbrack+\mathcal{O}(m^2)$, to obtain,
\begin{equation}
    \langle\mathbf{12}\rangle^{21}\langle\mathbf{23}\rangle^{11}\langle\mathbf{31}\rangle^{12} =
    -m_1^2 \frac{\langle23\rangle\lbrack12\rbrack\lbrack31\rbrack}{\lbrack23\rbrack^2} +
    \mathcal{O}(m^4).
\end{equation}
The numerator is zero at this order by conservation of momentum, $\langle23\rangle\lbrack31\rbrack=\mathcal{O}(m^2)$, so this term also does not contribute at this order. Putting this all together, we find that the vertex structure with coefficient $\mathcal{N}_1$ does not contribute at this order in the high-energy limit and thus we can not constrain $\mathcal{N}_1$ from the high-energy limit.

We now turn our attention to the vertex structure with coefficient $\mathcal{N}_2$.  Since we have not found any of our expected high-energy limit interactions yet, we expect to find them all in this and the related terms. 
There will be a great number of cases to keep track of.  In order to make where each term comes from clear, we will add them to Table~\ref{tab:<><>[] terms} as we compute them.
\begin{table}
    \centering
    \begin{tabular}{|ccc|c|c|c|}
        \hline
         \multicolumn{3}{|c|}{} & & & \\
         \multicolumn{3}{|c|}{Helicity} &
         $\, \langle\mathbf{12}\rangle \langle\mathbf{23}\rangle \lbrack\mathbf{31}\rbrack   + 
             \lbrack\mathbf{12}\rbrack \lbrack\mathbf{23}\rbrack \langle\mathbf{31}\rangle \,$ &
         $\, \langle\mathbf{12}\rangle \lbrack\mathbf{23}\rbrack \langle\mathbf{31}\rangle + 
             \lbrack\mathbf{12}\rbrack \langle\mathbf{23}\rangle \lbrack\mathbf{31}\rbrack \,$ &
         $\, \lbrack\mathbf{12}\rbrack \langle\mathbf{23}\rangle \langle\mathbf{31}\rangle   + 
             \langle\mathbf{12}\rangle \lbrack\mathbf{23}\rbrack \lbrack\mathbf{31}\rbrack \,$ \\
         \multicolumn{3}{|c|}{} & & &  \\
         \hline\hline
%         $-1$ & $-1$ & $-1$ &      $0$ & $0$ & $0$ \\
%         $-1$ & $-1$ & 
%                       $\phantom{+}0$ 
%                            &      $0$ & $0$ & $0$ \\
%         $-1$ & $\phantom{+}0$ 
%                     & $-1$ 
%                            &      $0$ & $0$ & $0$ \\
%         $\phantom{+}0$ 
%              & $-1$ & $-1$ &      $0$ & $0$ & $0$ \\
         $\phantom{+}0$ 
                     & $\phantom{+}0$ 
                     & $-1$ &      $ \displaystyle
                                     \frac{m_2 m_3}{4} \frac{\langle23\rangle \langle31\rangle}{\langle12\rangle}$ & 
                                   $ \displaystyle
                                     \frac{m_1 m_3}{4} \frac{\langle23\rangle \langle31\rangle}{\langle12\rangle}$ &
                                   $ \displaystyle
                                    -\frac{m_3^2}{4}   \frac{\langle23\rangle \langle31\rangle}{\langle12\rangle}$ \\
         $-1$ & 
                $\phantom{+}0$ 
                     & $\phantom{+}0$ 
                            &      $ \displaystyle
                                     \frac{m_1m_2}{4}  \frac{\langle12\rangle \langle31\rangle}{\langle23\rangle}$ & 
                                   $ \displaystyle
                                    -\frac{m_1^2}{4}   \frac{\langle12\rangle \langle31\rangle}{\langle23\rangle}$ &
                                   $ \displaystyle
                                     \frac{m_1m_3}{4}  \frac{\langle12\rangle \langle31\rangle}{\langle23\rangle}$ \\
         $\phantom{+}0$ 
              & $-1$ & $\phantom{+}0$ 
                            &      $ \displaystyle
                                    -\frac{m_2^2}{4}   \frac{\langle12\rangle\langle23 \rangle}{\langle31\rangle}$ & 
                                   $ \displaystyle
                                     \frac{m_1m_2}{4}  \frac{\langle12\rangle\langle23 \rangle}{\langle31\rangle}$ &
                                   $ \displaystyle
                                     \frac{m_2m_3}{4}  \frac{\langle12\rangle\langle23 \rangle}{\langle31\rangle}$ \\
         $+1$ & $-1$ & $-1$ &      $ 0$ &
                                   $ \displaystyle
                                    -m_2 m_3 \frac{\langle23\rangle^3}{\langle12\rangle \langle31\rangle}$
                                   & $ 0$ \\
         $-1$ & $-1$ & $+1$ &      $0$ & $0$ &
                                   $ \displaystyle
                                    -m_1 m_2 \frac{\langle12\rangle^3}{\langle23\rangle \langle31\rangle}$ \\
         $-1$ & $+1$ & $-1$ &
                                   $ \displaystyle
                                    -m_1 m_3 \frac{\langle31\rangle^3}{\langle12\rangle \langle23\rangle}$ &
                                   $0$ & $0$ \\
%         $\phantom{+}0$ 
%              & $\phantom{+}0$ 
%                     & $\phantom{+}0$ 
%                            &      $0$ & $0$ & $0$ \\
%         $+1$ & $\phantom{+}0$ 
%                     & $-1$ &      $0$ & $0$ & $0$ \\
%         $+1$ & $-1$ & $\phantom{+}0$ 
%                            &      $0$ & $0$ & $0$ \\
%         $-1$ & $\phantom{+}0$ 
%                     & $+1$ &      $0$ & $0$ & $0$ \\
%         $\phantom{+}0$ 
%              & $-1$ & $+1$ &      $0$ & $0$ & $0$ \\
%         $-1$ & $+1$ & $\phantom{+}0$ 
%                            &      $0$ & $0$ & $0$ \\
         $\phantom{+}0$ 
              & $+1$ & $-1$ &      $0$ & $0$ & $0$ \\
         $+1$ & $-1$ & $+1$ &      $ \displaystyle
                                    -m_1 m_3 \frac{\lbrack31\rbrack^3}{\lbrack12\rbrack \lbrack23\rbrack}$ &
                                   $0$ & $0$ \\
         $-1$ & $+1$ & $+1$ &      $0$ &
                                   $ \displaystyle
                                    -m_2 m_3 \frac{\lbrack23\rbrack^3}{\lbrack12\rbrack \lbrack31\rbrack}$ &
                                   $0$ \\
         $+1$ & $+1$ & $-1$ &      $0$ & $0$ &
                                   $ \displaystyle
                                    -m_1 m_2 \frac{\lbrack12\rbrack^3}{\lbrack23\rbrack \lbrack31\rbrack}$ \\
         $\phantom{+}0$ 
              & $+1$ & $\phantom{+}0$ 
                            &      $ \displaystyle
                                    -\frac{m_2^2}{4}  \frac{\lbrack12\rbrack \lbrack23\rbrack}{\lbrack31\rbrack}$ &
                                   $ \displaystyle
                                     \frac{m_1m_2}{4} \frac{\lbrack12\rbrack \lbrack23\rbrack}{\lbrack31\rbrack}$ &
                                   $ \displaystyle
                                     \frac{m_2m_3}{4} \frac{\lbrack12\rbrack \lbrack23\rbrack}{\lbrack31\rbrack}$ \\
         $+1$ & $\phantom{+}0$ 
                     & $\phantom{+}0$ 
                            &      $ \displaystyle
                                     \frac{m_1m_2}{4} \frac{\lbrack12\rbrack \lbrack31\rbrack}{\lbrack23\rbrack}$ &
                                   $ \displaystyle
                                    -\frac{m_1^2}{4}  \frac{\lbrack12\rbrack \lbrack31\rbrack}{\lbrack23\rbrack}$ &
                                   $ \displaystyle
                                     \frac{m_1m_3}{4} \frac{\lbrack12\rbrack \lbrack31\rbrack}{\lbrack23\rbrack}$ \\
         $\phantom{+}0$ 
              & $\phantom{+}0$ 
                     & $+1$ &      $ \displaystyle
                                     \frac{m_2m_3}{4} \frac{\lbrack23\rbrack \lbrack31\rbrack}{\lbrack12\rbrack}$ &
                                   $ \displaystyle
                                     \frac{m_1m_3}{4} \frac{\lbrack23\rbrack \lbrack31\rbrack}{\lbrack12\rbrack}$ &
                                   $ \displaystyle
                                    -\frac{m_3^2}{4}  \frac{\lbrack23\rbrack \lbrack31\rbrack}{\lbrack12\rbrack}$ \\
%         $\phantom{+}0$ 
%              & $+1$ & $+1$ &      $0$ & $0$ & $0$ \\
%         $+1$ & $\phantom{+}0$ 
%                     & $+1$ &      $0$ & $0$ & $0$ \\
%         $+1$ & $+1$ & $\phantom{+}0$ 
%                            &      $0$ & $0$ & $0$ \\
%         $+1$ & $+1$ & $+1$ &      $0$ & $0$ & $0$ \\
         \hline
    \end{tabular}
    \caption{Contributions of the spinor products in Eq.~(\ref{eq:3 spin-1 general}) in the high-energy limit.  Their calculation is described in Sec.~\ref{sec:3 spin-1}.  The left column gives the helicities of the three particles. We have only listed helicity combinations that have non-zero contributions.}
    \label{tab:<><>[] terms}
\end{table}
We first find $\langle\mathbf{12}\rangle^{11}\langle\mathbf{23}\rangle^{11}\lbrack\mathbf{31}\rbrack^{11} + \lbrack\mathbf{12}\rbrack^{11}\lbrack\mathbf{23}\rbrack^{11}\langle\mathbf{31}\rangle^{11} = \mathcal{O}(m^2)=\mathcal{A}(-1,-1,-1)+\mathcal{O}(m^2)$.  We next begin incrementing the index on each particle, starting with particle~$1$.  We find,
\begin{align}
   &\frac{1}{2}
    \Big(
      \langle\mathbf{12}\rangle^{11}\langle\mathbf{23}\rangle^{11}\lbrack\mathbf{31}\rbrack^{12}
     +\langle\mathbf{12}\rangle^{21}\langle\mathbf{23}\rangle^{11}\lbrack\mathbf{31}\rbrack^{11}
     +\lbrack\mathbf{12}\rbrack^{11}\lbrack\mathbf{23}\rbrack^{11}\langle\mathbf{31}\rangle^{12}
     +\lbrack\mathbf{12}\rbrack^{21}\lbrack\mathbf{23}\rbrack^{11}\langle\mathbf{31}\rangle^{11}
    \Big) \nonumber \\
    & \qquad
    = - \frac{m_3}{2\sqrt{2E_3}}
    \langle12\rangle\langle23\rangle\lbrack\tilde{\zeta}^+_31\rbrack + \mathcal{O}(m^3).
\end{align}
We use momentum conservation $\lbrack\tilde{\zeta}^+_31\rbrack\langle12\rangle=-\lbrack\tilde{\zeta}^+_33\rbrack\langle32\rangle+\mathcal{O}(m^2)=-\sqrt{2E_3}\langle23\rangle+\mathcal{O}(m^2)$.  We then multiply by $\lbrack31\rbrack/\lbrack31\rbrack$ and use momentum conservation $\langle23\rangle\lbrack31\rbrack=\mathcal{O}(m^2)$ to see that this term vanishes at this order.  We find a similar result if we increment the index on particles $1$ or $2$ giving us,
\begin{eqnarray}
\frac{1}{2}\left(
    \langle\mathbf{12}\rangle^{11}\langle\mathbf{23}\rangle^{11}\lbrack\mathbf{31}\rbrack^{12} +
    \langle\mathbf{12}\rangle^{21}\langle\mathbf{23}\rangle^{11}\lbrack\mathbf{31}\rbrack^{11} +
    \lbrack\mathbf{12}\rbrack^{11}\lbrack\mathbf{23}\rbrack^{11}\langle\mathbf{31}\rangle^{12} +
    \lbrack\mathbf{12}\rbrack^{21}\lbrack\mathbf{23}\rbrack^{11}\langle\mathbf{31}\rangle^{11}
    \right) &=& \mathcal{O}(m^3), \\
\frac{1}{2}\left(    
    \langle\mathbf{12}\rangle^{11}\langle\mathbf{23}\rangle^{21}\lbrack\mathbf{31}\rbrack^{11} +
    \langle\mathbf{12}\rangle^{12}\langle\mathbf{23}\rangle^{11}\lbrack\mathbf{31}\rbrack^{11} +
    \lbrack\mathbf{12}\rbrack^{11}\lbrack\mathbf{23}\rbrack^{21}\langle\mathbf{31}\rangle^{11} +
    \lbrack\mathbf{12}\rbrack^{12}\lbrack\mathbf{23}\rbrack^{11}\langle\mathbf{31}\rangle^{11}
    \right) &=& \mathcal{O}(m^3), \\
\frac{1}{2}\left(
    \langle\mathbf{12}\rangle^{11}\langle\mathbf{23}\rangle^{11}\lbrack\mathbf{31}\rbrack^{21} +
    \langle\mathbf{12}\rangle^{11}\langle\mathbf{23}\rangle^{12}\lbrack\mathbf{31}\rbrack^{11} +
    \lbrack\mathbf{12}\rbrack^{11}\lbrack\mathbf{23}\rbrack^{11}\langle\mathbf{31}\rangle^{21} +
    \lbrack\mathbf{12}\rbrack^{11}\lbrack\mathbf{23}\rbrack^{12}\langle\mathbf{31}\rangle^{11}
    \right) &=& \mathcal{O}(m^3) ,
\end{eqnarray}
agreeing with $\mathcal{A}(0,-1,-1)=\mathcal{A}(-1,0,-1)=\mathcal{A}(-1,-1,0)=0$.
To find a nonzero result, we have to increment two indices, either the same one or different ones.  Let's begin by incrementing two different indices, starting with particles $1$ and $2$.
\begin{align}
   &\frac{1}{4}
    \Big(
      \langle\mathbf{12}\rangle^{11} \langle\mathbf{23}\rangle^{21} \lbrack\mathbf{31}\rbrack^{12}
     +\langle\mathbf{12}\rangle^{21} \langle\mathbf{23}\rangle^{21} \lbrack\mathbf{31}\rbrack^{11}
     +\langle\mathbf{12}\rangle^{12} \langle\mathbf{23}\rangle^{11} \lbrack\mathbf{31}\rbrack^{12}      
     +\langle\mathbf{12}\rangle^{22} \langle\mathbf{23}\rangle^{11} \lbrack\mathbf{31}\rbrack^{11} \nonumber \\
   & +\lbrack\mathbf{12}\rbrack^{11} \lbrack\mathbf{23}\rbrack^{21} \langle\mathbf{31}\rangle^{12}         
     +\lbrack\mathbf{12}\rbrack^{21} \lbrack\mathbf{23}\rbrack^{21} \langle\mathbf{31}\rangle^{11}
     +\lbrack\mathbf{12}\rbrack^{12} \lbrack\mathbf{23}\rbrack^{11} \langle\mathbf{31}\rangle^{12} 
     +\lbrack\mathbf{12}\rbrack^{22} \lbrack\mathbf{23}\rbrack^{11} \langle\mathbf{31}\rangle^{11}
    \Big) \nonumber \\ 
   &\displaystyle \qquad
    =-\frac{m_2m_3}{4\sqrt{4E_2E_3}}\langle12\rangle\langle\zeta^-_23\rangle\lbrack\tilde{\zeta}^+_31\rbrack
     -\frac{m_2m_3}{4\sqrt{4E_2E_3}}\langle1\zeta^-_2\rangle\langle23\rangle\lbrack\tilde{\zeta}^+_31\rbrack
     +\frac{m_2m_3}{4\sqrt{4E_2E_3}}\lbrack1\tilde{\zeta}^+_2\rbrack\lbrack2\tilde{\zeta}^+_3\rbrack\langle31\rangle 
     + \mathcal{O}(m^3) \nonumber \\
   &\displaystyle \qquad
    =\frac{m_2m_3}{4\sqrt{2E_2}}\langle23\rangle\langle\zeta^-_23\rangle
     +\frac{m_2m_3}{4\sqrt{2E_3}}\frac{\langle23\rangle\lbrack23\rbrack}{\lbrack31\rbrack}
     +\frac{m_2m_3}{4\sqrt{2E_3}}\langle23\rangle\lbrack2\tilde{\zeta}^+_3\rbrack 
     + \mathcal{O}(m^3) \nonumber \\
   &\displaystyle \qquad
    =-\frac{m_2m_3}{4}\frac{\lbrack12\rbrack\langle23\rangle}{\lbrack31\rbrack}
    +\frac{m_2m_3}{4}\frac{\langle23\rangle\langle31\rangle}{\langle12\rangle}
     + \mathcal{O}(m^3) \nonumber \\
   &\displaystyle \qquad
    =\frac{m_2m_3}{4}\frac{\langle23\rangle\langle31\rangle}{\langle12\rangle}
     + \mathcal{O}(m^3) =
    \frac{m_2m_3}{4}\mathcal{A}\left(0,0,-1\right)
     + \mathcal{O}(m^3),
\end{align}
where, for the first round of simplification, we used $\lbrack\tilde{\zeta}^+_31\rbrack\langle12\rangle = -\lbrack\tilde{\zeta}^+_33\rbrack\langle32\rangle+\mathcal{O}(m^2)=-\sqrt{2E_3}\langle23\rangle+\mathcal{O}(m^2)$, we multiplied the middle term by $\lbrack31\rbrack/\lbrack31\rbrack$ and used $\lbrack31\rbrack\langle1\zeta^-_2\rangle=-\lbrack32\rbrack\langle2\zeta^-_2\rangle+\mathcal{O}(m^2)=\sqrt{2E_2}\lbrack23\rbrack+\mathcal{O}(m^2)$, and $\langle31\rangle\lbrack1\tilde{\zeta}^+_2\rbrack=-\langle32\rangle\lbrack2\tilde{\zeta}^+_2\rbrack+\mathcal{O}(m^2)=\sqrt{2E_2}\langle23\rangle+\mathcal{O}(m^2)$.  For the second round of simplification, we multiplied the first term by $\lbrack31\rbrack/\lbrack31\rbrack$ and used $\langle\zeta^-_23\rangle\lbrack31\rbrack=-\langle\zeta^-_22\rangle\lbrack21\rbrack+\mathcal{O}(m^2)=-\sqrt{2E_2}\lbrack12\rbrack+\mathcal{O}(m^2)$.  The second term contains $\langle23\rangle\lbrack23\rbrack=-2p_2\cdot p_3+\mathcal{O}(m^2)=-(p_2+p_3)^2+\mathcal{O}(m^2)=-p_1^2+\mathcal{O}(m^2)=\mathcal{O}(m^2)$ placing it at higher order.  We multiply the third term by $\langle12\rangle/\langle12\rangle$ and use $\langle12\rangle\lbrack2\tilde{\zeta}^+_3\rbrack=-\langle13\rangle\lbrack3\tilde{\zeta}^+_3\rbrack+\mathcal{O}(m^2)=\sqrt{2E_3}\langle31\rangle+\mathcal{O}(m^2)$.  For the final round of simplification, we note that $\lbrack12\rbrack\langle23\rangle=\mathcal{O}(m^2)$.  We see that this gives us the helicity amplitude for a helicity-$0$ particle~$1$ and $2$ and a $-1$-helicity particle~$3$, as expected and agrees with Eq.~(\ref{eq:massless:0,0,1}).  This, of course, is an interaction with two Goldstone bosons.  The factor $\mathcal{N}_2$ will be determined by an interaction without Goldstones in it, so we wait to determine the coefficient.  We can find the case where we increment indices on particles $2$ and $3$ by interchanging $1$ and $3$ and introducing a minus sign on both sides.
\begin{align}
   &\frac{1}{4}
    \Big(
     \langle\mathbf{12}\rangle^{11}\langle\mathbf{23}\rangle^{21}\lbrack\mathbf{31}\rbrack^{21} +
     \langle\mathbf{12}\rangle^{11}\langle\mathbf{23}\rangle^{22}\lbrack\mathbf{31}\rbrack^{11} + 
     \langle\mathbf{12}\rangle^{12}\langle\mathbf{23}\rangle^{11}\lbrack\mathbf{31}\rbrack^{21} +
     \langle\mathbf{12}\rangle^{12}\langle\mathbf{23}\rangle^{12}\lbrack\mathbf{31}\rbrack^{11} + \nonumber \\
   & \qquad
     \lbrack\mathbf{12}\rbrack^{11}\lbrack\mathbf{23}\rbrack^{21}\langle\mathbf{31}\rangle^{21} +
     \lbrack\mathbf{12}\rbrack^{11}\lbrack\mathbf{23}\rbrack^{22}\langle\mathbf{31}\rangle^{11} +
     \lbrack\mathbf{12}\rbrack^{12}\lbrack\mathbf{23}\rbrack^{11}\langle\mathbf{31}\rangle^{21} +
     \lbrack\mathbf{12}\rbrack^{12}\lbrack\mathbf{23}\rbrack^{12}\langle\mathbf{31}\rangle^{11}
    \Big) \nonumber \\
   &\displaystyle \qquad =
    \frac{m_1m_2}{4}
    \frac{\langle12\rangle\langle31\rangle}{\langle23\rangle} + \mathcal{O}(m^3) =
    \frac{m_1m_2}{4}\mathcal{A}\left(-1,0,0\right) + \mathcal{O}(m^3),
\end{align}
again agreeing with Eq.~(\ref{eq:massless:0,0,1}).  The case where we increment indices on particles 1 and 3 will require more work.  
\begin{align}
   &\frac{1}{4}
    \Big(
      \langle\mathbf{12}\rangle^{11}\langle\mathbf{23}\rangle^{11}\lbrack\mathbf{31}\rbrack^{22} +
      \langle\mathbf{12}\rangle^{21}\langle\mathbf{23}\rangle^{11}\lbrack\mathbf{31}\rbrack^{21} + 
      \langle\mathbf{12}\rangle^{11}\langle\mathbf{23}\rangle^{12}\lbrack\mathbf{31}\rbrack^{12} +
      \langle\mathbf{12}\rangle^{21}\langle\mathbf{23}\rangle^{12}\lbrack\mathbf{31}\rbrack^{11} + \nonumber \\
   &  \qquad
      \lbrack\mathbf{12}\rbrack^{11}\lbrack\mathbf{23}\rbrack^{11}\langle\mathbf{31}\rangle^{22} +
      \lbrack\mathbf{12}\rbrack^{21}\lbrack\mathbf{23}\rbrack^{11}\langle\mathbf{31}\rangle^{21} +
      \lbrack\mathbf{12}\rbrack^{11}\lbrack\mathbf{23}\rbrack^{12}\langle\mathbf{31}\rangle^{12} +
      \lbrack\mathbf{12}\rbrack^{21}\lbrack\mathbf{23}\rbrack^{12}\langle\mathbf{31}\rangle^{11}
    \Big) \nonumber \\
   & \displaystyle \qquad
     = \frac{1}{4}
         \left(1-\frac{m_1^2}{4E_1^2}-\frac{m_2^2}{4E_2^2}-\frac{m_3^2}{4E_3^2}
         \right)\langle12\rangle\langle23\rangle\lbrack31\rbrack 
      -\frac{m_1^2}{8E_1}\langle\zeta^-_12\rangle\langle23\rangle\lbrack3\tilde{\zeta}^+_1\rbrack \nonumber \\
   & \qquad \qquad
     \displaystyle
      -\frac{m_3^2}{8E_3}\langle12\rangle\langle2\zeta^-_3\rangle\lbrack\tilde{\zeta}^+_31\rbrack
      +\frac{m_2^2}{8E_2}\lbrack1\tilde{\zeta}^+_2\rbrack\lbrack\tilde{\zeta}^+_23\rbrack\langle31\rangle
      +\mathcal{O}(m^3) \nonumber \\
   & \displaystyle \qquad
     = -\frac{m_1^2}{4\sqrt{2E_1}}\langle\zeta^-_12\rangle\langle12\rangle
       +\frac{m_3^2}{4\sqrt{2E_3}}\langle2\zeta^-_3\rangle\langle23\rangle
       -\frac{m_2^2}{4\sqrt{2E_2}}\lbrack1\tilde{\zeta}^+_2\rbrack\langle12\rangle
       +\mathcal{O}(m^3) \nonumber \\
   & \displaystyle \qquad
     =  \frac{m_1^2}{4}\frac{\langle12\rangle\lbrack31\rbrack}{\lbrack23\rbrack}
       +\frac{m_3^2}{4}\frac{\langle23\rangle\lbrack31\rbrack}{\lbrack12\rbrack}
       -\frac{m_2^2}{4}\frac{\langle12\rangle\langle23\rangle}{\langle31\rangle}
       +\mathcal{O}(m^3) \nonumber \\
   & \displaystyle \qquad
     = -\frac{m_2^2}{4}\frac{\langle12\rangle\langle23\rangle}{\langle31\rangle}
       +\mathcal{O}(m^3) =
       -\frac{m_2^2}{4}\mathcal{A}\left(0,-1,0\right)
       +\mathcal{O}(m^3).
\end{align}
In the first round of simplification, we multiply the first term by $\lbrack31\rbrack/\lbrack31\rbrack$ and note that both $\lbrack31\rbrack\langle12\rangle=\mathcal{O}(m^2)$ and $\langle23\rangle\lbrack31\rbrack=\mathcal{O}(m^2)$ putting this term at $\mathcal{O}(m^4)$ and not contributing at this order.  For the second term, we used $\langle23\rangle\lbrack3\tilde{\zeta}^+_1\rbrack=-\langle21\rangle\lbrack1\tilde{\zeta}^+_1\rbrack+\mathcal{O}(m^2)=\sqrt{2E_1}\langle12\rangle+\mathcal{O}(m^2)$.  For the third term, we use $\lbrack\tilde{\zeta}^+_31\rbrack\langle12\rangle=-\lbrack\tilde{\zeta}^+_33\rbrack\langle32\rangle+\mathcal{O}(m^2)=-\sqrt{2E_3}\langle23\rangle+\mathcal{O}(m^2)$.  For the fourth term, we use $\lbrack\tilde{\zeta}^+_23\rbrack\langle31\rangle=-\lbrack\tilde{\zeta}^+_22\rbrack\langle21\rangle+\mathcal{O}(m^2)=-\sqrt{2E_2}\langle12\rangle+\mathcal{O}(m^2)$.  For the second round of simplifcation, we multiply the first term by $\lbrack23\rbrack/\lbrack23\rbrack$ and use $\langle\zeta^-_12\rangle\lbrack23\rbrack=-\langle\zeta^-_11\rangle\lbrack13\rbrack+\mathcal{O}(m^2)=-\sqrt{2E_1}\lbrack31\rbrack+\mathcal{O}(m^2)$.  For the second term, we multiply by $\lbrack12\rbrack/\lbrack12\rbrack$ and use $\lbrack12\rbrack\langle2\zeta^-_3\rangle=-\lbrack13\rbrack\langle3\zeta^-_3\rangle+\mathcal{O}(m^2)=\sqrt{2E_3}\lbrack31\rbrack+\mathcal{O}(m^2)$.  For the third term, we multiply by $\langle31\rangle/\langle31\rangle$ and use $\langle31\rangle\lbrack1\tilde{\zeta}^+_2\rbrack=-\langle32\rangle\lbrack2\tilde{\zeta}^+_2\rbrack+\mathcal{O}(m^2)=\sqrt{2E_2}\langle23\rangle+\mathcal{O}(m^2)$.  Finally, in the last simplification, we use that $\langle12\rangle\lbrack31\rbrack=\mathcal{O}(m^2)$and $\langle23\rangle\lbrack31\rbrack=\mathcal{O}(m^2)$ to see that the first two terms are higher order.  As usual, we find that our results agrees with the massless vertices given in Eq.~(\ref{eq:massless:0,0,1}).

We next increment the same particle's index twice.  We begin by incrementing particle~$1$.  We obtain,
\begin{eqnarray}
    \langle12\rangle^{21}\langle23\rangle^{11}\lbrack31\rbrack^{12} +
    \lbrack12\rbrack^{21}\lbrack23\rbrack^{11}\langle31\rangle^{12} &=&
    -\frac{m_1m_3}{\sqrt{4E_1E_3}}
     \langle\zeta^-_12\rangle \langle23\rangle \lbrack\tilde{\zeta}^+_31\rbrack +
    \mathcal{O}(m^3) \nonumber \\
    &=& 
    -\frac{m_1m_3}{\sqrt{2E_3}}
    \frac{\lbrack31\rbrack\langle23\rangle\lbrack\tilde{\zeta}^+_31\rbrack}{\lbrack23\rbrack}
    +\mathcal{O}(m^3) = \mathcal{O}(m^3).
\end{eqnarray}
We simplified by multiplying by $\lbrack23\rbrack/\lbrack23\rbrack$ and using $\langle\zeta^-_12\rangle\lbrack23\rbrack = -\langle\zeta^-_11\rangle \lbrack13\rbrack + \mathcal{O}(m^2) = -\sqrt{2E_1}\lbrack31\rbrack + \mathcal{O}(m^2)$.  Finally, we note that $\langle23\rangle\lbrack31\rbrack=\mathcal{O}(m^2)$ showing us that this is higher order.  We must find the contribution for helicities $(+1,-1,-1)$ from a different term.  Likewise, we find that if we increment the index on particle~$3$ twice, we find,
\begin{equation}
    \langle12\rangle^{11}\langle23\rangle^{12}\lbrack31\rbrack^{21}+
    \lbrack12\rbrack^{11}\lbrack23\rbrack^{12}\langle31\rangle^{21} =
    \mathcal{O}(m^3) .
\end{equation}
On the other hand, if we increment particle~$2$ twice, we obtain,
\begin{eqnarray}
    \langle12\rangle^{12}\langle23\rangle^{21}\lbrack31\rbrack^{11}+
    \lbrack12\rbrack^{12}\lbrack23\rbrack^{21}\langle31\rangle^{11} &=&
    \frac{m_1m_3}{\sqrt{4E_1E_3}}\lbrack\tilde{\zeta}^+_12\rbrack\lbrack2\tilde{\zeta}^+_3\rbrack\langle31\rangle + \mathcal{O}(m^3) \nonumber\\
    &=& -m_1m_3\frac{\langle31\rangle^3}{\langle23\rangle\langle12\rangle} + \mathcal{O}(m^3).
\end{eqnarray}
Simplifying the first time involves mutliplying by $\langle23\rangle/\langle23\rangle$ and $\langle12\rangle/\langle12\rangle$ and using $\lbrack\tilde{\zeta}^+_12\rbrack\langle23\rangle=-\lbrack\tilde{\zeta}^+_11\rbrack\langle13\rangle+\mathcal{O}(m^2)=-\sqrt{2E_1}\langle31\rangle+\mathcal{O}(m^2)$ and $\langle12\rangle\lbrack2\tilde{\zeta}^+_3\rbrack=-\langle13\rangle\lbrack3\tilde{\zeta}^+_3\rbrack+\mathcal{O}(m^2)=\sqrt{2E_3}\langle31\rangle+\mathcal{O}(m^2)$.  As we can see this result agrees with the massless vertex $\mathcal{A}(-1,+1,-1)$ from Eq.~(\ref{eq:3-point gluon amplitude}).

We next consider incrementing indices three times.  We expect to find zero at this order for all these cases since the spins won't add up to $\pm1$.  We begin by incrementing the index on all three particles once.  We obtain,
\begin{align}
   &\frac{1}{8}
    \Big(
        \langle12\rangle^{11}\langle23\rangle^{21}\lbrack31\rbrack^{22} 
       +\langle12\rangle^{11}\langle23\rangle^{22}\lbrack31\rbrack^{12} 
       +\langle12\rangle^{12}\langle23\rangle^{11}\lbrack31\rbrack^{22} 
       +\langle12\rangle^{12}\langle23\rangle^{12}\lbrack31\rbrack^{12} \nonumber \\
   & \qquad \qquad 
       +\langle12\rangle^{21}\langle23\rangle^{21}\lbrack31\rbrack^{21} 
       +\langle12\rangle^{21}\langle23\rangle^{22}\lbrack31\rbrack^{11}
       +\langle12\rangle^{22}\langle23\rangle^{11}\lbrack31\rbrack^{21}
       +\langle12\rangle^{22}\langle23\rangle^{12}\lbrack31\rbrack^{11}
       +\langle\rangle \leftrightarrow \lbrack\rbrack
    \Big) \nonumber \\
   & \displaystyle =
     \frac{m_2}{8\sqrt{2E_2}}\langle12\rangle\langle\zeta^-_23\rangle\lbrack31\rbrack +
     \frac{m_2}{8\sqrt{2E_2}}\langle1\zeta^-_2\rangle\langle23\rangle\lbrack31\rbrack  \nonumber \\
   & \qquad \qquad 
    -\frac{m_2}{8\sqrt{2E_2}}\lbrack1\tilde{\zeta}^+_2\rbrack\lbrack23\rbrack\langle31\rangle -
     \frac{m_2}{8\sqrt{2E_2}}\lbrack12\rbrack\lbrack\tilde{\zeta}^+_23\rbrack\langle31\rangle 
    +\mathcal{O}(m^3) = \mathcal{O}(m^3),
\end{align}
where the $1/8$ is due to symmetrizing three indices.  However, all of these are zero at this order due to momentum conservation such as $\lbrack31\rbrack\langle12\rangle = \mathcal{O}(m^2)$.  
We next try incrementing the index on particle~$1$ twice and another once.  For example, 
\begin{eqnarray}
    \frac{1}{2}\left(
        \langle12\rangle^{21}\langle23\rangle^{21}\lbrack31\rbrack^{12} +
        \langle12\rangle^{22}\langle23\rangle^{11}\lbrack31\rbrack^{12} +
        \lbrack12\rbrack^{21}\lbrack23\rbrack^{21}\langle31\rangle^{12} +
        \lbrack12\rbrack^{22}\lbrack23\rbrack^{11}\langle31\rangle^{12}
    \right) = \mathcal{O}(m^3), 
\end{eqnarray}
and,
\begin{align}
   &\frac{1}{2}
     \left(
        \langle12\rangle^{21}\langle23\rangle^{11}\lbrack31\rbrack^{22} +
        \langle12\rangle^{21}\langle23\rangle^{12}\lbrack31\rbrack^{12} +
        \lbrack12\rbrack^{21}\lbrack23\rbrack^{11}\langle31\rangle^{22} +
        \lbrack12\rbrack^{21}\lbrack23\rbrack^{12}\langle31\rangle^{12}
     \right) \nonumber \\
   & \qquad \qquad
     \displaystyle
     = \frac{m_1}{2\sqrt{2E_1}}\langle\zeta^-_12\rangle\langle23\rangle\lbrack31\rbrack + \mathcal{O}(m^3)
     = \mathcal{O}(m^3),
\end{align}
where we have used $\langle23\rangle \lbrack31\rbrack = \mathcal{O}(m^2)$.  If we increment particle~$3$ twice, we find the same results.  If we increment particle~$2$ twice, we find,
\begin{align}
   &\frac{1}{2}
      \left(
        \langle12\rangle^{12}\langle23\rangle^{21}\lbrack31\rbrack^{12} +
        \langle12\rangle^{22}\langle23\rangle^{21}\lbrack31\rbrack^{11} +
        \lbrack12\rbrack^{12}\lbrack23\rbrack^{21}\langle31\rangle^{12} +
        \lbrack12\rbrack^{22}\lbrack23\rbrack^{21}\langle31\rangle^{11}
      \right) \nonumber \\
   & \qquad \qquad
     = - \frac{m_3}{2\sqrt{2E_3}}\lbrack12\rbrack\lbrack2\tilde{\zeta}^+_3\rbrack\langle31\rangle + \mathcal{O}(m^3) 
     = \mathcal{O}(m^3), \\
   &\frac{1}{2}
      \left(
        \langle12\rangle^{12}\langle23\rangle^{21}\lbrack31\rbrack^{21} +
        \langle12\rangle^{12}\langle23\rangle^{22}\lbrack31\rbrack^{11} +
        \lbrack12\rbrack^{12}\lbrack23\rbrack^{21}\langle31\rangle^{21} +
        \lbrack12\rbrack^{12}\lbrack23\rbrack^{22}\langle31\rangle^{11}
      \right) \nonumber \\
   & \qquad \qquad 
     = -\frac{m_1}{2\sqrt{2E_1}}\lbrack\tilde{\zeta}^+_12\rbrack\lbrack23\rbrack\langle31\rangle + \mathcal{O}(m^3) 
     = \mathcal{O}(m^3),
\end{align}
where we have used $\langle31\rangle\lbrack12\rbrack=\mathcal{O}(m^2)$ and $\lbrack23\rbrack\langle31\rangle=\mathcal{O}(m^2)$.

Next, we increment the indices four times.  We begin by incrementing two indices twice.  We expect to only find a nonzero result when we increment index $1$ and $3$.  For this case we find,
\begin{eqnarray}
    \langle12\rangle^{21}\langle23\rangle^{12}\lbrack31\rbrack^{22} +
    \lbrack12\rbrack^{21}\lbrack23\rbrack^{12}\langle31\rangle^{22} &=&
    \frac{m_1m_3}{\sqrt{4E_1E_3}}\langle\zeta^-_12\rangle\langle2\zeta^-_3\rangle\lbrack31\rbrack + \mathcal{O}(m^3)\nonumber\\
    &=& -m_1m_3\frac{\lbrack31\rbrack^3}{\lbrack12\rbrack\lbrack23\rbrack} 
    +\mathcal{O}(m^3),
\end{eqnarray}
where we have multiplied by $\lbrack23\rbrack/\lbrack23\rbrack$ and $\lbrack12\rbrack/\lbrack12\rbrack$ and used $\langle\zeta^-_12\rangle\lbrack23\rbrack = -\langle\zeta^-_11\rangle\lbrack13\rbrack + \mathcal{O}(m^2) = -\sqrt{2E_1}\lbrack31\rbrack + \mathcal{O}(m^2)$ and $\lbrack12\rbrack\langle2\zeta^-_3\rangle = -\lbrack13\rbrack\langle3\zeta^-_3\rangle + \mathcal{O}(m^2)=\sqrt{2E_3}\lbrack31\rbrack + \mathcal{O}(m^2)$.  This result agrees with $\mathcal{A}(+1,-1,+1)$ given in Eq.~(\ref{eq:3-point gluon amplitude}).  We next consider incrementing the indices on particles $1$ and $2$ twice.  We find,
\begin{eqnarray}
    \langle12\rangle^{22}\langle23\rangle^{21}\lbrack31\rbrack^{12} +
    \lbrack12\rbrack^{22}\lbrack23\rbrack^{21}\langle31\rangle^{12} &=&
    -\frac{m_1m_3}{\sqrt{4E_1E_3}}\lbrack12\rbrack\lbrack2\tilde{\zeta}^+_3\rbrack\langle3\zeta^-_1\rangle + \mathcal{O}(m^3) = \mathcal{O}(m^3).
\end{eqnarray}
We have simplified by multiplying by $\langle12\rangle/\langle12\rangle$ and used $\langle12\rangle\lbrack2\tilde{\zeta}^+_3\rbrack = -\langle13\rangle\lbrack3\tilde{\zeta}^+_3\rbrack + \mathcal{O}(m^2)=\sqrt{2E_3}\langle31\rangle + \mathcal{O}(m^2)$.  However, we then have $\langle31\rangle\lbrack12\rbrack = \mathcal{O}(m^2)$ putting this at higher order.  The same is true for,
\begin{equation}
    \langle12\rangle^{12}\langle23\rangle^{22}\lbrack31\rbrack^{21} +
    \lbrack12\rbrack^{12}\lbrack23\rbrack^{22}\langle31\rangle^{21} =
    \mathcal{O}(m^3).
\end{equation}
We have included these results in Table~\ref{tab:<><>[] terms}. We continue to increment four total times, but now we increment one index twice and the other two once.  We expect all of these to be nonzero and contribute to Goldstone interactions.  We begin by incrementing the index on particle~$2$ twice.  
\begin{align}
   &\frac{1}{4}
      \Big(
        \langle12\rangle^{12}\langle23\rangle^{21}\lbrack31\rbrack^{22} +
        \langle12\rangle^{12}\langle23\rangle^{22}\lbrack31\rbrack^{12} +
        \langle12\rangle^{22}\langle23\rangle^{21}\lbrack31\rbrack^{21} +
        \langle12\rangle^{22}\langle23\rangle^{22}\lbrack31\rbrack^{11} +
        \langle\rangle \leftrightarrow \lbrack\rbrack
      \Big) \nonumber\\
   & \qquad 
     = \frac{m_2^2}{8E_2}\langle1\zeta^-_2\rangle\langle\zeta^-_23\rangle\lbrack31\rbrack 
      -\frac{m_3^2}{8E_3}\lbrack12\rbrack\lbrack2\tilde{\zeta}^+_3\rbrack\langle\zeta^-_31\rangle
      +\mathcal{O}(m^3)
     = \frac{m_2^2}{4\sqrt{2E_2}}\lbrack23\rbrack\langle\zeta^-_23\rangle
      +\frac{m_3^2}{4\sqrt{2E_3}}\lbrack23\rbrack\lbrack2\tilde{\zeta}^+_3\rbrack
      +\mathcal{O}(m^3) \nonumber\\
   & \qquad 
     =-\frac{m_2^2}{4}\frac{\lbrack23\rbrack\lbrack12\rbrack}{\lbrack31\rbrack} 
      +\mathcal{O}(m^3).
\end{align}
In the first round of simplifications, we used $\lbrack31\rbrack\langle1\zeta^-_2\rangle = -\lbrack32\rbrack\langle2\zeta^-_2\rangle + \mathcal{O}(m^2) = \sqrt{2E_2}\lbrack23\rbrack + \mathcal{O}(m^2)$ and $\langle\zeta^-_31\rangle\lbrack12\rbrack = -\langle\zeta^-_33\rangle\lbrack32\rbrack + \mathcal{O}(m^2) = -\sqrt{2E_3}\lbrack23\rbrack + \mathcal{O}(m^2)$.  In the second round of simplifications, we multiply the first term by $\lbrack31\rbrack/\lbrack31\rbrack$ and use $\langle\zeta^-_23\rangle\lbrack31\rbrack = -\langle\zeta^-_22\rangle\lbrack21\rbrack + \mathcal{O}(m^2) = -\sqrt{2E_2}\lbrack12\rbrack$ and we multiply the second term by $\langle12\rangle/\langle12\rangle$ and use $\langle12\rangle\lbrack2\tilde{\zeta}^+_3\rbrack = -\langle13\rangle\lbrack3\tilde{\zeta}^+_3\rbrack + \mathcal{O}(m^2) = \sqrt{2E_3}\langle31\rangle + \mathcal{O}(m^2)$ and note that $\lbrack23\rbrack\langle31\rangle=\mathcal{O}(m^2)$.  We next increment the index on particle~$1$ twice.  
\begin{align}
   &\frac{1}{4}
      \Big(
        \langle12\rangle^{21}\langle23\rangle^{21}\lbrack31\rbrack^{22} +
        \langle12\rangle^{21}\langle23\rangle^{22}\lbrack31\rbrack^{12} +
        \langle12\rangle^{22}\langle23\rangle^{11}\lbrack31\rbrack^{22} +
        \langle12\rangle^{22}\langle23\rangle^{12}\lbrack31\rbrack^{12} +
        \langle\rangle \leftrightarrow \lbrack\rbrack
      \Big) \nonumber \\
   & \qquad 
     =  \frac{m_1m_2}{4\sqrt{4E_1E_2}}\langle\zeta^-_12\rangle\langle\zeta^-_23\rangle\lbrack31\rbrack
       +\frac{m_1m_2}{16E_1E_2}\lbrack12\rbrack\langle23\rangle\lbrack31\rbrack 
       -\frac{m_1m_2}{4\sqrt{4E_1E_2}}\lbrack1\tilde{\zeta}^+_2\rbrack\lbrack23\rbrack\langle3\zeta^-_1\rangle
       \nonumber \\ 
   & \qquad \qquad
     - \frac{m_1m_2}{4\sqrt{4E_1E_2}}\lbrack12\rbrack\lbrack\tilde{\zeta}^+_23\rbrack\langle3\zeta^-_1\rangle
     + \mathcal{O}(m^3) \nonumber \\
   & \qquad 
     = -\frac{m_1m_2}{4\sqrt{2E_1}}\lbrack12\rbrack\langle\zeta^-_12\rangle
       -\frac{m_1m_2}{4\sqrt{2E_2}}\lbrack12\rbrack\lbrack1\tilde{\zeta}^+_2\rbrack
       +\mathcal{O}(m^3)
     =  \frac{m_1m_2}{4}\frac{\lbrack12\rbrack\lbrack31\rbrack}{\lbrack23\rbrack}
       +\mathcal{O}(m^3).
\end{align}
In the first round of simplification, we use $\langle\zeta^-_23\rangle\lbrack31\rbrack = -\langle\zeta^-_22\rangle\lbrack21\rbrack + \mathcal{O}(m^2) = -\sqrt{2E_2}\lbrack12\rbrack+\mathcal{O}(m^2)$, $\lbrack12\rbrack\langle23\rangle = \mathcal{O}(m^2)$, $\lbrack23\rbrack\langle3\zeta^-_1\rangle = -\lbrack21\rbrack\langle1\zeta^-_1\rangle + \mathcal{O}(m^2) = \sqrt{2E_1}\lbrack12\rbrack + \mathcal{O}(m^2)$, and multiply the last term by $\langle31\rangle/\langle31\rangle$ and use $\lbrack\tilde{\zeta}^+_23\rbrack\langle31\rangle = -\lbrack\tilde{\zeta}^+_22\rbrack\langle21\rangle + \mathcal{O}(m^2)$ followed by $\lbrack12\rbrack\langle21\rangle=\mathcal{O}(m^2)$.  In the second round of simplification, we multiply the first term by $\lbrack23\rbrack/\lbrack23\rbrack$ and use $\langle\zeta^-_12\rangle\lbrack23\rbrack = -\langle\zeta^-_11\rangle\lbrack13\rbrack + \mathcal{O}(m^2) = -\sqrt{2E_1}\lbrack31\rbrack + \mathcal{O}(m^2)$ and multiply the second term by $\langle31\rangle/\langle31\rangle$ and use $\langle31\rangle\lbrack1\tilde{\zeta}^+_2\rbrack = -\langle32\rangle\lbrack2\tilde{\zeta}^+_2\rbrack + \mathcal{O}(m^2) = \sqrt{2E_2}\langle23\rangle+\mathcal{O}(m^2)$ followed by $\lbrack12\rbrack\langle23\rangle = \mathcal{O}(m^2)$.  Similarly we find,
\begin{align}
   &\frac{1}{4}
      \Big(
        \langle12\rangle^{11}\langle23\rangle^{22}\lbrack31\rbrack^{22} +
        \langle12\rangle^{12}\langle23\rangle^{12}\lbrack31\rbrack^{22} +
        \langle12\rangle^{21}\langle23\rangle^{22}\lbrack31\rbrack^{21} +
        \langle12\rangle^{22}\langle23\rangle^{12}\lbrack31\rbrack^{21} +
        \langle\rangle \leftrightarrow \lbrack\rbrack
      \Big) \nonumber \\
   & \qquad \qquad = \frac{m_2m_3}{4}\frac{\lbrack23\rbrack\lbrack31\rbrack}{\lbrack12\rbrack}
    +\mathcal{O}(m^3).
\end{align}
We have included all these results in Table~\ref{tab:<><>[] terms}.

We now increment the indices five times, leaving only one particle at $0$-helicity.  We expect all of these to be zero at this order.  We begin with particle~$1$ being $0$-helicity and find
\begin{align}
   &\frac{1}{2}
      \Big(
        \langle12\rangle^{12}\langle23\rangle^{22}\lbrack31\rbrack^{22} +
        \langle12\rangle^{22}\langle23\rangle^{22}\lbrack31\rbrack^{21} +
        \lbrack12\rbrack^{12}\lbrack23\rbrack^{22}\langle31\rangle^{22} +
        \lbrack12\rbrack^{22}\lbrack23\rbrack^{22}\langle31\rangle^{21}
       \Big) \nonumber \\
   & \qquad \qquad
     = \frac{m_3}{2\sqrt{2E_3}}\lbrack12\rbrack\lbrack23\rbrack\langle\zeta^-_31\rangle
     + \mathcal{O}(m^3).
\end{align}
We use $\langle\zeta^-_31\rangle\lbrack12\rbrack = -\langle\zeta^-_33\rangle\lbrack32\rbrack + \mathcal{O}(m^2) = -\sqrt{2E_3}\lbrack23\rbrack + \mathcal{O}(m^2)$.  We then multiply by $\langle31\rangle/\langle31\rangle$ and note that $\lbrack23\rbrack\langle31\rangle = \mathcal{O}(m^2)$ to see that this term is higher order.  The other cases are similar giving us,
\begin{eqnarray}
    \frac{1}{2}\left(
        \langle12\rangle^{12}\langle23\rangle^{22}\lbrack31\rbrack^{22} +
        \langle12\rangle^{22}\langle23\rangle^{22}\lbrack31\rbrack^{21} +
        \lbrack12\rbrack^{12}\lbrack23\rbrack^{22}\langle31\rangle^{22} +
        \lbrack12\rbrack^{22}\lbrack23\rbrack^{22}\langle31\rangle^{21}
    \right) &=&
    \mathcal{O}(m^3), \\
    \frac{1}{2}\left(
        \langle12\rangle^{22}\langle23\rangle^{12}\lbrack31\rbrack^{22} +
        \langle12\rangle^{21}\langle23\rangle^{22}\lbrack31\rbrack^{22} +
        \lbrack12\rbrack^{22}\lbrack23\rbrack^{12}\langle31\rangle^{22} +
        \lbrack12\rbrack^{21}\lbrack23\rbrack^{22}\langle31\rangle^{22}
    \right) &=&
    \mathcal{O}(m^3), \\
    \frac{1}{2}\left(
        \langle12\rangle^{22}\langle23\rangle^{22}\lbrack31\rbrack^{12} +
        \langle12\rangle^{22}\langle23\rangle^{21}\lbrack31\rbrack^{22} +
        \lbrack12\rbrack^{22}\lbrack23\rbrack^{22}\langle31\rangle^{12} +
        \lbrack12\rbrack^{22}\lbrack23\rbrack^{21}\langle31\rangle^{22}
    \right) &=&
    \mathcal{O}(m^3).
\end{eqnarray}
Finally, if we increment six times so that all particles are $1$-spin, we find
\begin{equation}
    \langle12\rangle^{22}\langle23\rangle^{22}\lbrack31\rbrack^{22} +
    \lbrack12\rbrack^{22}\lbrack23\rbrack^{22}\langle31\rangle^{22} =
    \frac{m_1m_3}{4E_1E_3}\lbrack12\rbrack\lbrack23\rbrack\lbrack31\rbrack
    +\mathcal{O}(m^3).
\end{equation}
However, multiplying this by $\langle23\rangle/\langle23\rangle$ and using $\lbrack12\rbrack\langle23\rangle=\mathcal{O}(m^2)$ shows that this is higher order leaving us with
\begin{equation}
    \langle12\rangle^{22}\langle23\rangle^{22}\lbrack31\rbrack^{22} +
    \lbrack12\rbrack^{22}\lbrack23\rbrack^{22}\langle31\rangle^{22} =
    \mathcal{O}(m^3)
\end{equation}
All of these results have been included in Table~\ref{tab:<><>[] terms}.

The other two columns of Table~\ref{tab:<><>[] terms} can be filled in using simple symmetry arguments.  We will just focus on the non-zero terms, since the zeros in the other two columns are straight-forward to see.  For the nonzero terms, we simply rotate the particles once or twice.  We begin by considering $1\to2\to3\to1$.  In this case,
\begin{equation}
    \langle\mathbf{12}\rangle\langle\mathbf{23}\rangle\lbrack\mathbf{31}\rbrack
    \to
    \lbrack\mathbf{12}\rbrack\langle\mathbf{23}\rangle\langle\mathbf{31}\rbrack.
\end{equation}
Next, if we look at the first non-zero entry in the first column, we find,
\begin{equation}
    \frac{m_2m_3}{4}\frac{\langle23\rangle\langle31\rangle}{\langle12\rangle}
    \to 
    \frac{m_1m_3}{4}\frac{\langle12\rangle\langle31\rangle}{\langle23\rangle}.
\end{equation}
We also need to keep in mind that the helicities of the particles also rotate, so we have spins $(0,0,-1)$ goes to $(-1,0,0)$.  This makes sense since this is the massless result for $\mathcal{A}(-1,0,0)$, so we add this to the table. The other Goldstone interactions are similar and we have entered them all in Table~\ref{tab:<><>[] terms}.  We will also do a non-Goldstone term.  We find,
\begin{equation}
    -m_1m_3\frac{\langle31\rangle^3}{\langle12\rangle\langle23\rangle}
    \to 
    -m_1m_2\frac{\langle12\rangle^3}{\langle23\rangle\langle31\rangle}
\end{equation}
and the helicities $(-1,+1,-1)$ goes to $(-1,-1,+1)$.

Now that we have calculated the high-energy limit for all the terms, we must determine the normalization coefficients $\mathcal{N}_2, \mathcal{N}_3$ and $\mathcal{N}_4$.  As usual, we will do this using the non-Goldstone terms so that the interactions with all helicities being $\pm1$ should be continuous in the massless limit.  We see from Table~\ref{tab:<><>[] terms} that we should choose,
\begin{equation}
    \mathcal{N}_1=-\frac{1}{m_1m_3},\quad
    \mathcal{N}_2=-\frac{1}{m_2m_3} \quad
    \mbox{and}\quad
    \mathcal{N}_3=-\frac{1}{m_1m_2}.
\end{equation}
For the high-energy limit of the Goldstone boson terms, we must take the masses to zero at the same rate.
Taking these coefficients into account, we have added these vertices to Table~\ref{tab:ew gauge} for the $WWZ$ vertex.  We find all the massless interactions accounted for including both with $\pm 1$-helicity bosons by themselves and with Goldstone bosons and agree with the vertices in Eqs.~(\ref{eq:3-point gluon amplitude}) and (\ref{eq:massless:0,0,1}).

One may wonder where the Goldstone interactions are that include two particles of $\pm 1$-helicity or the interactions with all helicity-$0$ particles.  We expect that these are higher order in the expansion.  In both cases, there are not any combinations of helicity spinors that give the right transformation properties and the right mass dimension of $1$.  But, both these become possible when the mass dimension of the vertex is shared between the helicity-spinors and the masses of the particles.  We leave these interactions, which are outside the scope of the present paper, to a future work.

\subsection{\label{subsec:Higgs}The Higgs Couplings}
In this section, we work out the final vertices, those of the $0$-spin Higgs boson.  We begin with the interaction with massive fermions.  We will take the Higgs boson as the third particle.  This vertex will contain one spin-spinor of either type for particle~$1$ and the same for particle~$2$.  We have only one possibility for this vertex.  It is,
\begin{equation}
    \langle\mathbf{12}\rangle+\lbrack\mathbf{12}\rbrack =
    \left(\begin{array}{cc}
    \langle12\rangle&0\\
    0&\lbrack12\rbrack
    \end{array}\right)
    + \mathcal{O}(m_f).
    \label{eq:H,1/2,1/2}
\end{equation}
The top-left term corresponds with both the fermion and anti-fermion having $-1/2$-helicity while the bottom-right term is for when they both have $+1/2$-helicity.  We have included this vertex in Table \ref{tab:Higgs}.

We next consider the vertex with two $1$-spin bosons, appropriate to the interactions with $W$ and $Z$-boson.  Each of particle~$1$ and $2$ have two spin-spinors of either type.  They cannot be contracted with each other, so we must contract them with spin-spinors for the other particle.  Therefore, this vertex is,
\begin{equation}
    \mathcal{N}_1\langle\mathbf{12}\rangle\lbrack\mathbf{12}\rbrack +
    \mathcal{N}_2\left(\langle\mathbf{12}\rangle\langle\mathbf{12}\rangle+\lbrack\mathbf{12}\rbrack\lbrack\mathbf{12}\rbrack\right),
    \label{eq:HWW general}
\end{equation}
where $\mathcal{N}_1$ and $\mathcal{N}_2$ are unrelated to the normalization constants of previous sections and have inverse mass dimension of $1$.  Since these terms have two indices, they can be written in matrix form. We use Eqs.~(\ref{eq:<12>^2 expanded in m}) and (\ref{eq:<12>[12] expanded in m}) and keep terms to linear order in the masses. 
\begin{equation}
    \langle\mathbf{12}\rangle\lbrack\mathbf{12}\rbrack =
    \left[
    \begin{array}{ccc} \displaystyle
       0
     & \displaystyle
        -\frac{m_1}{2\sqrt{2E_1}}\langle 12\rangle\lbrack \tilde{\zeta}_1^+2\rbrack
     & \displaystyle
       0 \\ \displaystyle
       -\frac{m_2}{2\sqrt{2E_2}}\langle 12\rangle\lbrack 1\tilde{\zeta}_2^+\rbrack
     & \displaystyle \langle12\rangle\lbrack12\rbrack
     & \displaystyle
     \frac{m_2}{2\sqrt{2E_2}}\lbrack 12\rbrack\langle 1\zeta_2^-\rangle\\ \displaystyle
       0 
     & \displaystyle
     \frac{m_1}{2\sqrt{2E_1}}\lbrack 12\rbrack\langle\zeta_1^-2\rangle & \displaystyle
     0
    \end{array}
    \right]
    +\mathcal{O}(m^2).
\end{equation}
The center term is $-2p_1\cdot p_2+\mathcal{O}(m^2)$.   We see that this is $-2p_1 \cdot p_2 = -(p_1+p_2)^2 + \mathcal{O}(m^2) = -p_3^2 + \mathcal{O}(m^2) = \mathcal{O}(m^2)$ and so does not contribute at this order.  We next focus on the left entry.  We multiply it by $\langle31\rangle/\langle31\rangle$ and use $\langle31\rangle\lbrack1\tilde{\zeta}_2^+\rbrack = \sqrt{2E_2}\langle23\rangle + \mathcal{O}(m^2)$.  We multiply the right term by $\lbrack31\rbrack/\lbrack31\rbrack$ and use $\lbrack31\rbrack\langle1\zeta_2^-\rangle = \sqrt{2E_2}\lbrack23\rbrack+\mathcal{O}(m^2)$.  Next, we multiply the top term by $\langle23\rangle/\langle23\rangle$ and use $\lbrack\tilde{\zeta}_1^+2\rbrack\langle23\rangle = -\sqrt{2E_1}\langle31\rangle + \mathcal{O}(m^2)$.  We multiply the bottom term by $\lbrack23\rbrack/\lbrack23\rbrack$ and use $\langle\zeta_1^-2\rangle\lbrack23\rbrack = -\sqrt{2E_1}\lbrack31\rbrack + \mathcal{O}(m^2)$.  All together we have,
\begin{equation}
    \langle\mathbf{12}\rangle\lbrack\mathbf{12}\rbrack =
    \left[\begin{array}{ccc} \displaystyle
         0 & \displaystyle
         \frac{m_1}{2}\frac{\langle12\rangle\langle31\rangle}{\langle23\rangle} &
        0 \\ \displaystyle
         -\frac{m_2}{2}\frac{\langle12\rangle\langle23\rangle}{\langle31\rangle}& \displaystyle
        0 
         & \displaystyle
        \frac{m_2}{2}\frac{\lbrack12\rbrack\lbrack23\rbrack}{\lbrack31\rbrack} \\ \displaystyle
        0 & \displaystyle
        -\frac{m_1}{2}\frac{\lbrack12\rbrack\lbrack31\rbrack}{\lbrack23\rbrack}& 0
    \end{array}\right]
    +\mathcal{O}(m^2).
    \label{eq:HWW general HE 1}
\end{equation}
We see that in all these terms, particle~$3$ has helicity-$0$, as it must.  In the center row, we see that particle~$1$ is also helicity-$0$.  It is a Goldstone boson.  Particle~$2$ is a $-1$-helicity particle in the left column and a $+1$-helicity particle in the right column.  The middle column is the opposite case.  Particle~$2$ is a helicity-$0$ Goldstone boson while particle~$1$ has $-1$-helicity in the top row and $+1$-helicity in the bottom row.  As before, these agree with the massless vertices, $\mathcal{A}(0,\pm1,0)$ in the center row and $\mathcal{A}(\pm1,0,0)$ in the middle column, as given in the appendix.

It is not clear what $\mathcal{N}_1$ should be here.  But, we suspect it should be either the Higgs mass or the vacuum expectation value $v$ for the Higgs.  We will take the latter, $\mathcal{N}_1=1/v$.  Since the masses are present only when $v \neq 0$, we do not expect these terms to be continuous unless we take both the mass and $v$ to zero simultaneously.  

Next, we consider,
\begin{equation}
    \langle\mathbf{12}\rangle\langle\mathbf{12}\rangle =
    \left[
    \begin{array}{ccc} \displaystyle
    \langle 12 \rangle^2 & \displaystyle
    \frac{m_2}{\sqrt{2E_2}} \langle 12 \rangle \langle 1 \zeta_2^- \rangle & \displaystyle
    0 \\ \displaystyle
    \frac{m_1}{\sqrt{2E_1}}\langle 12\rangle\langle\zeta_1^- 2\rangle & \displaystyle
    0 & 0 \\ \displaystyle
    0 & 0 & 0
    \end{array}\right]
    +\mathcal{O}(m^2)
\end{equation}
We multiply the top-left term by $\lbrack23\rbrack/\lbrack23\rbrack$ and see that it is zero at this order. We multiply the top-middle term by $\lbrack31\rbrack/\lbrack31\rbrack$ and use $\lbrack31\rbrack\langle1\zeta_2^-\rangle=\sqrt{2E_2}\lbrack23\rbrack+\mathcal{O}(m^2)$.  We follow this with $\langle12\rangle\lbrack23\rbrack=\mathcal{O}(m^2)$ to see that none of these terms contribute at this order, 
\begin{equation}
    \langle\mathbf{12}\rangle\langle\mathbf{12}\rangle =
    \mathcal{O}(m^2).
    \label{eq:HWW general HE 2}
\end{equation}
The same is true if we interchange angle and square brackets,
\begin{equation}
    \lbrack\mathbf{12}\rbrack\lbrack\mathbf{12}\rbrack =
    \mathcal{O}(m^2).
    \label{eq:HWW general HE 3}
\end{equation}
So, $\mathcal{N}_2$ appears to be unconstrained by the high-energy limit, at least at this order.  We have included this vertex in Table \ref{tab:Higgs}.

There are no fundamental constructive $3$-point vertices with two Higgs and one $Z$-boson because it would involve two spin-spinors for the $Z$-boson and nothing else.  They would have to be contracted with each other resulting in an antisymmetry in their SU($2$) index, which is symmetrized.  

We are left with only a vertex with all Higgs bosons.  Of course, since there are no spinors to work with, this vertex is simply a constant.  We have included this vertex in Table \ref{tab:Higgs}.

\section{\label{sec:conclusions}Conclusion}
In this paper, we have constructed the full set of minimal $3$-point vertices for the massive SM using only the symmetry properties of the S-Matrix, the mass dimension and the high-energy behavior.  We have done this in terms of on-shell particles without any recourse to fields or their gauge redundancies.  A powerful feature of this constructive approach is that the gravitational vertices are no more complicated than the other SM vertices and, so, we have included them as well.  In this section, we will first summarize some of our results.  We will follow this with a discussion of some of the open questions and possible future directions for research.

In Sec.~\ref{app:masses={m,m,0}}, we considered $3$-point vertices with one massless particle and two massive particles, appropriate to the QED, QCD and gravitational vertices.  Since these vertices only include one linearly independent helicity spinor, we followed \cite{Arkani-Hamed:2017jhn} by introducing the $x$ factor in Eq.~(\ref{eq:app:x definition}), which we used to construct the vertices for positive helicity.  We found it convenient to also introduce a $\tilde{x}$ factor in Eq.~(\ref{eq:app:xtilde definition}), which gave a more minimal and convenient form for the negative helicity vertices.  We then gave the vertex for a $+1$-helicity and $-1$-helicity particle interacting with two massive $1/2$-spin fermions in Eqs.~(\ref{eq:QQg:x<12>}) and (\ref{eq:QQg:xtilde[12]}), respectively.  These vertices were applicable to the interactions of a photon with charged fermions as well as a gluon with quarks.  We worked out the high-energy behavior of both of these, given in Eqs.~(\ref{eq:x<12> HE limit}) and (\ref{eq:xtilde[12] HE limit}), respectively, and showed that they agreed with the massless vertices given in Eq.~(\ref{eq:massless:1/2,-1/2,1}).  After finishing with the $1/2$-spin fermions, we went on to consider the interactions of a photon with two $1$-spin $W$-bosons.  We gave the vertices for $+1$-helicity and $-1$-helicity photons in Eqs.~(\ref{eq:x<12>^2/MW}) and (\ref{eq:xtilde[12]^2/MW}), respectively, and again worked out their high-energy behavior in Eqs.~(\ref{eq:x<12>^2/MW HE}) and (\ref{eq:xtilde[12]^2/MW HE}), showing that they also agreed with the massless vertices given in Eq.~(\ref{eq:3-point gluon amplitude}).  We further collect all these QED and QCD vertices, along with their high-energy behavior, in Tables~\ref{tab:qed} and \ref{tab:qcd}, respectively.  Finally, although the gravitational vertices are typically considered outside the scope of the SM, we show that in this constructive formalism, they are equally simple as all the other SM vertices.  We begin with the interactions with a $1/2$-spin fermion.  We give the vertices for a $+2$-helicity and $-2$-helicity graviton in Eq.~(\ref{eq:x^2mf<12>}), and work out their high-energy behavior in Eqs.~(\ref{eq:x^2mf<12> HE}) and (\ref{eq:xtilde^2mf[12] HE}), respectively, showing that they agreed with the massless vertices given in Eq.~(\ref{eq:A(1/2,-1/2,2)}).  For $1$-spin bosons, we give the vertices for $+2$-helicity and $-2$-helicity in Eq.~(\ref{eq:x^2<12>^2}), with their high-energy limit in Eqs.~(\ref{eq:x^2<12>^2 HE}) and (\ref{eq:xtilde^2[12]^2 HE}), respectively, agreeing with the massless vertices in Eq.~(\ref{eq:A(1,-1,2)}).  We also consider the gravitational interaction of the Higgs.  In Eq.~(\ref{eq:x^2Mh^2}), we give the Higgs interaction with a $+2$-helicity graviton and its high-energy behavior in Eq.~(\ref{eq:x^2Mh^2 HE}).  In Eq.~(\ref{eq:xtilde^2Mh^2}), we give the interaction with a $-2$-helicity graviton, together with its high-energy behavior. In both cases, we note that the high-energy behavior agrees with the massless vertices given in Eq.~(\ref{eq:A(0,0,2)}).  We collect all the gravitational vertices, along with their high-energy behavior, in Table~\ref{tab:gravitational}.  Again, we comment on how impressive it is that such a simple structure spans such a large subset of the SM.  It covers all the QED, QCD and gravitational vertices.

Although the previous structure covered an impressive fraction of the SM, it did not cover everything.  The next vertex structure we consider is that of two massless particles and one massive particle, covering the weak interaction of the $Z$-boson and the neutrinos.  We give the vertex for a $-1/2$-helicity neutrino and a $+1/2$-helicity anti-neutrino in Eq.~(\ref{eq:<3b1>[23b]}), followed by its high-energy behavior in Eq.~(\ref{eq:<3b1>[23b] HE}), which agrees with the all-massless vertex given in Eq.~(\ref{eq:massless:1/2,-1/2,1}).  Notwithstanding the fact that we know the neutrinos have a mass, we do not yet have a complete, verified theory of neutrino masses.  Therefore, we do not consider massive neutrinos in this paper.  However, adding massive neutrinos when the right-chiral sector is understood, should not be difficult.  We have included these neutrino interactions, along with their high-energy behavior, in Table~\ref{tab:ew gauge}.

In Sec.~\ref{sec: 1 massless and 2 massive with different masses}, we turn our attention to vertices with one massless and two massive particles, where the masses are different, which will apply to the $W$-boson interaction with leptons.  Unlike the previous cases, this vertex structure is not uniquely determined by the symmetry and mass dimension alone.  There are two structures that must be considered as given in Eq.~(\ref{eq:Wlnubar}) for the $+1/2$-helicity anti-neutrino and in Eq.~(\ref{eq:Wlnu}) for the $-1/2$-helicity neutrino.  In an attempt to uniquely resolve these structures, we look at each of their high-energy behaviors.  We find that one of the two structures vanishes in the high-energy limit (its leading-order behavior is $m/E$) in Eq.~(\ref{eq:Wlnu}).  As a result, the coefficient of this term cannot be determined by a comparison with massless vertices.  This poses a challenge for this constructive method since this term cannot currently be determined solely based on symmetry, dimension and high-energy behavior, without considering field theory or Feynman vertices.  We consider the determination of this coefficient an open problem.  On the other hand, the other structure does have a nonzero high-energy limit, which we show in Eqs.~(\ref{eq:Wlnubar HE 2}) for the anti-neutrino and (\ref{eq:Wlnu HE 2}) for the neutrino.  Both of these agree with the massless vertices given in Eq.~(\ref{eq:A(0,1/2,1/2)}) for the Goldstone boson and Eq.~(\ref{eq:massless:1/2,-1/2,1}) for the $\pm 1$-helicity boson.  We include these vertices, along with their high-energy behavior in Table~\ref{tab:ew gauge}.

There is one important case left, the vertex with three massive particles, and we attack this in Sec.~\ref{sec: 3 massive}.  In Subsec.~\ref{sec:3massive 1/2,1/2,1}, we consider the case of two $1/2$-spin fermions and one $1$-spin boson, appropriate to the interactions of the $Z$-boson and fermions as well as the $W$-boson and quarks.  As in the previous section, we find that there are two independent structures that must be considered, as seen in Eq.~(\ref{eq:1/2,1/2,1 general}).  Also, as in the previous section, we find that one of the structures vanishes in the high-energy limit, as seen in Eqs.~(\ref{eq:1/2,1/2,1 general vanish 1}), (\ref{eq:1/2,1/2,1 general vanish 2}), (\ref{eq:1/2,1/2,1 general vanish 3}) and (\ref{eq:1/2,1/2,1 general vanish 4}) , and therefore, its coefficient cannot be determined from the structure of the massless vertices.  On the other hand, the other structure does have a nonzero high-energy limit, as we show in Eqs.~(\ref{eq:ffZ:A+--}), (\ref{eq:ffZ:A-+-}), (\ref{eq:ffZ:A--0}), (\ref{eq:ffZ:A++0}), (\ref{eq:ffZ:A+-+}) and (\ref{eq:ffZ:A-++}).  All of these agree with the massless vertices given in Eq.~(\ref{eq:A(0,1/2,1/2)}) for the Goldstone boson interaction and Eq.~(\ref{eq:massless:1/2,-1/2,1}) for the $\pm 1$-helicity interaction.  We include these vertices along with their high-energy limit in Table~\ref{tab:ew gauge}.  We next move onto Subsection~\ref{sec:3 spin-1}, where we consider the interactions of three $1$-spin bosons, applicable to the interaction of a $Z$-boson and two $W$-bosons.  This vertex has the greatest ambiguity with four independent terms, as shown in Eq.~(\ref{eq:3 spin-1 general}).  We find that the first of these vanishes at high energy in Eq.~(\ref{eq:3 spin-1 general HE 1}) and so its coefficient is not determined by the massless vertices.  However, the other three have nonzero terms at high energy.  We display these high-energy limits in Table~\ref{tab:<><>[] terms}.  We find that the high-energy limit of these corresponds with the massless vertices given in Eq.~(\ref{eq:massless:0,0,1}) for the Goldstone boson interactions and Eq.~(\ref{eq:3-point gluon amplitude}) for the pure $\pm 1$-helicity boson interactions, as we expect.  We have also included this vertex and its high-energy behavior in Table~\ref{tab:ew gauge}.  

In the last subsection, Subsection~\ref{subsec:Higgs}, we consider the interactions of the Higgs boson, a $0$-spin particle.  We begin with its interaction with $1/2$-spin fermions.  This vertex, along with its high-energy limit, is given in Eq.~(\ref{eq:H,1/2,1/2}) and agrees with the massless vertex in Eq.~(\ref{eq:A(0,1/2,1/2)}).  We then consider the Higgs interaction with $1$-spin bosons, such as the $W$ and $Z$-bosons.  This $3$-point vertex is given in Eq.~(\ref{eq:HWW general}) and has two independent terms that should be determined.  We find the high-energy limit of the first term in Eq.~(\ref{eq:HWW general HE 1}) and find that it agrees with the massless vertices in Eq.~(\ref{eq:massless:0,0,1}).  The other term, on the other hand, vanishes in the high-energy limit, as seen in Eqs.~(\ref{eq:HWW general HE 2}) and (\ref{eq:HWW general HE 3}), and so is unconstrained by the massless vertices.  We have included all these Higgs vertices in Table~\ref{tab:Higgs}.

Although we have been able to construct an incredible fraction of the minimal SM by simply considering the symmetry, mass dimensions and high-energy limit of the S-Matrix, several open questions remain.  First of all, although the majority of the minimal $3$-point vertices were  uniquely determined by considering the high-energy limit and comparing to the massless vertices, there were a few terms that vanished in this limit, and are therefore, currently left ambiguous. One could determine these by comparing with Feynman vertices, but that goes against the spirit of the constructive approach. We could also compare with experiment, but we wonder if there is something more fundamental to guide us.  At this point, we do not know.

In this paper, we have also only considered the 3-point vertices, but it is well known that some theories, such as $0$-spin theories, also have $4$-point ``contact" vertices that must be included when calculating $4$-point amplitudes and beyond.  This is beyond the scope of the present paper, but we intend to include it in a later work.  It remains to be seen whether they can all be determined based purely on the properties of the S-matrix, including analyticity and unitarity, or whether comparison with Feynman diagrams must be done in order to successfully construct all of the vertices of a given theory.  However, our expectation is that the perturbative unitarity of $2\to2$ scattering will either require the presence of or require the absence of 4-point contact terms for most particle combinations.  For those processes where the contact term is required, we expect that the high-energy limit of the contact term along with its required cancellation of the high-energy growth in that process will determine which form it should take, much as the high-energy limit of the 3-point amplitudes fixed the couplings of most of the vertices in this paper.  For those processes where the contact term is not required but, nevertheless, is allowed, we expect the coupling to be a free parameter.  We expect the Higgs' 4-point amplitude to fall under this latter scenario.

Although \cite{Arkani-Hamed:2017jhn} have shown how to combine these $3$-point vertices together with on-shell propagators to form $4$-point amplitudes, it is still unclear what the general BCFW-like rule is for constructing higher-point amplitudes in complete generality.  We need a consistent set of rules that can be followed to calculate any amplitude with any number of particles.  We do not yet know what that rule is.  Presumably, it has something in common with the BCFW rules.  Moreover, a consistent set of rules for constructing amplitudes with any number of loops still needs to be worked out. We envision that in the near future,  the tools to construct any particle theory and calculate any scattering amplitude, in complete generality, without any direct influence by fields or gauge symmetries and their inherent redundancies will be a realistic option for high-energy physicists.

Once an amplitude is worked out, it must be appropriately squared and summed (or averaged) over the spins in order to obtain  a desired decay rate or  scattering cross-section.  It appears to us that after multiplying by the conjugate and summing over spins, we will find pieces like $|\mathbf{i}\rangle_\alpha^J\lbrack\mathbf{i}|_{\dot{\beta}J}$ which we will replace with the momentum matrix $p_{\alpha\dot{\beta}}$.  This will create chains of 2x2 matrices that are traced similar to gamma matrices in Feynman diagrams, but only 2 dimensional.  This appears to be straight forward to us, but needs to be worked out in detail.  We intend to do this in an upcoming work.  Of course, the helicity amplitudes could also be numerically calculated for each helicity, and only afterwards squared and added together.  Presumably, there will be situations where each is advantageous.

\section{Acknowledgements}
We would like to thank Nima Arkani-Hamed for answering our questions about \cite{Arkani-Hamed:2017jhn} during the early stages of this work.

\appendix

\section{\label{app:conventions}Conventions}
In this appendix, we state our full set of conventions and work out some important results that are used throughout this paper.  Our conventions are the same as those in \cite{Arkani-Hamed:2017jhn}.  We use the mostly negative metric so that $p^2=m^2$.  The SL(2,$\mathbb{C}$) indices are raised and lowered with,
\begin{equation}
\epsilon_{\alpha\beta}=-\epsilon^{\alpha\beta}=\epsilon_{\dot{\alpha}\dot{\beta}}=-\epsilon^{\dot{\alpha}\dot{\beta}}=\left(\begin{array}{cc}0&-1\\1&0\end{array}\right).
\end{equation}
The momenta are written with SL(2,$\mathbb{C}$) indices as,
\begin{equation}
    p_{\alpha\dot{\beta}}=%p_\mu\sigma^\mu_{\alpha\dot{\beta}}=
    \left(\begin{array}{cc}p^0+p^3&p^1-i p^2\\p^1+i p^2&p^0-p^3\end{array}\right),
    \label{eq:p_alphabetadot}
\end{equation}
\begin{equation}
    p^{\dot{\alpha}\beta}=%p_\mu\bar{\sigma}^{\mu\dot{\alpha}\beta}=
    \left(\begin{array}{cc}p^0-p^3&-p^1+i p^2\\-p^1-i p^2&p^0+p^3\end{array}\right),
    \label{eq:p^alphadotbeta}
\end{equation}
where the determinant of both of these satisfies,
\begin{equation}
    \det(p_{\alpha\dot{\beta}})=\det(p^{\dot{\alpha}\beta}) = p^2=m^2.
\end{equation}
Next, we state the properties of our spin-spinors [representations of SU(2)$\times$SL(2,$\mathbb{C}$)].  To simplify our discussion, we will only use the angle- and square-bracket notation rather than introducing them along with $\lambda$ and $\tilde{\lambda}$.  We hope this will make it easier for the reader to keep track of the calculations and not have to switch back and forth between the two.  Explicitly, we can write our momentum fully in terms of our spin-spinors as,
\begin{eqnarray}
    p_{i\alpha\dot{\beta}}&=&\epsilon_{IJ}|\mathbf{i}\rangle_\alpha^I\lbrack \mathbf{i}|_{\dot{\beta}}^J,
\end{eqnarray}
for the momentum of particle~$i$, where the greek letters represent the SL($2,\mathbb{C}$) indices and the capital latin letters represent the SU($2$) little group indices.  In the massless limit, this definition holds if we unbold the $\mathbf{i}$ and drop the SU($2$) index.  The momentum would become rank $1$ and these spin-spinors become helicity spinors, as we will shortly see. 

We normally suppress these indices except when they are needed for clarity.  Instead, we rely on the type and direction of the brackets to determine the type and position of the SL($2,\mathbb{C}$) indices.  So, 
\begin{eqnarray}
    |\mathbf{i}\rangle &:=& |\mathbf{i}\rangle_\alpha^I, \qquad
    \langle\mathbf{i}|  :=  \langle\mathbf{i}|^{\alpha I}, \\
    |\mathbf{i}\rbrack &:=& |\mathbf{i}\rbrack^{\dot{\alpha}I}, \qquad
    \lbrack\mathbf{i}|  :=  \lbrack\mathbf{i}|_{\dot{\alpha}}^{I},
\end{eqnarray}
where we have used $\alpha$ and $I$ as examples of the above indices.  They are, of course, unique to each product.  In the amplitudes we discuss, the SU($2$) indices are never contracted but are fully symmetrized. The SL($2,\mathbb{C}$) indices, on the other hand, are always fully contracted in amplitudes making SL($2,\mathbb{C}$) invariant objects that still transform under the little group as our constructed amplitudes must.  With this notation, the indices can always be put back in afterwards. 
To completely illustrate our sample calculations, it is helpful to have explicit examples for clarity.  So, we give a few here, for instance, 
\begin{equation}
    \langle \mathbf{i j}\rangle = 
    \langle\mathbf{i}|^{\alpha I}|\mathbf{j}\rangle_\alpha^{J}, \qquad
    \lbrack \mathbf{i j}\rbrack = 
    \lbrack\mathbf{i}|_{\dot{\alpha}}^{I}|\mathbf{j}\rbrack^{\dot{\alpha} J},
    \label{eq:<ij> and [ij] def}
\end{equation}
where the $I$ and $J$ would be symmetrized if they corresponded with the same particle ($i=j$).  Here, we also see the usual convention that we can sum \textit{descending} undotted indices and \textit{ascending} dotted ones, which are then usually suppressed.  Another example with a momentum sandwiched in between is given by,
\begin{eqnarray}
    \langle\mathbf{i}|p_k|\mathbf{j}\rbrack &=& \langle\mathbf{i}|^{\alpha I}p_{k\alpha\dot{\beta}}|\mathbf{j}\rbrack^{\dot{\beta}J}, \\
    \lbrack\mathbf{i}|p_k|\mathbf{j}\rangle &=& \lbrack\mathbf{i}|_{\dot{\alpha}}^I p_{k}^{\dot{\alpha}\beta}|\mathbf{j}\rangle_\beta^J.
\end{eqnarray}
One more example with two momenta should be sufficient to illustrate all our needed properties (additional properties of two-component spinors are cataloged in Ref~\cite{Dreiner:2008tw}),
\begin{eqnarray}
    \langle\mathbf{i}|p_k p_l|\mathbf{j}\rangle &=&
    \langle\mathbf{i}|^{\alpha I} p_{k\alpha\dot{\beta}}p_l^{\dot{\beta}\omega}|\mathbf{j}\rangle_{\omega}^J,
    \label{eq:<i|pkpl|j>}\\
    \lbrack\mathbf{i}|p_k p_l|\mathbf{j}\rbrack &=&
    \lbrack\mathbf{i}|_{\dot{\alpha}}^{I} p_{k}^{\dot{\alpha}\beta}p_{l\beta\dot{\omega}}|\mathbf{j}\rbrack^{\dot{\omega}J}.
    \label{eq:[i|pkpl|j]}
\end{eqnarray}
At times we will want to switch the order of the momenta.  This can be done using the Clifford algebra property of the Pauli matrices but can also be worked out explicitly with the matrices given in Eqs.~(\ref{eq:p_alphabetadot}) and (\ref{eq:p^alphadotbeta}) resulting in,
\begin{eqnarray}
    p_{k}^{\dot{\alpha}\beta}p_{l\beta\dot{\omega}}+p_{l}^{\dot{\alpha}\beta}p_{k\beta\dot{\omega}} &=& 2p_k\cdot p_l \delta^{\dot{\alpha}}_{\dot{\omega}} \, ,\\
    p_{k\alpha\dot{\beta}}p_l^{\dot{\beta}\omega}+p_{l\alpha\dot{\beta}}p_k^{\dot{\beta}\omega} &=& 2p_k\cdot p_l \delta_{\alpha}^{\omega} \, ,
    \label{eq:p1p2+p2p1}
\end{eqnarray}
which are very convenient when we simplify amplitudes.  

Now that we understand the implicit indices, we can discuss a few more properties of these spinors.  When they represent massive particles, they satisfy the usual Dirac property,
\begin{eqnarray}
    \langle\mathbf{i}|p_{i}&=&m_i\lbrack\mathbf{i}|\label{eq:<i|pi}, \\
    p_{i}|\mathbf{i}\rbrack &=&-m_i|\mathbf{i}\rangle\label{eq:pi|i]}.
\end{eqnarray}
These can be solved for a standard momentum and then boosted into other reference frames.  The standard form given in \cite{Arkani-Hamed:2017jhn} is,
\begin{eqnarray} 
    |\mathbf{i}\rangle_\alpha^I &=& \sqrt{E+p}
                                 \; \zeta_\alpha^+(p) \; \zeta^{-I}(k)
                                 +  \sqrt{E-p}
                                 \; \zeta_\alpha^-(p) \; \zeta^{+I}(k), \label{eq:|i> expanded def} 
                                 \\ 
    \lbrack\mathbf{i}|_{\dot{\alpha}I} &=& \sqrt{E+p} 
                                       \; \tilde{\zeta}_{\dot{\alpha}}^-(p)
                                       \; \zeta^{+}_{I}(k) 
                                        + \sqrt{E-p}
                                       \; \tilde{\zeta}_{\dot{\alpha}}^+(p)
                                       \; \zeta^{-}_{I}(k), \label{eq:|i] expanded def}
\end{eqnarray}
where $E$ and $p$ are the energy and momentum of the $i$th particle and,
\begin{eqnarray}
    \zeta_\alpha^+(p) = \left(\begin{array}{c}c\\s\end{array}\right), 
     &\quad& 
    \tilde{\zeta}_{\dot{\alpha}}^-(p) = \left(\begin{array}{c}c\\s^*\end{array}\right), \\
    \zeta_\alpha^-(p) = \left(\begin{array}{c}-s^*\\c\end{array}\right), 
    &\quad& 
    \tilde{\zeta}_{\dot{\alpha}}^+(p) = \left(\begin{array}{c}-s\\c\end{array}\right),
\end{eqnarray}
with,
\begin{equation}
    c \equiv \cos\left(\frac{\theta}{2}\right),  \quad  
    s \equiv \sin\left(\frac{\theta}{2}\right)e^{i\phi},
\end{equation}
where $\theta$ is the polar angle and $\phi$ is the azimuthal angle.  Plugging these definitions in, we obtain,
\begin{eqnarray}
    |\mathbf{i}\rangle^I_\alpha &=& 
    \left(\begin{array}{cc}\sqrt{E+p}\ c & -\sqrt{E-p}\ s^* \\
    \sqrt{E+p}\ s                        & \sqrt{E-p}\ c\end{array}\right), \label{eq:|i> def} \\
    \lbrack\mathbf{i}|_{\dot{\alpha}I} &=& 
    \left(\begin{array}{cc}\sqrt{E+p}\ c & -\sqrt{E-p}\ s \\
    \sqrt{E+p}\ s^*                      & \sqrt{E-p}\ c\end{array}\right),
    \label{eq:|i] def}
\end{eqnarray}
where the SU($2$) index $I$ gives the column and the SL($2,\mathbb{C}$) indices $\alpha$ and $\dot{\alpha}$ give the row.  We can check that the product of these gives the momentum. Indeed, 
\begin{eqnarray}
    |\mathbf{j}\rangle^I_{\alpha}\lbrack\mathbf{j}|_{\dot{\beta}I} 
    &=&
    \left(\begin{array}{cc}
    E_j+p_j\cos\theta_j         & p_j\sin\theta_j e^{-i\phi_j} \\
    p_j\sin\theta_j e^{i\phi_j} & E_j-p_j\cos\theta_j\end{array}\right) \nonumber \\
    &=& 
    \left(\begin{array}{cc}
    E_j+p_{jz}                  & p_{jx} - i p_{jy} \\
    p_{jx}+i p_{jy}             & E_j - p_{jz} \end{array} \right) \nonumber \\
    &=& 
    \left(\begin{array}{cc}
    p^0 + p^3                   & p^1 - i p^2 \\
    p^1 + i p^2                 & p^0 - p^3 \end{array} \right)
    \equiv p_{\alpha\dot{\beta}},
\end{eqnarray}
of particle~$j$.

Before we can explicitly calculate other expressions, we must explicitly define our convention for raising and lowering indices using the $\epsilon$-tensors.  For example, there is a sign difference whether we define raising the spin index using $\epsilon^{\alpha\beta}|\mathbf{j}\rangle_\alpha^I$ or $\epsilon^{\beta\alpha}|\mathbf{j}\rangle_\alpha^I$.  In order to have a helpful mnemonic to remember the convention, we will do it in the order that we would expect from matrix operations.  That is, we will define, 
\begin{eqnarray}
    \langle\mathbf{i}|^{\alpha I} & = & 
       \epsilon^{\alpha\beta}|\mathbf{i}\rangle_\beta^I, \qquad
    |\mathbf{i}\rangle_{\alpha I}   = 
       |\mathbf{i}\rangle_\alpha^J\epsilon_{JI}, \\
    |\mathbf{i}\rbrack^{\dot{\alpha}}_I & = & 
       \epsilon^{\dot{\alpha}\dot{\beta}} \lbrack \mathbf{i} |_{\dot{\beta}I}, \qquad
    \lbrack\mathbf{i}|_{\dot{\alpha}}^I   = 
       \lbrack \mathbf{i} |_{\dot{\alpha}J} \epsilon^{JI},
\end{eqnarray}
and so on.  This should remove any ambiguity from our calculations.  

With these definitions, we also find the following useful identities,
\begin{eqnarray}
    |\mathbf{i}\rbrack_I^{\dot{\alpha}}\langle\mathbf{i}|^{\alpha I} &=& p^{\dot{\alpha}\alpha}, \\
    \langle\mathbf{jj}\rangle^{IK} &=& -m_j\epsilon^{IK} \label{eq:<jj>}, \\
    \lbrack\mathbf{jj}\rbrack_{IK} &=& -m_j\epsilon_{IK} \label{eq:[jj]},
\end{eqnarray}
leading to,
\begin{eqnarray}
    \langle\mathbf{j}|^{\alpha K}p_{j\alpha\dot{\alpha}} &= \langle\mathbf{jj}\rangle^{KI}\lbrack\mathbf{j}|_{\dot{\alpha}I} =& m_j \lbrack\mathbf{j}|_{\dot{\alpha}}^K
    \label{eq:<j|j>[j|},\\
    p_{j\alpha\dot{\alpha}}|\mathbf{j}\rbrack^{\dot{\alpha}}_K &=|\mathbf{j}\rangle_\alpha^I\lbrack\mathbf{jj}\rbrack_{IK} =& -m_j |\mathbf{j}\rangle_{\alpha K},
    \label{eq:|j>[j|j]}
\end{eqnarray}
which agrees with Eqs.~(\ref{eq:<i|pi}) and (\ref{eq:pi|i]}).  Finally, we find,
\begin{eqnarray}
    p_j^{\dot{\alpha}\alpha}|\mathbf{j}\rangle_\alpha^K &=
    |\mathbf{j}\rbrack_I^{\dot{\alpha}}\langle\mathbf{jj}\rangle^{IK} &= -m_j|\mathbf{j}\rbrack^{\alpha K},\\
    \lbrack\mathbf{j}|_{\dot{\alpha}K}p_j^{\dot{\alpha}\alpha} &=
    \lbrack\mathbf{jj}\rbrack_{KI}\langle\mathbf{j}|^{\alpha I} &=
    m_j\langle\mathbf{j}|_K^{\alpha},
\end{eqnarray}
which can be written more minimally with our implicit index structure as,
\begin{eqnarray}
    p_j|\mathbf{j}\rangle &=& -m_j|\mathbf{j}\rbrack
    \label{eq:pi|i>}, \\
    \lbrack\mathbf{j}|p_j &=& +m_j\langle\mathbf{j}|,
    \label{eq:[i|pi}
\end{eqnarray}
where it is understood that the SU($2$) index is at the same height on both sides of the equation. So, now we have a mnemonic for remembering where the signs are.  When a momentum acts on an angle or square bracket from the left, we find a minus sign.  When a momentum acts on either bracket from the right, we find a plus sign.  

We take all particles as in-going, so momentum conservation is given by $\sum p_i=0$.

\subsection{High-Energy Limit}
We would now like to consider the high-energy limit of the conventions in this appendix and how to find the massless limit of the amplitudes calculated within this scheme.  We first note that the determinant of these matrices are identically zero (in the massless limit).  Because of this, they are rank 1 and now their construction only requires one independent spinor.  Looking back at Eqs.~(\ref{eq:|i> def}) and (\ref{eq:|i] def}), we see that we can simply use the first column of these matrices in this limit giving us the helicity-spinors,
\begin{equation}
    |i\rangle_{\alpha} = \sqrt{2E}\left(\begin{array}{c}c\\s\end{array}\right), \qquad
    \lbrack i|_{\dot{\alpha}} = \sqrt{2E}\left(\begin{array}{c}c\\s^*\end{array}\right),
\end{equation}
where we have removed the bold from the $i$ to signify that these are helicity-spinors rather than spin-spinors, and therefore, we do not require a spin index.  The massless momentum is now simply given by,
\begin{equation}
p_{i\alpha\dot{\beta}} = |i\rangle_\alpha \lbrack i|_{\dot{\beta}}.
\end{equation}
When we use all of our above properties and some trigonometry, we find,
\begin{equation}
    p_{i\alpha\dot{\beta}} = 
    E_i \left(
        \begin{array}{cc} 
          1 + \cos\theta_i         & \sin\theta_i e^{-i\phi_i} \\
          \sin\theta_i e^{i\phi_i} & 1 - \cos\theta_i\end{array}
        \right),
\end{equation}
as expected. We also find,
\begin{equation}
    \langle i j \rangle \lbrack j i \rbrack = 2 p_i\cdot p_j .
\end{equation}

Therefore, we see that, by defining these helicity-spinors to be the massless limit of the massive spin-spinors, we sometimes only need unbold the spin-spinors (dropping all the spin indices) and drop all the masses to obtain the massless limit when this process is smooth (when we do not need to worry about encountering an ambiguous $0/0$).  In fact, this was the design of \cite{Arkani-Hamed:2017jhn}.  In particular, we find the well known property of helicity-spinors, \begin{eqnarray}
    \langle j j\rangle = 0 \quad\mbox{and}\quad
    \lbrack j j\rbrack = 0,
\end{eqnarray}
as it must since the same helicity-spinor is contracted with the antisymmetric $\epsilon$-tensor. 

However, when an amplitude is not smooth (for example, if there is an $m$ in the denominator), we must be a bit more careful and expand to a higher-order in mass.  To quadratic order in $m$, we can write Eqs.~(\ref{eq:|i> expanded def}) and (\ref{eq:|i] expanded def}) as,
\begin{eqnarray} \nonumber
    |\mathbf{i}\rangle_\alpha^I &=& 
       \sqrt{2E_i} \left(1-\frac{m_i^2}{8E_i^2}\right) \; \zeta_\alpha^+(p) \; \zeta^{-I}(k)
     + \frac{m_i}{\sqrt{2E_i}} \; \zeta_\alpha^-(p) \; \zeta^{+I}(k) \nonumber \\
    &=& \sqrt{2E_i}
    \left[\begin{array}{ll} \displaystyle
      ( 1-m_i^2/8E_i^2 ) c & -(m_i/2E_i) s^* \\ 
      ( 1-m_i^2/8E_i^2 ) s & +(m_i/2E_i) c
    \end{array}\right], \label{eq:|i>^I expanded in m}\\
    \lbrack\mathbf{i}|_{\dot{\alpha}I} 
    &=& \sqrt{2E_i} 
        \left(1-\frac{m_i^2}{8E_i^2}\right)
        \; \tilde{\zeta}_{\dot{\alpha}}^-(p) \; \zeta^{+}_{I}(k) \nonumber
     + \frac{m_i}{\sqrt{2E_i}} 
        \; \tilde{\zeta}_{\dot{\alpha}}^+(p) \; \zeta^{-}_{I}(k) \nonumber \\
    &=& \sqrt{2E_i}\left[\begin{array}{ll}
     ( 1 - m_i^2/8E_i^2 )c   & -(m_i/2E_i) s \\ 
     ( 1 - m_i^2/8E_i^2 )s^* & +(m_i/2E_i) c
    \end{array}\right], \label{eq:[i|^I expanded in m}
\end{eqnarray}
where the expressions in square brackets are matrices in $\alpha$ and $I$ and $\dot{\alpha}$ and $I$, respectively. From this, we obtain, 
\begin{eqnarray} 
    \langle\mathbf{i j}\rangle \! &=& \!\!
    \left[\begin{array}{cc} \displaystyle
    \left(1-\frac{m_i^2}{8E_i^2}-\frac{m_j^2}{8E_j^2}\right)\langle i j\rangle & \displaystyle
    \frac{m_j}{\sqrt{2E_j}}\langle i\zeta^-_j\rangle \\ \displaystyle
    \frac{m_i}{\sqrt{2E_i}}\langle \zeta^-_i j\rangle & \displaystyle
    \frac{m_im_j}{4E_iE_j}\lbrack i j\rbrack
    \end{array}\right],
    \label{eq:<ij> expanded in m} \\
    \lbrack\mathbf{i j}\rbrack_{IJ} \! &=& \!\!
    \left[\begin{array}{cc} \displaystyle
    \left(1-\frac{m_i^2}{8E_i^2}-\frac{m_j^2}{8E_j^2}\right)\lbrack i j\rbrack & \displaystyle
    \frac{m_j}{\sqrt{2E_j}}\lbrack i\tilde{\zeta}^+_j\rbrack \\ \displaystyle
    \frac{m_i}{\sqrt{2E_i}}\lbrack \tilde{\zeta}^+_i j\rbrack & \displaystyle
    \frac{m_im_j}{4E_iE_j}\langle i j\rangle
    \end{array}\right],
\end{eqnarray}
also, to quadratic order, where the matrices have two SU($2$) indices. Of course, if the $i$th-particle is massless, then we just have the first row and if $j$ is massless, we just have the first column.   Higher-order expansions may also prove important when loop diagrams are considered. The spin indices on the $\langle\mathbf{ij}\rangle$ are implicit and upper, however we have made the spin indices on the $\lbrack\mathbf{ij}\rbrack$ explicit, because they are still lower and we will need to raise them to satisfy our convention of the implicit spin indices being upper and also to find correct expressions when vertices include both angle and square brackets.  The spin indices can be raised with epsilon tensors as usual, giving us $\lbrack\mathbf{ij}\rbrack^{IJ}=\lbrack\mathbf{ij}\rbrack_{KL}\epsilon^{KI}\epsilon^{LJ}$ or
\begin{equation}
    \lbrack\mathbf{i j}\rbrack =
    \left[\begin{array}{cc} \displaystyle
        \frac{m_im_j}{4E_iE_j}\langle ij\rangle & \displaystyle
        -\frac{m_i}{\sqrt{2E_i}}\lbrack\tilde{\zeta}_i^+j\rbrack \\ \displaystyle
        -\frac{m_j}{\sqrt{2E_j}}\lbrack i\tilde{\zeta}^+_j\rbrack & \displaystyle
        \left(1-\frac{m_i^2}{8E_i^2}-\frac{m_j^2}{8E_j^2}\right)\lbrack i j\rbrack
    \end{array}\right].
    \label{eq:[ij] expanded in m}
\end{equation}

In some cases, we will have these brackets to a higher power (in the $W^+W^-\gamma$ vertex, for example).  So, we note that,
\begin{equation}
    \langle\mathbf{12}\rangle^2 = 
      \frac{1}{2} 
      \left(
         \langle \mathbf{12} \rangle^{I_1 J_1}
         \langle \mathbf{12} \rangle^{I_2 J_2} + \langle\mathbf{12}\rangle^{I_1 J_2}
         \langle \mathbf{12} \rangle^{I_2 J_1}
      \right).
\end{equation}
The right side is what is implicitly meant by the left side, SU($2$) indices are implicitly symmetrized.  Since this has two independent indices, we can write it as a matrix.  Because each type of spin index is symmetrized, it takes 3 independent values.  Therefore, this can be written as a $3\times3$ matrix.  These rows and columns are for the three spins of each $1$-spin boson. We find,
\begin{equation}
    \langle\mathbf{i j}\rangle^2 =
    \left[
    \begin{array}{ccc} \displaystyle
    \left( 1 - \frac{m_i^2}{4E_i^2} - \frac{m_j^2}{4E_j^2}
    \right) \langle ij \rangle^2 & \displaystyle
    \frac{m_j}{\sqrt{2E_j}} \langle ij \rangle \langle i \zeta_j^- \rangle & \displaystyle
    \frac{m_j^2}{2E_j} \langle i \zeta_j^- \rangle^2 \\ \displaystyle
    \frac{m_i}{\sqrt{2E_i}}\langle ij\rangle\langle\zeta_i^- j\rangle & \displaystyle
    \frac{m_im_j}{8E_iE_j} \langle ij \rangle \lbrack ij \rbrack
    +\frac{m_im_j}{2\sqrt{4E_iE_j}}\langle i\zeta_j^-\rangle\langle\zeta_i^- j\rangle & 0 \\ \displaystyle
    \frac{m_i^2}{2E_i}\langle\zeta_i^-j\rangle^2 & 0 & 0
    \end{array}\right],
    \label{eq:<12>^2 expanded in m}
\end{equation}
again, to quadratic order.  We find a similar result for $\lbrack\mathbf{ij}\rbrack^2$, switching angle and square brackets and making the replacement $\zeta^-\to\tilde{\zeta}^+$. We then need to rearrange to find the spin indices to be upstairs, giving
\begin{equation}
    \lbrack\mathbf{i j}\rbrack^2 =
    \left[
    \begin{array}{ccc} \displaystyle
    0 & \displaystyle
    0 & \displaystyle
    \frac{m_i^2}{2E_i}\lbrack\tilde{\zeta}_i^+j\rbrack^2 \\ \displaystyle
    0 & \displaystyle
    \frac{m_im_j}{8E_iE_j} \langle ij \rangle \lbrack ij \rbrack
    +\frac{m_im_j}{2\sqrt{4E_iE_j}}\lbrack i\tilde{\zeta}_j^+\rbrack\lbrack\tilde{\zeta}_i^+ j\rbrack & \displaystyle
    -\frac{m_i}{\sqrt{2E_i}}\lbrack ij\rbrack\lbrack\tilde{\zeta}_i^+ j\rbrack\\ \displaystyle
    \frac{m_j^2}{2E_j} \lbrack i \tilde{\zeta}_j^+ \rbrack^2 & \displaystyle
    -\frac{m_j}{\sqrt{2E_j}} \lbrack ij \rbrack \lbrack i \tilde{\zeta}_j^+ \rbrack& \displaystyle
    \left( 1 - \frac{m_i^2}{4E_i^2} - \frac{m_j^2}{4E_j^2}
    \right) \lbrack ij \rbrack^2
    \end{array}\right].
    \label{eq:[12]^2 expanded in m}
\end{equation}

On the other hand, if we multiply $\langle\mathbf{ij}\rangle$ by $\lbrack\mathbf{ij}\rbrack$, we have four separate terms in our symmetrization, 
\begin{equation}
    \langle\mathbf{ij}\rangle\lbrack\mathbf{ij}\rbrack =
    \frac{1}{4}\left(
        \langle\mathbf{ij}\rangle^{I_1J_1}\lbrack\mathbf{ij}\rbrack^{I_2J_2} +
        \langle\mathbf{ij}\rangle^{I_1J_2}\lbrack\mathbf{ij}\rbrack^{I_2J_1} +
        \langle\mathbf{ij}\rangle^{I_2J_1}\lbrack\mathbf{ij}\rbrack^{I_1J_2} +
        \langle\mathbf{ij}\rangle^{I_2J_2}\lbrack\mathbf{ij}\rbrack^{I_1J_1}
    \right).
\end{equation}
With this, we obtain the $3 \times 3$ matrix,
\begin{equation}
    \langle\mathbf{ij}\rangle\lbrack\mathbf{ij}\rbrack =
    \left[
    \begin{array}{ccc} \displaystyle
       \frac{m_im_j}{4E_iE_j} \langle ij \rangle^2
     & \displaystyle
        -\frac{m_i}{2\sqrt{2E_i}}\langle ij\rangle\lbrack \tilde{\zeta}_i^+j\rbrack
     & \displaystyle
       -\frac{m_im_j}{\sqrt{4E_iE_j}}\langle i\zeta_j^-\rangle\lbrack\tilde{\zeta}_i^+j\rbrack \\ \displaystyle
       -\frac{m_j}{2\sqrt{2E_j}}\langle ij\rangle\lbrack i\tilde{\zeta}_j^+\rbrack
     & \begin{array}{c} \displaystyle
         -\frac{m_i^2}{8E_i}
          \langle \zeta_i^- j\rangle 
          \lbrack \tilde{\zeta}_i^+ j \rbrack
         -\frac{m_j^2}{8E_j}
          \langle i\zeta_j^- \rangle
          \lbrack i\tilde{\zeta}_j^+ \rbrack \\ \displaystyle \qquad
        +\left(
          1 - \frac{m_i^2}{4E_i^2} - \frac{m_j^2}{4E_j^2}
         \right) \langle ij \rangle \lbrack ij \rbrack \\ 
       \end{array}
     & \displaystyle
     \frac{m_j}{2\sqrt{2E_j}}\lbrack ij\rbrack\langle i\zeta_j^-\rangle\\ \displaystyle
       -\frac{m_im_j}{\sqrt{4E_iE_j}}\langle\zeta_i^-j\rangle\lbrack i\tilde{\zeta}_j^+\rbrack 
     & \displaystyle
     \frac{m_i}{2\sqrt{2E_i}}\lbrack ij\rbrack\langle\zeta_i^-j\rangle & \displaystyle
     \frac{m_im_j}{4E_iE_j}\lbrack ij\rbrack^2
    \end{array}
    \right],
    \label{eq:<12>[12] expanded in m}
\end{equation}
to quadratic order in the masses. It is, of course, imperative when implementing these spinor combinations within a scattering amplitude, that these symmetry factors are not neglected. 

%%%%%%%%%%%%%%%%%%%%%%%%%%%%%%%%%%%%%%%%%%%%%%%%%%%%%%%%%%
%       High-Energy Limit Expectation for Vertices
%%%%%%%%%%%%%%%%%%%%%%%%%%%%%%%%%%%%%%%%%%%%%%%%%%%%%%%%%%
\section{\label{app:HE 3-point}Massless 3-Point Vertices}
In this appendix, we briefly review what we expect to find for the various 3-point vertices in the high-energy limit.  They can be constructed purely based on their transformation properties and the dimensionality of the vertex.  

Note that because all three particles are massless in the high-energy limit, we have,
\begin{equation}
p_1^2 = p_2^2 = p_3^2 = 0.
\end{equation}
But, because of momentum conservation $p_1+p_2+p_3=0$, with all momenta ingoing (or all outgoing).  Therefore, $p_i=-(p_j+p_k)$, where $i,j,k$ are any combination of $1,2,3$.  Then,
\begin{equation}
p_i^2 = (p_j+p_k)^2 = p_j^2+p_k^2 + 2p_j\cdot p_k.
\end{equation}
But, since $p_i^2=p_j^2=p_k^2=0$, we have,
\begin{equation}
p_j\cdot p_k = 0,
\end{equation}
for all $j$ and $k$.  However, since $-2p_j\cdot p_k=\langle jk\rangle\lbrack jk\rbrack$ at this order, we have,
\begin{equation}
\langle j k\rangle [j k] = 0,
\end{equation}
for any massless 3-point amplitude.  This means that either $\langle j k\rangle=0$ or $[j k]=0$.  We have three momenta and we need to make the choice for each combination, but it turns out that there is only one choice,
\begin{equation}
\langle 12\rangle = \langle23\rangle = \langle13\rangle = 0 \quad \mbox{or} \quad [12] = [23] = [13] = 0.
\label{eq:3-point:<ij>=0}
\end{equation}
To see that there is only one choice, note that $\langle i j\rangle=0$ implies that $|i\rangle\propto|j\rangle$ and similarly for the the square brackets.  So, suppose we had $\langle 12\rangle = \langle23\rangle = 0$ and $[13] = 0$.  Well, $\langle 12\rangle = 0$ implies that $|1\rangle\propto|2\rangle$ and $\langle23\rangle = 0$ implies $|2\rangle\propto|3\rangle$ so we actually have the first of Eq.~(\ref{eq:3-point:<ij>=0}).  All other cases can be seen in a similar way.  

So, each massless 3-point amplitude can \textit{only} be constructed from either the angle brackets \textit{or} the square brackets, but not both.

So, there are only two choices for a given 3-point amplitude.  
Either it is,
\begin{equation}
\mathcal{A}(h_1,h_2,h_3) = \langle12\rangle^{c_{12}}\langle23\rangle^{c_{23}}\langle31\rangle^{c_{31}},
\end{equation}
or,
\begin{equation}
\mathcal{A}(h_1,h_2,h_3) = [12]^{\tilde{c}_{12}}[23]^{\tilde{c}_{23}}[31]^{\tilde{c}_{31}},
\end{equation}
We know that the amplitude must transform as,
\begin{equation}
\mathcal{A}(h_1,h_2,h_3) \rightarrow e^{+i\left(h_1\theta_1+h_2\theta_2+h_3\theta_3\right)} \mathcal{A}(h_1,h_2,h_3).
\end{equation}
And, we know how the brackets transform
\begin{eqnarray}
\langle i j\rangle &\rightarrow& e^{i\left(-\theta_1/2 -\theta_2/2\right)} \langle i j\rangle,
\label{eq:<> transformation rule}\\
\lbrack i j\rbrack &\rightarrow& e^{i\left(+\theta_1/2 +\theta_2/2\right)} \lbrack i j\rbrack.\nonumber
\end{eqnarray}
So, we can solve this to obtain the values of $c_{12}$, etc.  We find that either,
\begin{equation}
\mathcal{A}(h_1,h_2,h_3) = \langle12\rangle^{h_3-h_1-h_2}\langle23\rangle^{h_1-h_2-h_3}\langle31\rangle^{h_2-h_3-h_1},
\end{equation}
or,
\begin{equation}
\mathcal{A}(h_1,h_2,h_3) = [12]^{h_1+h_2-h_3}[23]^{h_2+h_3-h_1}[31]^{h_3+h_1-h_2},
\end{equation}
up to a constant (the coupling constant).  

There is one more property that the amplitude must satisfy, the dimension (i.e. units).  If the amplitude has $n$ particles, it must be dimension $4-n$.  This is required to find the right dimensions for the scattering cross section which is measured.  So, for our $3$-point amplitude, it must have dimension 1.  Each bracket, whether $\langle\, \rangle$ or $\lbrack\, \rbrack$ has dimension $1$.  This means that the dimensions of these two amplitudes are,
\begin{equation}
d=-(h_1+h_2+h_3) \quad \mbox{and} \quad d=h_1+h_2+h_3
\end{equation}
Since this must be $1$, for non-gravitational vertices with no dimensionful parameter, we finally have the amplitude for $3$ particles,
\begin{equation}
\mathcal{A}(h_1,h_2,h_3) = 
\begin{cases} \displaystyle
      \langle12\rangle^{1+2h_3}\langle23\rangle^{1+2h_1}\langle31\rangle^{1+2h_2} \\ \displaystyle
      \hspace{1in}\text{if}\ \sum_i h_i=-1 \\ \displaystyle
      \lbrack12\rbrack^{1-2h_3}\lbrack23\rbrack^{1-2h_1}\lbrack31\rbrack^{1-2h_2} \\ \displaystyle
      \hspace{1in}\text{if}\ \sum_i h_i=1 \\ \displaystyle
      0 \hspace{0.95in} \text{otherwise}.
    \end{cases}
\end{equation}
On the other hand, for gravitational vertices, we must divide by the Planck mass, therefore, the dimension of the brackets must be $2$.  This gives us,
\begin{equation}
\mathcal{A}_{\textrm{grav}}(h_1,h_2,h_3) = 
\begin{cases} \displaystyle
      \langle12\rangle^{2+2h_3}\langle23\rangle^{2+2h_1}\langle31\rangle^{2+2h_2} \\ \displaystyle
      \hspace{1in}\text{if}\ \sum_i h_i=-2 \\ \displaystyle
      \lbrack12\rbrack^{2-2h_3}\lbrack23\rbrack^{2-2h_1}\lbrack31\rbrack^{2-2h_2} \\ \displaystyle
      \hspace{1in}\text{if}\ \sum_i h_i=2 \\ \displaystyle
      0 \hspace{0.95in} \text{otherwise}.
    \end{cases}
\end{equation}

We will also need to consider some vertices involving Goldstone bosons that are only non-zero at linear order in the mass.  Therefore, we also consider vertices that have one dimensionful parameter and only allow it to the first power in the vertex so that the vertex must have dimension 0.  In this case, we have,
\begin{equation}
\mathcal{A}_{m^1}(h_1,h_2,h_3) = m
\begin{cases} \displaystyle
      \langle12\rangle^{2h_3}\langle23\rangle^{2h_1}\langle31\rangle^{2h_2}
      \pm
      \lbrack12\rbrack^{-2h_3}\lbrack23\rbrack^{-2h_1}\lbrack31\rbrack^{-2h_2}
      \hspace{0.25in}\text{if}\ \sum_i h_i=0 \\ \displaystyle
      0 \hspace{3.15in} \text{otherwise}.
    \end{cases}
\end{equation}

We now enumerate the high-energy limit vertices required for the SM.  First, the non-gravitational vertices that have dimension 1. We find,
\begin{equation}
    \mathcal{A}(0,0,+1) = \frac{[23][31]}{[12]}, \quad 
    \mathcal{A}(0,0,-1) = \frac{\langle23\rangle\langle31\rangle}{\langle12\rangle},
    \label{eq:massless:0,0,1}
\end{equation}
\begin{equation}
    \mathcal{A}\left(0,+\frac{1}{2},+\frac{1}{2}\right) = [23], \quad 
    \mathcal{A}\left(0,-\frac{1}{2},-\frac{1}{2}\right) = \langle23\rangle,
    \label{eq:A(0,1/2,1/2)}
\end{equation}
\begin{equation}
    \mathcal{A}\left(+\frac{1}{2},-\frac{1}{2},+1\right) = \frac{[31]^2}{[12]}, \quad \mathcal{A}\left(-\frac{1}{2},+\frac{1}{2},-1\right) = \frac{\langle31\rangle^2}{\langle12\rangle},
    \label{eq:massless:1/2,-1/2,1}
\end{equation}
\begin{equation}
    \mathcal{A}(+1,+1,-1) = \frac{[12]^3}{[23][31]}, \quad 
    \mathcal{A}(-1,-1,+1) = \frac{\langle12\rangle^3}{\langle23\rangle\langle31\rangle}.
    \label{eq:3-point gluon amplitude}
\end{equation}

For the gravitational vertices,
\begin{equation}
    \mathcal{A}(0,0,+2) =\frac{1}{M_{P}} \frac{\lbrack23\rbrack^2\lbrack31\rbrack^2}{\lbrack12\rbrack^2},
    \quad
    \mathcal{A}(0,0,-2) =\frac{1}{M_{P}}
    \frac{\langle23\rangle^2\langle31\rangle^2}{\langle12\rangle^2},
    \label{eq:A(0,0,2)}
\end{equation}
\begin{equation}
    \mathcal{A}\left(+\frac{1}{2},-\frac{1}{2},+2\right) =
                \frac{1}{M_{P}} \frac{\lbrack23\rbrack\lbrack31\rbrack^3}{\lbrack12\rbrack^2},
    \quad
    \mathcal{A}\left(-\frac{1}{2},+ \frac{1}{2},-2\right) =\frac{1}{M_{P}}
    \frac{\langle23\rangle\langle31\rangle^3}{\langle12\rangle^2},
    \label{eq:A(1/2,-1/2,2)}
\end{equation}
\begin{equation}
    \mathcal{A}(+1,-1,+2) =\frac{1}{M_{P}} \frac{\lbrack31\rbrack^4}{\lbrack12\rbrack^2},
    \quad
    \mathcal{A}(-1,+1,-2) =\frac{1}{M_{P}}
    \frac{\langle31\rangle^4}{\langle12\rangle^2},
    \label{eq:A(1,-1,2)}
\end{equation}
\begin{equation}
    \mathcal{A}(+2,-2,+2) =\frac{1}{M_{P}} \frac{\lbrack31\rbrack^6}{\lbrack12\rbrack^2\lbrack23\rbrack^2},
    \quad
    \mathcal{A}(-2,+2,-2) =\frac{1}{M_{P}}
    \frac{\langle31\rangle^6}{\langle12\rangle^2\langle23\rangle^2}.
\end{equation}
As expected, the gravitational vertices are the square of the non-gravitational vertices with half the helicities.

Finally, for the non-gravitational vertices that are only non-zero at first order in a dimensionful parameter.  We use $m$ to represent the appropriate scale.  We have,
\begin{equation}
    \mathcal{A}(0,0,0) = m,
\end{equation}
\begin{equation}
    \mathcal{A}\left(+\frac{1}{2},-\frac{1}{2},0\right) = m
    \left(\frac{\lbrack31\rbrack}{\lbrack23\rbrack} \pm
        \frac{\langle23\rangle}{\langle31\rangle}\right),
\end{equation}
\begin{equation}
    \mathcal{A}(+1,-1,0) = m
    \left(\frac{\lbrack31\rbrack^2}{\lbrack23\rbrack^2} \pm
        \frac{\langle23\rangle^2}{\langle31\rangle^2}\right).
\end{equation}
Interestingly, the terms are the square of the case with half the helicities.  

Before ending this appendix, we review how the rules discussed here allow the determination of the helicity of the particles involved in a massless vertex.  As described above, each massless vertex will include either angle brackets or square brackets, but not both.  Furthermore, we know that angle brackets transform like a $-1/2$-helicity particle while square brackets transform like a $+1/2$-helicity particle [see Eq.~(\ref{eq:<> transformation rule})].  Since this transformation is a phase, we see that if a particle appears in both the numerator and denominator of a vertex, the full transformation for that vertex involves the difference of the two powers.  For example,
\begin{equation}
    \frac{\langle i j \rangle^p}{\langle k i \rangle^q} \to e^{-i(p-q)\theta_i/2}e^{-i p\theta_j/2}e^{+i q\theta_k/2} \frac{\langle i j \rangle^p}{\langle k i \rangle^q}
    \quad \mbox{and} \quad
    \frac{\lbrack i j \rbrack^p}{\lbrack k i \rbrack^q} \to e^{+i(p-q)\theta_i/2}e^{+i p\theta_j/2}e^{-i q\theta_k/2} \frac{\lbrack i j \rbrack^p}{\lbrack k i \rbrack^q} .
\end{equation}
Focusing on particle~$i$, we see that this ratio transforms as particle with helicity $h=-(p-q)/2$ for angle brackets and $h=+(p-q)/2$ for square brackets.  So, for each massless vertex, we simply take the power of each particle in the numerator minus the power of that particle in the denominator and multiply by $-1/2$ if angle brackets or by $+1/2$ is square brackets, and this gives the helicity of the particle.  To clarify this process, consider the vertex,
\begin{equation}
    \frac{\lbrack23\rbrack\lbrack31\rbrack}{\lbrack12\rbrack}.
\end{equation}
Beginning with particle~$1$, we see that its power in the numerator is $1$ and its power in the denominator is $1$, so that the difference is $0$.  From this, we learn that particle~$1$ must have helicity-$0$.  We find exactly the same feature for particle~$2$, so it is also helicity-$0$.  Particle~$3$, on the other hand, has two powers in the numerator but none in the denominator.  Since this vertex uses square brackets, we learn that the helicity of particle~$3$ is $h_3=+(2-0)/2=+1$.  This agrees with our description of Eq.~(\ref{eq:massless:0,0,1}).  We will consider one more example.  Consider the vertex,
\begin{equation}
    \frac{\langle23\rangle\langle31\rangle^3}{\langle12\rangle^2}.
\end{equation}
We see particle~$1$ to the third power in the numerator and the second power in the denominator.  Since this vertex uses angle brackets, we find that the helicity of particle~$1$ is $h_1=-(3-2)/2=-1/2$.  We find particle~$2$ to the first power in the numerator and the second power in the denominator, giving us $h_2=-(1-2)/2=+1/2$.  Finally, we find particle~$3$ to the fourth power in the numerator and not at all in the denominator, giving $h_3=-(4-0)/2=-2$.  All together, this agrees with our description of Eq.~(\ref{eq:A(1/2,-1/2,2)}).

\end{widetext}

\end{document}